\documentclass[a4paper,11pt,titlepage]{article}
\usepackage[left=2.5cm,right=2.5cm,top=1cm,bottom=1cm,includeheadfoot]{geometry}
\usepackage{ngerman}
\usepackage[latin1]{inputenc}
\usepackage{graphicx}
\usepackage{subfig}
\usepackage{longtable}
\usepackage{paralist}

\begin{document}
\selectlanguage{german}

\begin{figure}[htbp]
\begin{minipage}[t]{3.3cm}
\vspace{0pt}
\begin{flushleft}
\includegraphics[width=3.3cm]{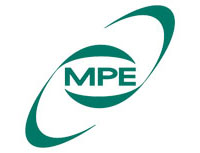}
\end{flushleft}
\end{minipage}
\hfill
\begin{minipage}[t]{5cm}
\end{minipage}
\hfill
\begin{minipage}[t]{1cm}
\vspace{3mm}
\includegraphics[width=3.3cm]{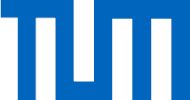}
\end{minipage}
\hfill
\begin{minipage}[t]{2cm}
\vspace{0pt}
\begin{flushright}
\includegraphics[width=2cm]{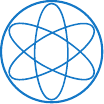}
\end{flushright}
\end{minipage}
\end{figure}

\begin{minipage}[t]{6cm}
\begin{flushleft} Max-Planck-Institut für Extraterrestrische Physik
\end{flushleft}
\end{minipage}
\hfill
\begin{minipage}[t]{7cm}
\begin{flushright} Technische Universität München Fakultät für Physik
\end{flushright}
\end{minipage}

\vspace{3cm}

\begin{center}\Large Abschlussarbeit im Bachelorstudiengang Physik\end{center}
\vspace{2cm}
\begin{center}
\LARGE{Messung der Masse des Schwarzen Loches in LMC X-3}\\
\vspace{1cm}
\Large{Measuring the mass of the black hole in LMC X-3}\\\vspace{3cm}\Large{Dominik Gräff}\\\vspace{7cm} 
\large{Abgabetermin: 24.07.2014}
\end{center}
\thispagestyle{empty}
\newpage~

\vspace{22cm}
\begin{flushleft} {Erstgutachter (Themensteller): PD \ Dr.\ J. Greiner\\Zweitgutachter: Prof.  Dr.\ S. Schönert}
\end{flushleft}
\thispagestyle{empty}
\newpage

\clearpage
\pagenumbering{Roman}

\begin{center}
\section*{Zusammenfassung}
\end{center}

Das Röntgen-Doppelsternsystem LMC X-3 ist auf Grund seiner hohen Gesamtmasse ein Kandidat für ein Doppelsternsystem mit Schwarzem Loch als Hauptkomponente. Die Masse dieses Schwarzen Lochs und die Bahnneigung des Gesamtsystems zu bestimmen, ist Ziel dieser Arbeit. Dazu wurden Daten des GROND Instruments am MPI/ESO 2.2m Teleskop auf dem Berg La Silla (Chile) verwendet, welche im Zeitraum vom 20. Oktober 2007 bis zum 13. Dezember 2013 aufgenommen wurden. Auf Grundlage dieser Daten, galt es ein Modell zu finden, welches LMC X-3 möglichst gut beschreibt und woraus sich die Masse und Bahnneigung ableiten lassen. Für die Modellbildung wurde das Programm \glqq XRbinary\grqq von Edward L. Robinson benutzt. Anhand dieses Modells wurde die Bahnneigung des Systems auf $(68_{-3}^{+2})°$ und die Masse des Schwarzen Lochs zu $(7.15\pm0.65) \ M_{\odot}$ bestimmt. Aus genauen Ergebnissen der Bahnneigung und schließlich der Masse lassen sich weiterführend Abschätzungen der Rotation des Schwarzen Lochs von LMC X-3 durchführen und die Massenverteilung der Schwarzen Löcher in Röntgen-Doppelsternsystemen weiter ergänzen.

\clearpage
\vspace*{-2cm}
\tableofcontents
\clearpage

\pagenumbering{arabic}

\section{Einleitung}
\subsection{Röntgen-Doppelsternsysteme}
Röntgen-Doppelsternsysteme gehören zu den hellsten Objekten am Himmel im Röntgenbereich und waren Ende der 70er Jahre mit dem Aufblühen der Röntgenastronomie die Objekte, auf denen der Fokus der Astronomen lag. Ein Röntgen-Doppelsternsystem besteht entweder aus einem Neutronenstern, oder einem Schwarzen Loch, welches Materie eines Begleitsterns akkretiert. Die erste extrasolare Röntgenquelle welche entdeckt wurde, war Sco X-1 im Sternbild Skorpion (Giacconi et al. 1962), die später als Röntgen-Doppelsternsystem erkannt wurde (Gursky et al. 1966).\\
Das erste Röntgen-Doppelsternsystem, welches als ein Schwarzes Loch mit Begleitstern identifiziert werden konnte, war dann Cyg X-1 im Sternbild Schwan (Webster et al. 1972). Seitdem ist die These, dass Röntgen-Doppelsternsysteme mit Schwarzem Loch (BHBs - Black-Hole-Binaries) existieren, stark untermauert worden.\\
Im Gegensatz zu  Doppelsternsystemen, welche aus Neutronensternen bestehen und harte Röntgenstrahlen emittieren, fallen die BHBs wegen ihres sehr weichen Röntgenspektrums auf (Tanaka et al. 1995). Solche BHBs zu beobachten, bringt Aufschlüsse über das Verhalten von Galaxien und ihren supermassiven Schwarzen Löchern im Zentrum, da das System Schwarzes Loch und Akkretionsscheibe (siehe Kapitel \ref{sec:Grundlagen}) quasi ein Modell einer Galaxie darstellt, bei dem die Winkelgeschwindigkeit am Innenrand der Scheibe circa $10^6$ mal größer ist, was somit wesentlich besser untersucht werden kann. Dies resultiert aus den Kepler'schen Gesetzen, welche mit der Masse skalieren. Somit ist die Zeitskala der Rotation wesentlich kleiner als bei Galaxien. Ebenfalls lässt sich das Verhalten von Materie in der Nähe eines Schwarzen Loches studieren, genauso wie auch die Krümmung von Licht.\\
Des Weiteren gibt eine eingehende Untersuchung der Akkretionsscheibe Aufschluss über die Entstehung von Sternen und Sternsystemen, welche unter Umständen auch Planeten beheimaten können, da die Verhältnisse in der Akkretionsscheibe einem jungen stellaren Objekt, bei dem ein protostellarer Kern von einer Akkretionsscheibe umgeben ist, ähneln.\\
Am wichtigsten für die Schwarzloch-Forschung ist allerdings, dass mit Hilfe der Ausdehnung der Akkretionsscheibe Aussagen über die Rotation des Schwarzen Lochs und die damit einhergehende Verformung des Ereignishorizonts auf Grund der Raumzeitkrümmung getroffen werden können. Bei einem nicht-rotierendem Schwarzen Loch reicht der Innenrand der Akkretionsscheibe maximal bis etwa sechs Schwarzschildradien an das Zentrum des Schwarzen Lochs heran, wohingegen bei einem rotierenden der Innenrand bis zum Ereignishorizont reicht (Bardeen et al. 1972).\\

\vspace{-0.5cm}

\subsection{Warum sind Massen von Schwarzen Löchern wichtig?}
Bei der Massenverteilung der beobachteten stellaren Schwarzen Löcher in Doppelsternsystemen fällt auf, dass die Häufigkeit nicht über alle theoretisch möglichen Massen konstant ist. Zwischen der Tolman-Oppenheimer-Volkoff-Grenze, welche bei circa drei Sonnenmassen liegt, und circa fünf Sonnenmassen finden sich fast keine Schwarzen Löcher. Dagegen hat die Häufigkeitsverteilung ihr Maximum zwischen sieben und acht Sonnenmassen und fällt schnell bei zehn Sonnenmassen ab (Farr et al. 2011). Das in dieser Arbeit beobachtete Objekt liegt an der unteren Grenze dieser Massenverteilung und hat eine Masse von ungefähr sieben Sonnenmassen. Der Grund für die auftretende Häufigkeitsverteilung ist bislang unbekannt (Özel et al. 2010). Er könnte in einem physikalischen Prinzip oder Vorgang liegen, oder schlicht mit den beobachteten Schwarzen Löchern zu tun haben. Dies herauszufinden und die Häufigkeitsverteilung durch mehr untersuchte Objekte zu ergänzen, ist ein wichtiger Punkt. Der andere ist, dass es bis heute keinen Test der allgemeinen Relativitätstheorie im Grenzfall eines starken Gravitationsfeldes gibt (Tanaka et al. 1995), weshalb es wichtig, ist die Masse von Schwarzen Löchern möglichst exakt zu bestimmen und in BHBs Aussagen über den Innenrand der Akkretionsscheibe zu machen, um die Vorhersagen der Relativitätstheorie zu bestätigen.

\section{Grundlagen}\label{sec:Grundlagen}
\subsection{Doppelsternsysteme mit Schwarzem Loch}

\subsubsection{Roche-Lobe}\label{sec:Roche}
\begin{figure}[h]
\begin{center}
\includegraphics[width=10cm]{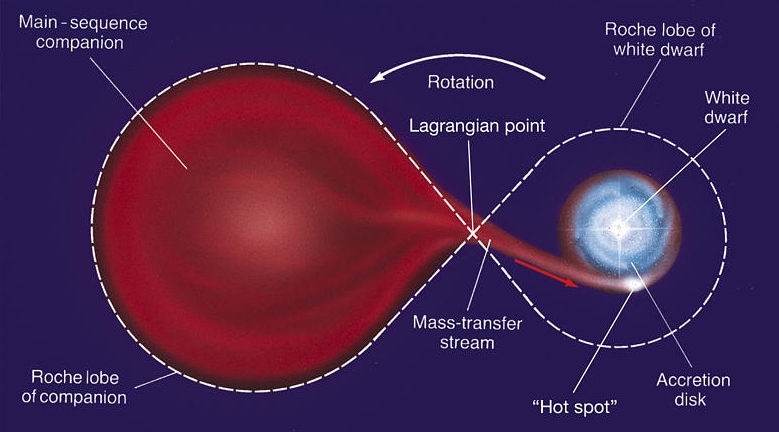}
\caption{Schematische Darstellung des Roche-Lobes eines Doppelsternsystems, wobei der Begleitstern hier mehr Masse als die Hauptkomponente hat und seinen Roche-Lobe ausfüllt. (Quelle: Ciardullo: http://www2.astro.psu.edu/users/rbc/a1/lec16n.html)}
\label{fig:Roche}
\end{center}
\end{figure}

Roche-Lobe, oder zu Deutsch Roche-Grenze, wird die Äquipotentialfläche in einem gravitativ wechselwirkendem Zweikörpersystem genannt, die durch den inneren Lagrange Punkt $L_1$ verläuft. Dieser liegt auf der Verbindungsachse zwischen den Schwerpunkten der beiden Himmelskörper. Der Roche-Lobe ist also sozusagen die, vom Zentrum der Himmelskörper aus gesehen, erste gemeinsame Potentialfläche. In der Fernfeldnäherung wirkt das Gravitationspotential der beiden Objekte dann wieder kugelsymmetrisch. Wenn der Begleitstern in einem Doppelsternsystem größer ist als sein Roche-Lobe (Der Stern ist dann nicht mehr kugelförmig, sondern hat die Form des Roche-Lobe.) kann Materie auf das kompakte Objekt überfließen (eng: Roche-lobe overflow). In Abbildung \ref{fig:Roche} ist schematisch für ein Doppelsternsystem mit einem weißen Zwerg als Hauptkomponente der Roche-Lobe eingezeichnet, wobei hier der Begleitstern schwerer als der weiße Zwerg ist und daher sein Roche-Lobe größer ist im Vergleich zum Roche Lobe des weißen Zwerg. Ebenfalls schön erkennbar sind die Komponenten eines Röntgen-Doppelsternsystems, wie sie im nächsten Kapitel noch genauer beschrieben werden. Zusätzlich kann man die Rotation ablesen, die natürlich aus Drehimpulserhaltungsgründen mit dem Drehsinn der Akkretionsscheibe einhergeht.

\subsubsection{Komponenten}
Doppelsternsysteme mit Schwarzem Loch stellen die Reste einer Supernova vom Typ II dar. Solche Röntgen-Doppelsternsysteme sind im Röntgenbereich oft nur vorübergehend für einige Tage aktiv und bleiben dann für einige hundert Tage in einem ruhigen Zustand (White et al. 1984). Im aktiven Zustand kann ihre Leuchtkraft um $10^4-10^5$ steigen. Diese Aktivitäten können aus einer Instabilität in der Akkretionsscheibe oder einem vermehrten Massenüberfluss vom Begleitstern stammen. Generell kann man BHBs vereinfacht durch vier Hauptkomponenten beschreiben. Diese sind:
\begin{itemize}
\item \textbf{Scharzes Loch:} (hier auch oft Primary genannt) Das schwarze Loch besitzt im Grunde nur zwei grundlegende Eigenschaften; seine Masse und sein Spin, was prinzipiell dem Drehimpuls entspricht (Reich 2013). Durch diese beiden Parameter lässt sich der Ereignishorizont berechnen, der, falls das Schwarze Loch schnell rotiert die Form einer gequetschten Kugel erhält. Der Spin des Schwarzen Lochs hat Auswirkung auf die Form der Akkretionsscheibe, beziehungsweise auf ihren inneren Radius
\item \textbf{Begleitstern:} (hier auch oft Secondary genannt) Der Begleitstern ist das Objekt, welches in den Röntgen-ruhigen Zuständen die Helligkeit des BHBs im sichtbaren Bereich dominiert. Er füllt meist mehr oder minder seinen Roche-Lobe aus und verliert über die Zeit an Masse, welche das Schwarze Loch akkretiert. Der Stern wird hauptsächlich über seine Spektralklasse, seine Masse und seinen Radius charakterisiert.
\item \textbf{Akkretionsscheibe:} (hier auch oft Scheibe genannt) Die Akkretionsscheibe ist der Teil des Systems, in dem die Masse, die vom Begleitstern abgegeben wird oder überfließt, sich sammelt, ehe sie ins Schwarze Loch stürzt. Sie hat aus Drehimpulserhaltungsgründen den gleichen Rotationssinn wie auch der Begleitstern um das Schwarze Loch und ist im Allgemeinen ein hochkomplexes Gebilde, welches sowohl optisch dicht als auch dünn und deren Ausdehnung durchaus unterschiedlich in verschiedenen Systemen sein kann. Bei Roche-Lobe füllenden Begleitsternen verläuft der Materiestrom vom Stern durch den inneren Lagrange-Punk, rotiert viskos um das Schwarze Loch und läuft schließlich spiralförmig hinein. Durch die Umwandlung von potentieller Energie im Gravitationsfeld entsteht die charakteristische Röntgenstrahlung. Der Außenrand der Scheibe wird durch Gezeitenkräfte limitiert. Der Temperaturverlauf in der Scheibe folgt einem Potenzgesetz $T=T_{max}R^{-\frac{3}{4}}$ mit dem Scheibenradius R und der Temperatur $T_{max}$ nahe dem Scheibeninnenrand.
\item \textbf{Hot Spot:} Der Hot Spot beschreibt die Stelle, an welcher der Massenstrom vom Stern auf die Akkretionsscheibe trifft. Er entsteht durch die Reibungsenergie des Materiestroms mit der Scheibe und emittiert je nach Akkretionsrate vom optischen bis in den UV-Bereich.
\end{itemize}

\begin{figure}[h!]
\begin{center}
\includegraphics[width=8cm]{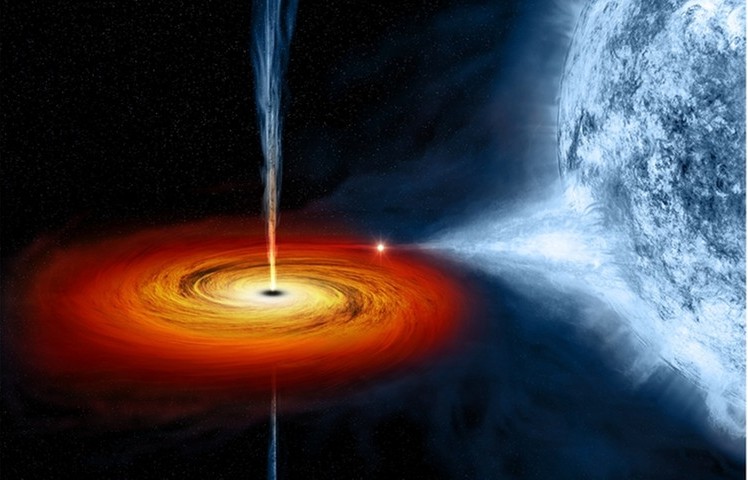}
\caption{Künstlerische Darstellung des Doppelsternsystems Cygnus X-1, welcher der erste Kandidat für ein Schwarzes Loch in einem Doppelsternsystem war. Der Begleitstern von LMC X-3 ist etwa ein Viertel mal so groß wie der von Cygnus X-1. (Quelle: Weiss: http://chandra.harvard.edu/photo/2011/cygx1/html)}
\label{fig:binary}
\end{center}
\end{figure}

In Abbildung \ref{fig:binary} ist eine künstlerische Darstellung eins Röntgen-Doppelsternsystems zu sehen. Ebenfalls lassen sich die oben beschriebenen Komponenten erkennen. Zusätzlich dazu ist ein Materie-Jet senkrecht zur Akkretionsscheibe zu sehen.

\vspace{-0.25cm}

\subsubsection{LMXBs und HMXBs}
\begin{figure}[h]
\begin{center}
\includegraphics[width=8cm]{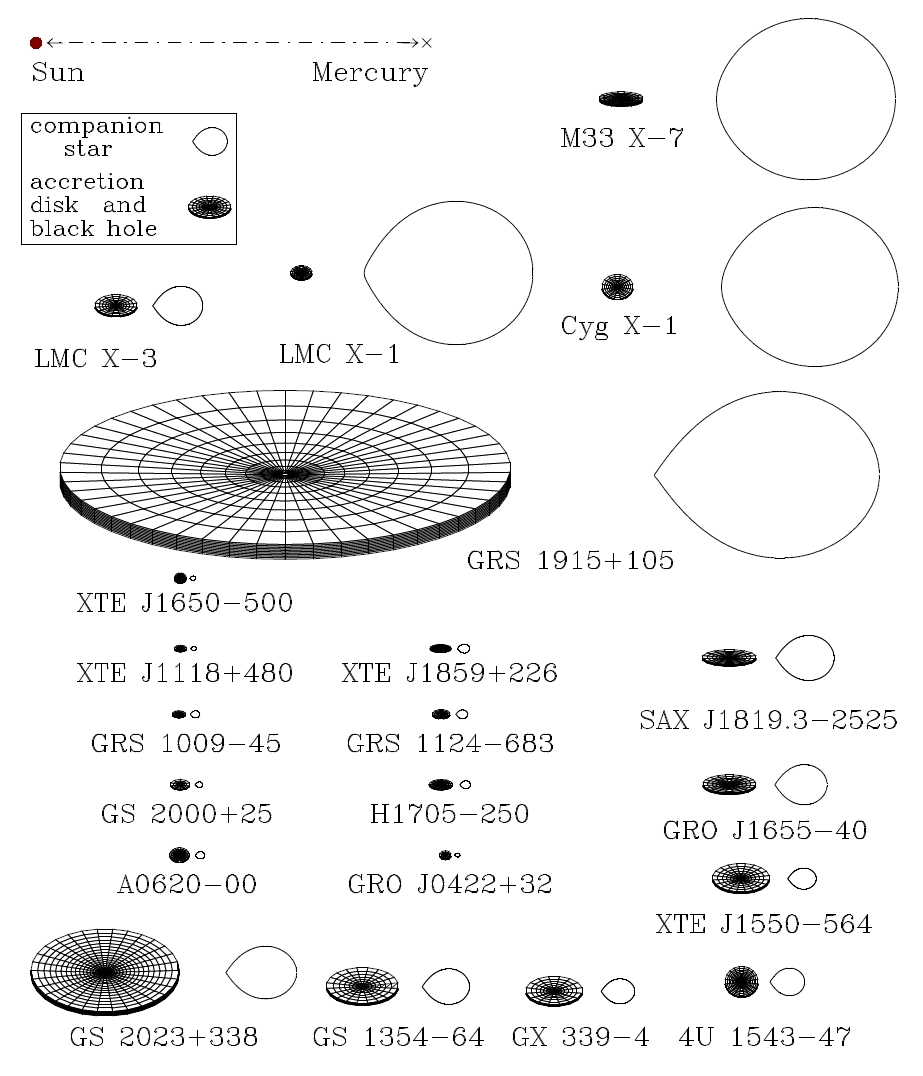}
\caption{Größenvergleich verschiedener Doppelsternsysteme mit Schwarzem Loch. Oben links ist zum Vergleich der Abstand Sonne-Merkur eingezeichnet; unter der Legende ist LMC X-3 zu finden. (Quelle: McClintock et al. 2011)}
\label{fig:Groessenvergleich}
\end{center}
\end{figure}

Man unterscheidet grob, und hierbei sind die Grenzen oft sehr verwaschen, zwischen sogenannten \glqq Low-Mass X-Ray Binaries\grqq \ (LMXBs), deren Begleitstern vom Spektraltyp A und später ist und \glqq High-Mass X-Ray Binaries\grqq \ (HMXBs), deren Begleitstern vom Spektral O oder B ist. Der spektrale Typ des Begleitsterns bestimmt die Art des Massentransfers zum Schwarzen Loch und seine direkte Umgebung (White et al. 1995).\\
\begin{itemize}
\item In LMXBs ist der stellare Wind nicht ausreichend um die beobachteten Röntgenstrahlen zu erklären. Demzufolge findet der Massentransfer hauptsächliche dadurch statt, dass der Begleitstern größer als sein Roche-Lobe, also die Fläche des kritischen Gravitationspotentials (siehe Kapitel \ref{sec:Roche}) ist, und deshalb Masse zum Schwarzen Loch überfließt. Das sogenannte \glqq X-Ray heating\grqq, also das Beleuchten und Heizen der Akkretionsscheibe und des Sterns durch die im Innern der Scheibe entstehende Röntgenstrahlung, dominiert die Helligkeit auch im Optischen, sodass LMXBs als schwach blaue Sterne erscheinen (Bradt et al. 1983).
\item In HMXBs ist die optische Leuchtkraft des Begleitsterns in der gleichen Größenordnung oder größer als jene, welche durch die Röntgenstrahlung erzeugt wird (Conti 1978). Bei diesem Typ ist der stellare Wind so kräftig, dass das Schwarze Loch, wenn es nahe genug ist, einen Großteil davon einfangen kann, um die beobachtete Röntgenstrahlung zu erzeugen. Durch die stellaren Winde werden die Röntgenstrahlen mit niedrigen Energien teilweise absorbiert. Ergänzend kann ebenfalls ein Massentransfer dadurch stattfinden, dass der Begleitstern größer als sein Roche-Lobe ist.
\end{itemize}
In Abbildung \ref{fig:Groessenvergleich} (McClintock et al. 2011) sind verschiedene bekannte Doppelsternsysteme mit Schwarzem Loch als Primärkomponente gezeigt. Man erkennt gut, dass die beiden Hauptkomponenten sich einander sehr nahe sind (meist weniger als der Abstand Sonne-Merkur), was die kurzen Periodendauern dieser Systeme von einigen Stunden bis wenigen Tagen erklärt. Ebenfalls ist das in dieser Arbeit untersuchte System LMC X-3 zu sehen, bei dem der Begleitstern annähernd so groß ist wie der Durchmesser der Akkretionsscheibe.

\vspace{2cm}
\subsubsection{Periodizität und Massenfunktion}
Das Schwarze Loch und der Begleitstern im BHB kreisen um ihren gemeinsamen Schwerpunkt. Kreisbewegungen können deshalb angenommen werden, da die Abstände im System sehr kurz im Vergleich zu den großen Massen sind, und Gezeitenkräfte wirken, die schließlich bei solch kurzen Perioden auf Kreisbahnen führen. Die Synchronisationszeit für eine Kreisbahn beträgt in etwa $10^6$ Jahre (Kuiper et al. 1988), was viel kleiner ist als die Lebensdauer eines Hauptreihensterns. Das System kann jedoch präzedieren, was aber für diese Arbeit vernachlässigt wird, da die Präzessionsbewegung sich vermutlich auf größeren Zeitskalen abspielt und außerdem für eine Präzessionsbewegung eine starre Disk angenommen werden muss, was eher nicht auf eine Akkretionsscheibe zutrifft (Brocksopp et al. 2001). Das Doppelsternsystem folgt also dem dritten Kepler'schen Gesetz, was besagt, dass die Quadrate der Umlaufzeiten im gleichen Verhältnis zueinander stehen wie die Kuben der großen Halbachsen. Die relative Bewegung der beiden Körper zu ihren Schwerpunkt kann durch eine Koordinatentransformation in eine elliptische Bewegung des Begleitsterns um das Schwarze Loch, welches sich in einem Brennpunkt der Ellipse befindet, reduziert werden. Somit kann eine große Halbachse des Systems definiert werden. Zusammen mit dem Gravitationsgesetz ergibt sich für das Zweikörpersystem der Zusammenhang
\begin{equation}
G\cdot (M_1 + M_2)= \frac{2\pi}{P}\cdot a^3
\end{equation}
mit der Gravitationskonstanten G, den beiden Massen $M_{BH}$ und $M_S$, der Periode des Systems P und der großen Halbachse a. Wenn man nun noch die Bahngeschwindigkeit des Begleitsterns mit einbringen will, muss auch die Bahnneigung i berücksichtigt werden, da im Allgemeinen die Beobachtungsrichtung nicht in der Bewegungsebene des Doppelsternsystems liegt. Somit ergibt sich die Massenfunktion für ein Doppelsternsystem zu
\begin{equation}
f(M_S)= \frac{M_{BH} \ sin^3i}{\left(1+\frac{M_S}{M_{BH}}\right)^2}= \frac{PK^3_S}{2\pi G}
\end{equation}
mit der Masse des Begleitsterns $M_S$, des Schwarzen Loches $M_{BH}$ und der gemessenen Bahngeschwindigkeit des Begleitsterns $K_S$, was der projizierten Geschwindigkeit auf die Beobachtungsebene entspricht. Diese Massenfunktion stellt wegen
\begin{equation}
f(M_S) \ = \ \frac{M_{BH} \ sin^3i}{\left(1+\frac{M_S}{M_{BH}}\right)^2} \ \leq \ \frac{M_{BH}^3 \ sin^3i}{M_{BH}^2} \ \leq \ M_{BH} \ sin^3i \ \leq \ M_{BH}
\end{equation}
eine untere Grenze für die Masse des Schwarzen Loches dar.
Die Periode des Systems kann über die periodische Helligkeitsschankung in seiner Lichtkurve (siehe Kapitel \ref{sec:ellipsoidal}) gemessen werden. Über spektrale Messungen der Seite des Begleitsterns, welche der Akkretionsscheibe abgewandt ist, kann man den spektralen Typ des Begleitsterns herausfinden und ihn im Hertzsprung-Russel-Diagramm einordnen, woraus man seine Masse ableiten kann. Für einen Stern der stark Masse verliert und daher nicht dem Spektrum eines Hauptreihensterns entspricht, ist es aber genauer die Masse über Oberflächengravitationsmessungen zu bestimmen.\\
Durch den relativistischen Dopplereffekt werden, je nach Relativgeschwindigkeit des Begleitsterns zur Erde, seine Spektrallinien verschoben. Die Frequenzverschiebung wird über eine volle Periode zu einem willkürlichen Nullpunkt gemessen. Es entsteht dadurch eine Sinusfunktion, welche die Frequenzverschiebung über der Phase angibt. Diese kann gefittet werden, wobei die halbe Amplitude des Sinus der maximalen Frequenzverschiebung entspricht, die über den Dopplereffekt
\begin{equation}
\omega'=\omega\left(1-\frac{u}{c}cos\theta\right)
\end{equation}
mit der Relativgeschwindigkeit u in nicht-relativistischer Näherung zwischen dem Beobachtungsstandpunkt auf der Erde und dem Begleitstern verknüpft ist. Da aus relativistischen Gründen der Standpunkt auf der Erde als in Ruhe angenommen werden kann, folgt direkt die auf die Beobachtungsebene projizierte Bahngeschwindigkeit des Begleitsterns $K_S$.\\

\subsubsection{Ellipsoidale Veränderliche}\label{sec:ellipsoidal}
\begin{figure}[h]
\begin{center}
\includegraphics[width=6cm]{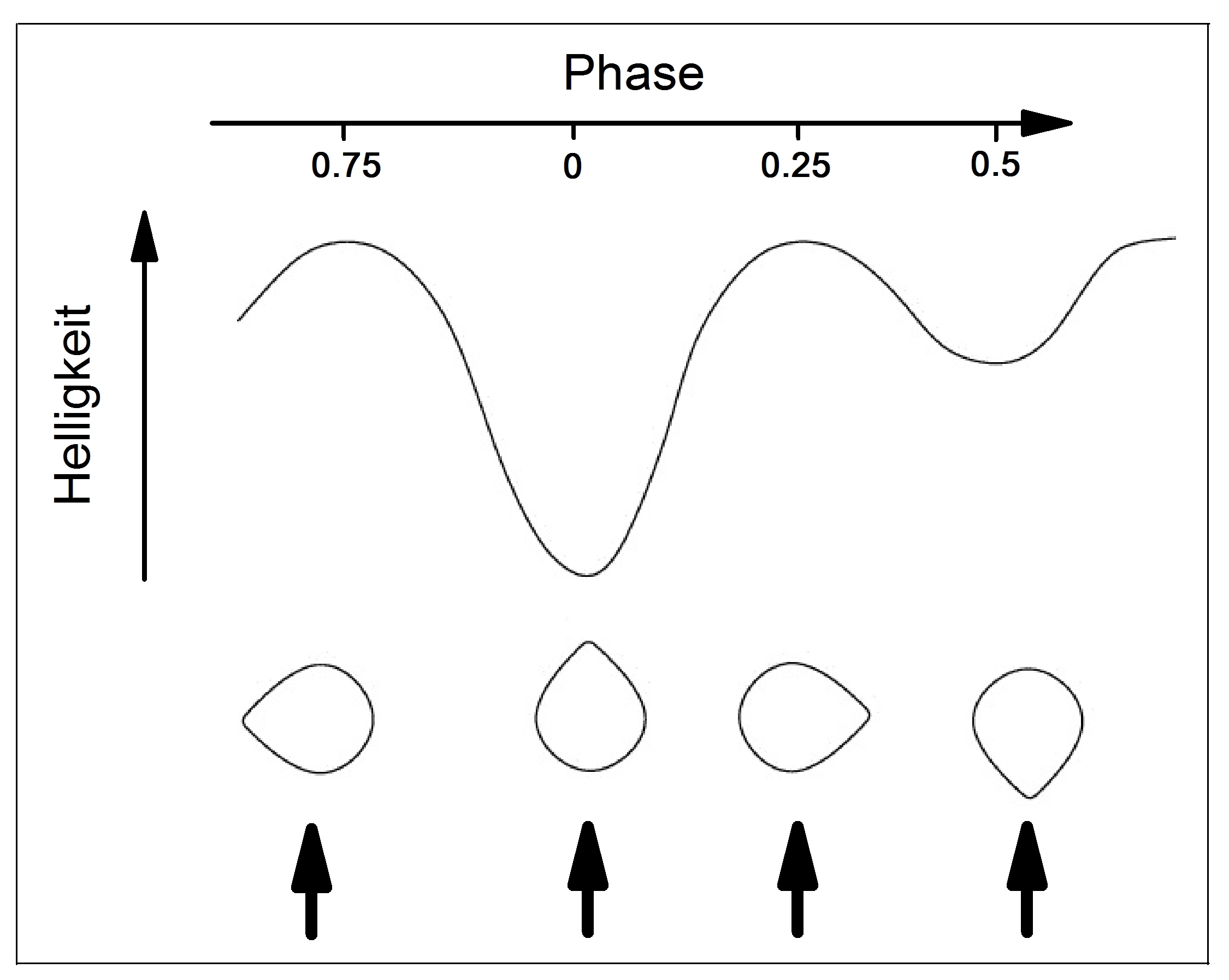}
\caption{Ellipsoidale Veränderliche zusammen mit der Blickrichtung auf den Begleitstern}
\label{fig:Ellips}
\end{center}
\end{figure}

Die Auflösung unserer optischen Teleskope ist zu gering um die Strukturen eines Doppelsternsystems auflösen zu können. Das beobachtete Objekt ist also für uns eine punktförmige Lichtquelle. Man beobachtet aber eine periodisch variierende Helligkeit des Doppelsternsystems während einer Phase der Kreisbewegung des Systems. Dies führt zu der sogenannten Ellipsoidalen Veränderlichen (engl. ellipsoidal variations). Der Grund für die Helligkeitsschwankung ist in erster Linie die Deformation des Begleitsterns. Da der Stern keine sphärische Symmetrie mehr aufweist, wird seine Energie anisotrop abgestrahlt. Wenn der Stern als Schwarzkörper angenommen wird, hängt der Fluss, welcher auf der Erde ankommt mit der relativen Fläche zusammen, auf welche man blickt. In Abbildung \ref{fig:Ellips} ist die Helligkeitsschwankung zu sehen, welche nur durch die Deformation des Begleitsterns verursacht wird, wobei in der Grafik angenommen wird, dass wir uns in der Ebene befinden, welche senkrecht auf dem Drehimpulsvektor des Systems steht (90°). Die vier dicken Pfeile am unteren Bildrand geben dabei die Blickrichtung an.

Wenn man auf die Rückseite des Sterns blickt ist also seine Helligkeit am geringsten. Die Maxima befinden sich jeweils in der Phasenlage, in welcher man den Stern beinahe als Kegel wahrnimmt. Das sanftere Minimum befindet sich hingegen in der Phasenlage, bei der man auf die \glqq Nase\grqq des Sterns schaut. Wenn nun das Doppelsternsystem um einen Winkel geneigt wird ($<90°$), der sich zwischen Drehimpulsvektor des Systems und Verbindungsachse Erde-Doppelsternsystem befindet, wird diese Schwankung im Fluss schwächer.\\
Nimmt man nun eine Akkretionsscheibe in diesem Modell hinzu, so wirkt dies quasi  in erster Näherung als konstanter Fluss, der die Minima auffüllt, also die Amplituden verringert. Genauso treten aber nun Bedeckungseffekte bei Winkeln um 90° auf, bei denen der Stern die Scheibe, oder die Scheibe den Stern verdeckt, je nachdem, welche Komponente die kältere ist.\\
Zusätzlich kann in der sogenannten Lichtkurve, welche die Helligkeit des Systems widerspiegelt der Hot Spot eine Erhöhung des Flusses mit sich bringen, je nach relativer Lage am Scheibenrand.\\
Die Nomenklatur wird im Folgenden so gewählt, dass Phase 0 dem tiefen Minimum entspricht (zweiter Pfeil von links in Abbildung \ref{fig:Ellips}), also der Situation, in welcher auf die Rückseite des Sterns geblickt wird; dem entsprechend ist Phase 0.5 der Fall in dem die Akkretionsscheibe sich zwischen Beobachter und Stern befindet. Die Phasen 0.25 und 0.75 liegen dementsprechend dazwischen.

\subsubsection{Korrelation zwischen Röntgenstrahlung und der optischen Lichtkurve}\label{sec:XRAY}
\begin{figure}[h]
\begin{center}
\includegraphics[width=7cm]{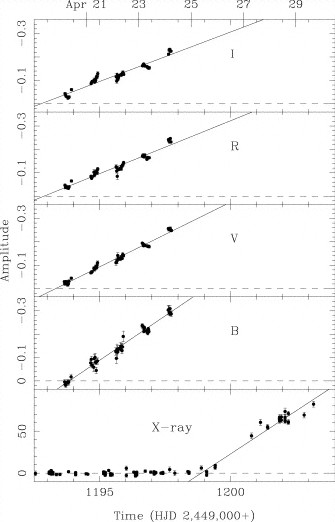}
\caption{Zeitversetzter Anstieg der Helligkeit im Röntgen-, bezogen auf den optischen Bereich bei GRO J1655-40 (Quelle: Orosz et al. 1997)}
\label{fig:Roentgenverhalten}
\end{center}
\end{figure}

Im Jahr 1997 stellten J.A. Orosz et al. eine direkte Korrelation zwischen der Röntgenaktivität eines Doppelsternsystems und einer Erhöhung der Strahlungsintensität im optischen fest. Sie beobachteten das Röntgen-Doppelsternsystem GROJ1644-40 welches sich auf Grund des geringen Röntgenflusses bis April 1996 in einem ruhigen Zustand befand. Dabei fanden sie heraus, dass sechs Tage vor dem Röntgenausbruch, welcher am 25. April startete sich die Strahlungsleistung im optischen ebenfalls stark erhöhte. Dabei begann der Anstieg in der Intensität zuerst im I-Band (20. April) und zuletzt im B-Band, also von rot nach blau. Dagegen ist die Steigung des Flusses gerade andersherum im B-Band am steilsten und im I-Band am flachsten (vgl. Abbildung \ref{fig:Roentgenverhalten} (J.A. Orosz et al. 1997)). Daraus schlossen sie , dass die Störung im System, welche diesen Röntgenausbruch verursacht hatte eine \glqq outside-in\grqq, also eine Störung sein musste, welche von von außen in die Akkretionsscheibe eindringt, da die Röntgenstrahlung erst im inneren der Akkretionsscheibe entstehen kann. Die Ursache muss also ein plötzlicher starker Massenfluss gewesen sein, welcher ins Zentrum der Scheibe wandert. Durch das freiwerden von potentieller Energie des Gravitationsfeldes und Reibung in der Akkretionsscheibe entstehen am Innenrand der Scheibe Röntgenstrahlen. Es kann also daraus geschlossen werden, dass der Zeitraum, den Materie, welche vom Begleitstern akkretiert, braucht um vom Außenrand der Scheibe an den Innenrand zu gelangen im Bereich von wenigen Tagen liegt. Um einen solchen Effekt auszuschließen, werden bei der Datenauswertung in Kapitel \ref{sec:Roentgen} die Röntgendaten von LMC X-3 analysiert.

\subsection{GROND}\label{sec:GROND}
\begin{figure}[h]
\center
\subfloat[\label{fig:GRONDAufbau}3D Modell des vereinfachten Aufbaus von GROND (Quelle: Greiner et al. 2008)]
{\includegraphics[width=9cm]{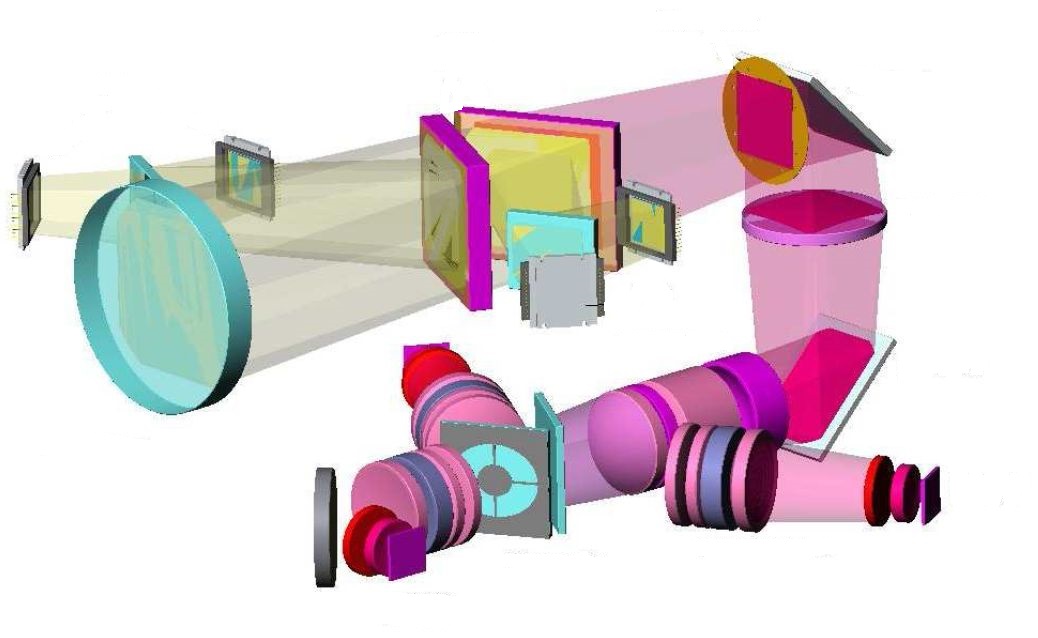}}
\qquad
\subfloat[\label{fig:GRONDFilter}Spektren der einzelnen GROND Filter (Quelle: Schiegg 2013)]
{\includegraphics[width=12cm]{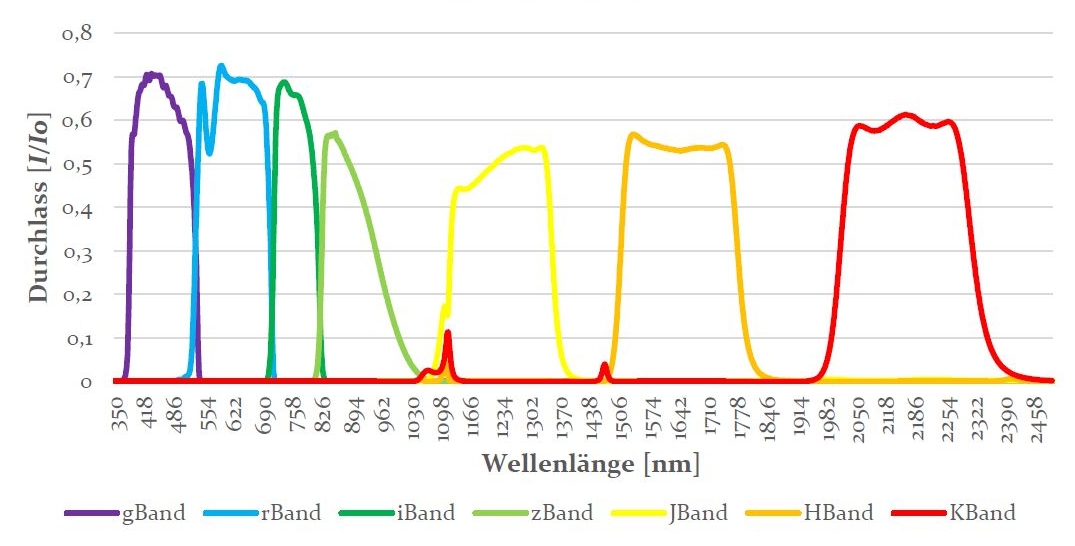}}
\end{figure}

Die, in dieser Arbeit verwendeten Daten stammen zum größten Teil von GROND (Gamma-Ray Burst Optical and Near-Infrared Detector) (J.Greiner et al. 2008), einer bildgebenden Apparatur auf CCD-Kamera Basis am MPI/ESO 2.2m Teleskop auf dem Berg La Silla in der chilenischen Atacamawüste. GROND wurde primär für Beobachtungen von Gamma-Ray-Bursts gebaut und ermöglicht die gleichzeitige Aufnahme in sieben verschiedenen Wellenlängenbereichen, nämlich in vier optischen (OPT) SDSS-Bändern (g',r',i',z'; im Folgenden ohne Striche) und drei Johnson-Bändern (J,H,K) des nahen Infrarotbereichs (NIR). Somit wird ein Wellenlängenbereich von etwa 400nm bis 2300nm abgedeckt (siehe Abbildung \ref{fig:GRONDFilter} (F. Schiegg 2013)). Der beobachtbare Bereich im aktuellen Aufbau beträgt in den optischen Bändern 5.4'x5.4' und 10'x10' im NIR-Bereich. Alle Detektoren von GROND sind kryostatisch gekühlt; die NIR-Sensoren auf 65K und die OPT-CCD Module auf 165K. Jede Aufnahme (Observation Block - OB) mit GROND ist aufgebaut aus verschiedenen Belichtungszeiten die kombiniert werden. Dabei besteht jeder OB aus vier, bzw. sechs Bildern pro Band mit leicht verschobenen Positionen am Himmel, den sog. "`telescope dithers (TD)"' um den Einfluss von Pixelfehler zu minimieren.  Die Aufnahmen in den optischen und infraroten Bändern unterscheiden sich stark und werden schließlich zu den TDs synchronisiert. Im optischen Auslesesystem, welches FIERA (Fast Imager Electronic Readout Assembly) genannt wird, ist die Belichtung im g und r Band unabhängig von der im i und z Band. Dies ermöglicht es im g und r Band länger zu belichten. Gleichzeitig werden im NIR-System, welches IRACE (Infrared Detetor High speed Array Control and processing Electronics) genannt wird, die NIR-Sensoren belichtet, wobei der Spiegel zum K-Band Sensor ebenfalls vor jeder Belichtung (sechsmal pro TD) leicht variiert wird, da der Himmel im K-Band Wellenlängenbereich deutlich heller ist im Vergleich zu den anderen Bändern. In der Zeit, in der sich dieser Spiegel bewegt, wird im J und H Band ebenfalls nicht belichtet, da sonst ein Hintergrundrauschen in den Bildern auftreten würde. Dabei werden für jedes TD mehrere Bilder im NIR gemacht. Der schematische Aufbau GRONDs ist in Abbildung \ref{fig:GRONDAufbau} zu sehen, wobei neben den Komponenten in GROND der Strahlgang dargestellt ist. Am Ende jedes Lichtbündels steht ein Sensor, der belichtet wird.

\subsection{\glqq XRbinary\grqq \ von Edward L. Robinson}\label{sec:XRbinary}
XRbinary ist ein Programm von Edward L. Robinson zur Simulation der Lichtkurven von Röntgen-Doppelsternsystemen (E.L. Robinson 2014). Für diese Arbeit wird die Version 2.3 verwendet, die leicht modifiziert wurde, um eine Begleitsterntemperatur von über 10000 Kelvin zu ermöglichen, wobei ein Schwarzkörperspektrum angewandt wird. Das Modell basiert auf einer kreisförmigen Bewegung von zwei Sternen um ihren gemeinsamen Schwerpunkt. Hierbei kann ein Stern dessen Ausdehnung sehr viel kleiner als jede andere Struktur im System sein soll (in diesem Fall ein schwarzes Loch), und \glqq primary star \grqq genannt wird von einer Akkretionsscheibe umgeben sein; der Begleitstern (\glqq secondary star \grqq) nicht. Das Programm besitzt eine Vielzahl von Parametern (einige zehn), die mehr oder minder großen Einfluss auf die Form der Lichtkurve haben. Hiervon werden für die späteren Simulationen nur wenige wichtige variiert (siehe: Kapitel \ref{sec:Parameter}).

\subsection{LMC X-3}
LMC X-3 ist ein helles Röntgen-Doppelsternsystem in der großen Magellan'schen Wolke (LMC - Large Magellanic Cloud), einer Satellitengalaxie der Milchstraße, welche ca. 50 kpc (G. Pietrzynsky 2013), also etwa 170 000 Lichtjahre von unserem Sonnensystem entfernt ist und ungefähr 15 Milliarden Sterne beheimatet. Sie ist die viertgrößte Galaxie in der lokalen Gruppe, hat einen Durchmesser von ca. 7.7 kpc und eine Masse von in etwa $10^{10} \  M_{\odot}$. LMC X-3 findet man unter der Position 05°38'56.63'' Rektaszension und -64°05'03.3'' Deklination (2000). Es wurde vom Uhuru-Satelliten, der den Himmel vollständig nach Röntgenquellen durchmusterte, entdeckt (Leong et al. 1971) und wurde als Kandidat für ein Schwarzes Loch Doppelsternsystem vermutet, nachdem die Periode des Orbits durch optische Spektroskopie auf ca. 1.7 Tage und die Bahnneigung auf unter 70° bestimmt werden konnte (Cowley et al. 1983). Somit war LMC X-3, chronologisch gesehen, der zweite Kandidat für ein Schwarzes Loch nach Cygnus X-1. Der Begleitstern des Schwarzen Loches in LMC X-3 wurde historisch gesehen verschiedenen Spektralklassen zugeordnet. Während in den Veröffentlichungen des letzten Jahrtausends durchgehend von einem Stern der Klasse B3V, also einem Hauptreihenstern ausgegangen wird, der seinen Roche-Lobe voll ausfüllt, sehen neuere Ergebnisse eine periodische Variation des Spektrums zwischen B3V und B5V (Val-Baker et al. 2007) und B5IV (Soria et al. 2001), also dem eines Unterreisen, was LMC X-3 zu den HMXBs einordnet. Auf Grund dieser maßgeblichen Unsicherheit im Spektraltyp und somit in der Temperatur des Begleitsterns gibt es auch in den neueren Veröffentlichungen folglich starke Diskrepanzen, was die Masse des Schwarzen Loches angeht.

\begin{figure}[ht]
\begin{center}
\includegraphics[width=6cm]{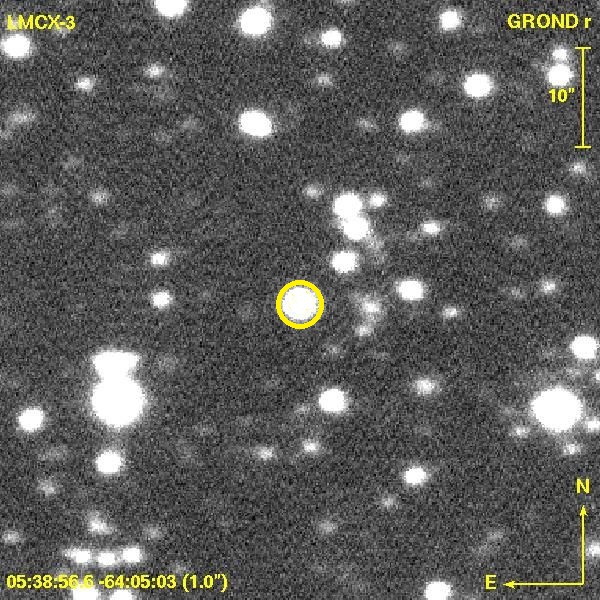}
\caption{GROND Aufnahme vom LMC X-3 (gelber Kreis herum) im g-Band}
\label{fig:LMCX3}
\end{center}
\end{figure}

\section{Parameterstudie zu LMC X-3}\label{sec:Parameterstudie}

\subsection{Abschätzung der Geometrie von LMC X-3}
Ziel der folgenden Abschätzung ist es, die Geometrie von LMC X-3 in etwa zu beschreiben. Dies geschieht mit den Ergebnissen von J.A. Orosz et al. 2014. Dort wurde die Masse des Schwarzen Lochs zu $M_{BH}=6.95\pm0.33$ $M_{\odot}$, die des Begleitsterns zu $M_{S}=3.70\pm0.24$ $M_{\odot}$ und die Periode zu $P=1.7048089$ Tage bestimmt.
Somit kann aus diesen drei Daten die Geometrie des Systems mit folgenden Formeln bestimmt werden (B. Warner 1995):\\
Der Radius des Roche-Lobe ist gegeben zu:
\begin{equation}
R_L=0.462 \cdot \left(1+\frac{M_{BH}}{M_{S}}\right)^{\frac{1}{3}} \cdot a
\end{equation}
wobei a die große Halbachse des Systems ist, die wiederum über das 3. Keppler'sche Gesetz berechnet werden kann zu:
\begin{equation}
a=\left(\frac{P^2 G(M_{BH}+M_{S})}{4\pi^2}\right)^{\frac{1}{3}}
\end{equation}
mit der Gravitationskonstante G.\\
Der Radius des Begleitsterns ist gegeben zu:
\begin{equation}
\frac{R_S}{R_{\odot}}=\left(\frac{M_S}{M_{\odot}}\right)^{0.88}
\label{eq:Begleitstern}
\end{equation}
Der Radius der Akkretionsscheibe ist gegeben zu:
\begin{equation}
r_D=\frac{0,6 \cdot a}{1+\frac{M_S}{M_{BH}}}
\end{equation}
Somit erhält man in der Abschätzung für LMC X-3 folgende Werte:\\
 \ \ - Große Halbachse $a=13.3$ $R_{\odot}$ , \ \ \ \ - Radius des Roche-Lobe $R_L=4.3$ $R_{\odot}$\\
 \ \ - Radius der Scheibe $r_{D}=5.1$ $R_{\odot}$, \ \  - Radius des Begleitsterns $R_S=3.2$ $R_{\odot}$ \\
Es ist zu sehen, dass der Radius des Begleitsterns circa einen Sonnenradius kleiner ist als sein Roche-Lobe.\\
Des Weiteren kann aus diesen Daten bestimmt werden, dass bei einem Neigungswinkel des Systems zwischen 72° und 90° bei Phase 0 der Begleitstern die Akkretionsscheibe teils verdeckt. \\
Es soll trotzdem hier erwähnt sein, dass dies nur eine sehr vage Abschätzung darstellt. Es ist vor allem fragwürdig, inwiefern Gleichung \ref{eq:Begleitstern} noch für einen B-Stern gilt, der auf Grund der höheren Temperatur eine kleinere Dichte und daher einen größeren Radius haben wird.

\subsection{Einfluss der Komponenten auf die Lichtkurve}
Im diesem Abschnitt wird anhand von simulierten Lichtkurven, welche mit XRbinary (siehe Kapitel \ref{sec:XRbinary}) erzeugt wurden, der Einfluss verschiedener Komponenten auf die Lichtkurve untersucht. Hierbei ist in erster Linie eine Auswirkung auf die Form der Lichtkurve und weniger auf die absolute Helligkeit interessant. Für eine sehr ausführliche und allgemeine Parameterstudie siehe auch: F. Knust 2013
\subsubsection{Bahnneigung}
Wie schon oben erwähnt wird bei Vergrößerung des Neigungswinkels (Winkel zwischen Beobachtungsrichtung und Drehimpulsvektor des Doppelsternsystems) die Amplitude und auch der Unterschied zwischen den beiden Minima größer, während es keinen Einfluss auf die beiden Maxima gibt auf Grund der Achsensymmetrie des Begleitsterns. Die Bahnneigung hat mit Abstand den größten Effekt auf die Form der Lichtkurve.
\subsubsection{Schwarzes Loch und Begleitstern}
Das Massenverhältnis zwischen Schwarzem Loch und Begleitstern ändert die Form der Lichtkurve nur sehr schwach. Je größer das Massenverhältnis $q=\frac{M_S}{M_{BH}}$ ist, desto kleiner wird die Amplitude und der Unterschied zwischen den beiden Minima, was aber verglichen mit dem Neigungswinkel des Systems nur einen Bruchteil ausmacht. Der Effekt rührt daher, dass durch das veränderte Massenverhältnis sich die Geometrie des Roche-Lobes ändert und daher die, bei Phase 0.25 und 0.75 beobachtete Fläche des Begleitsterns leicht variiert.\\
Eine Temperaturänderung des Begleitsterns hat keine geometrische Veränderung des Systems zur Folge, und wirkt sich deshalb nur auf die absolute Helligkeit aus, ebenfalls wie die Veränderung der Gesamtmasse.
\subsubsection{Akkretionsscheibe}
\begin{figure}[h!]
\center
\subfloat[\label{fig:TDVar}variable Scheibentemperatur $i=69.3°$]
{\includegraphics[width=8.4cm]{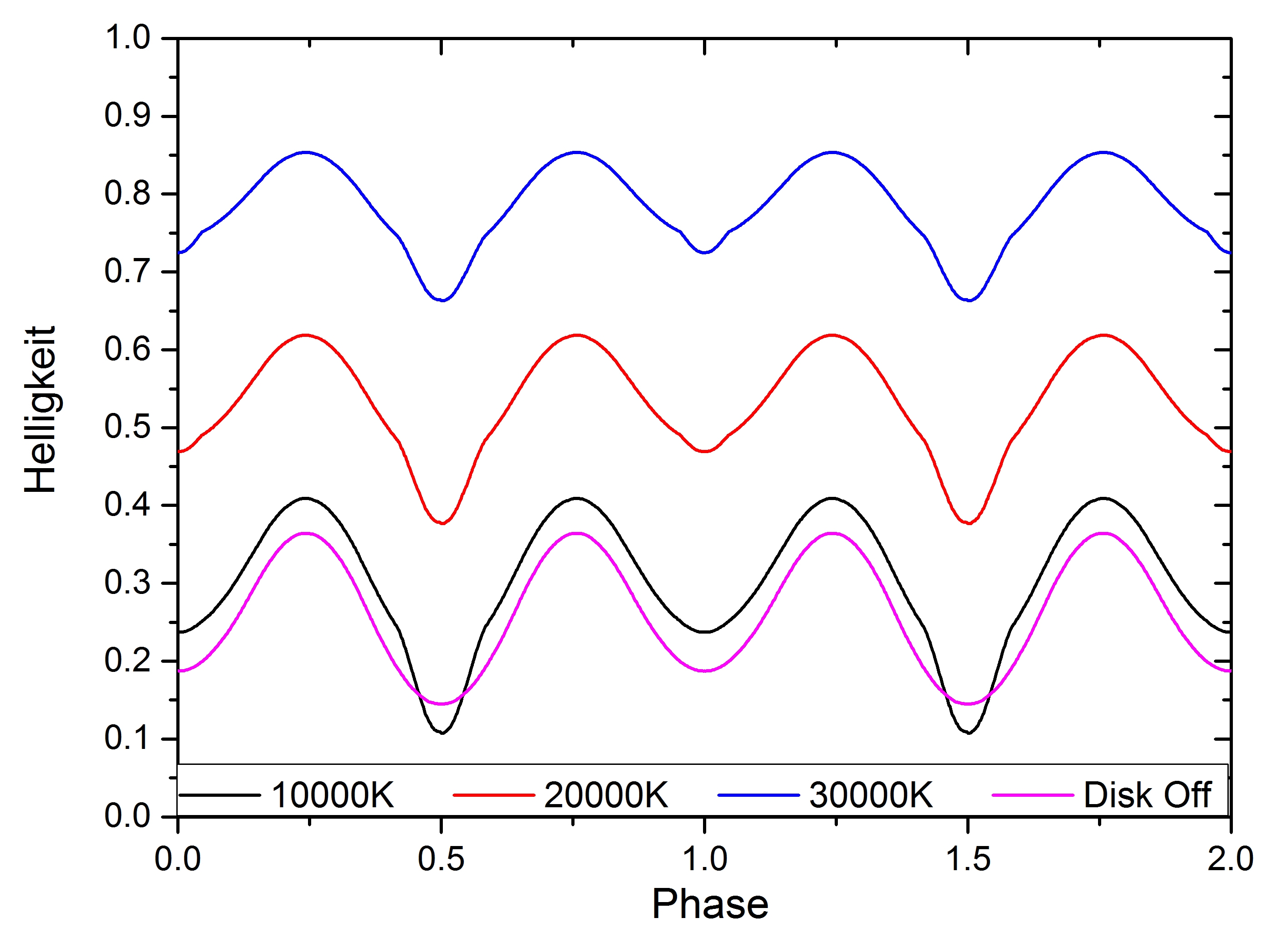}}
\subfloat[\label{fig:NWVar}variable Bahnneigung $T_D=10000 \ K$]
{\includegraphics[width=8.4cm]{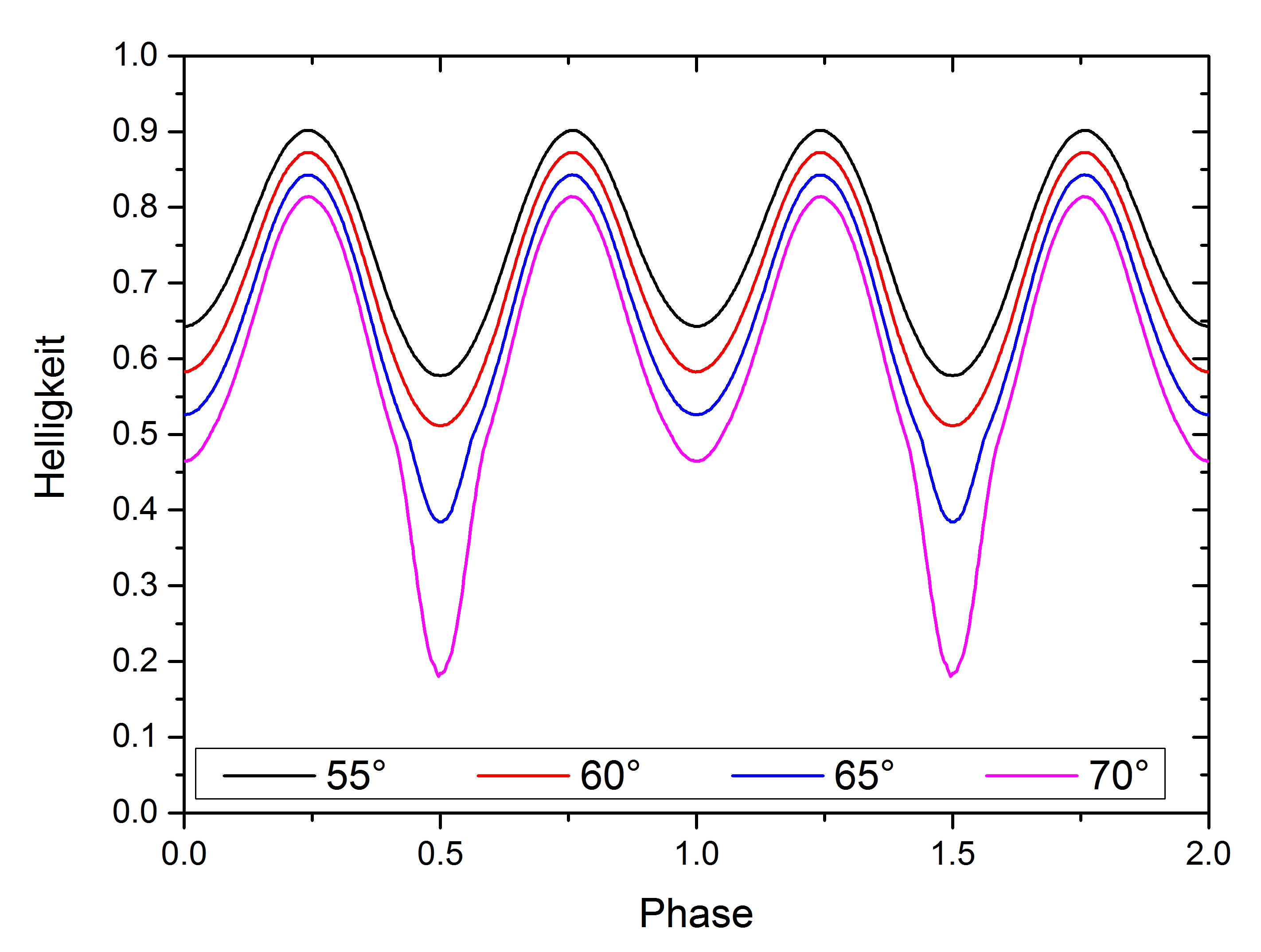}}
\caption{Simulierte Lichtkurven ($M_{BH}=6.95 \ M_{\odot}$, $M_S=3.7 \ M_{\odot}$, $T_S=15500K$)}
\end{figure}

\begin{figure}[h!]
\center
\subfloat[\label{fig:TD10NWVar}Akkretionsscheibentemperatur: 10000K]
{\includegraphics[width=8.4cm]{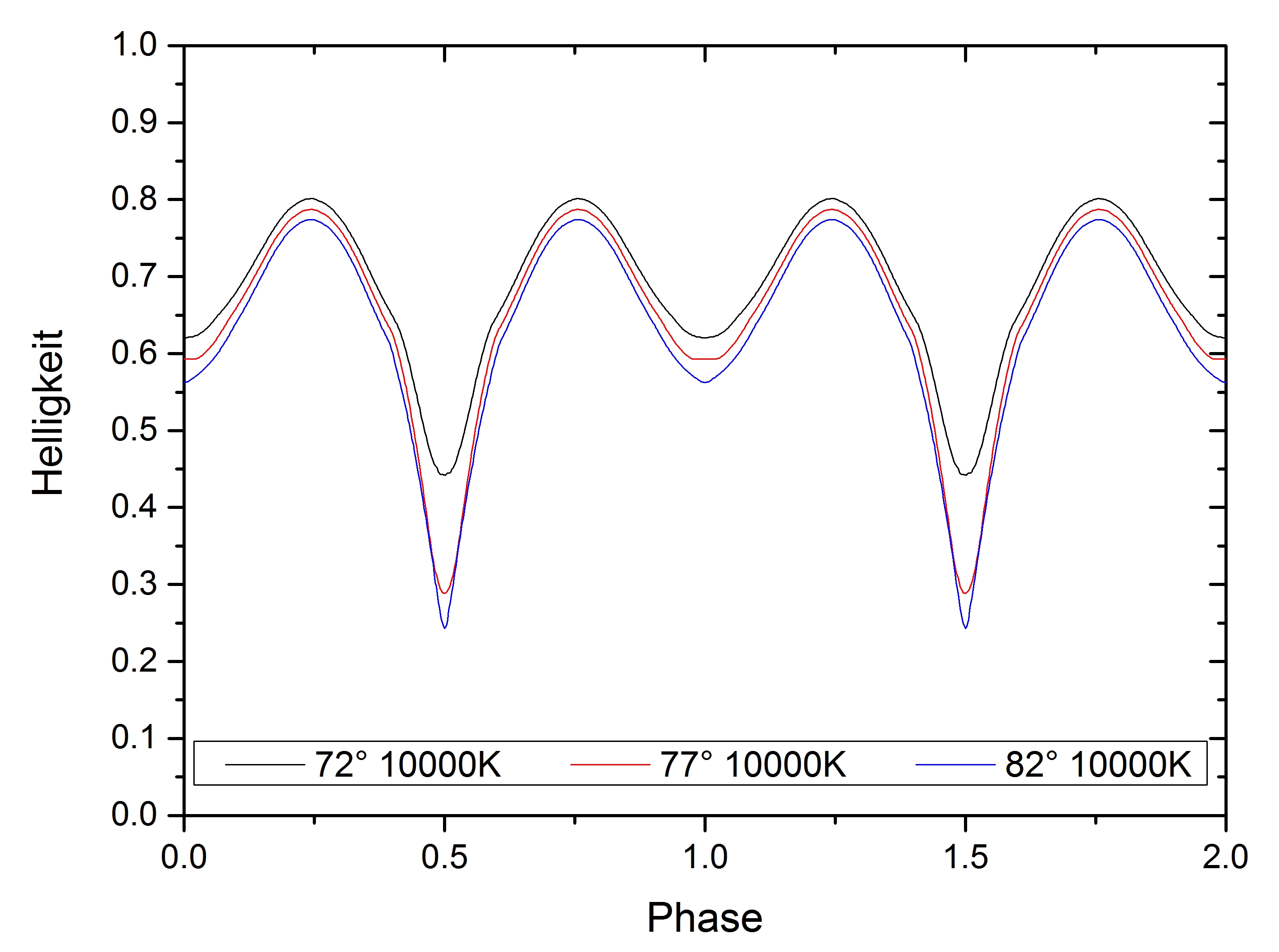}}
\subfloat[\label{fig:TD30NWVar}Akkretionsscheibentemperatur: 30000K]
{\includegraphics[width=8.4cm]{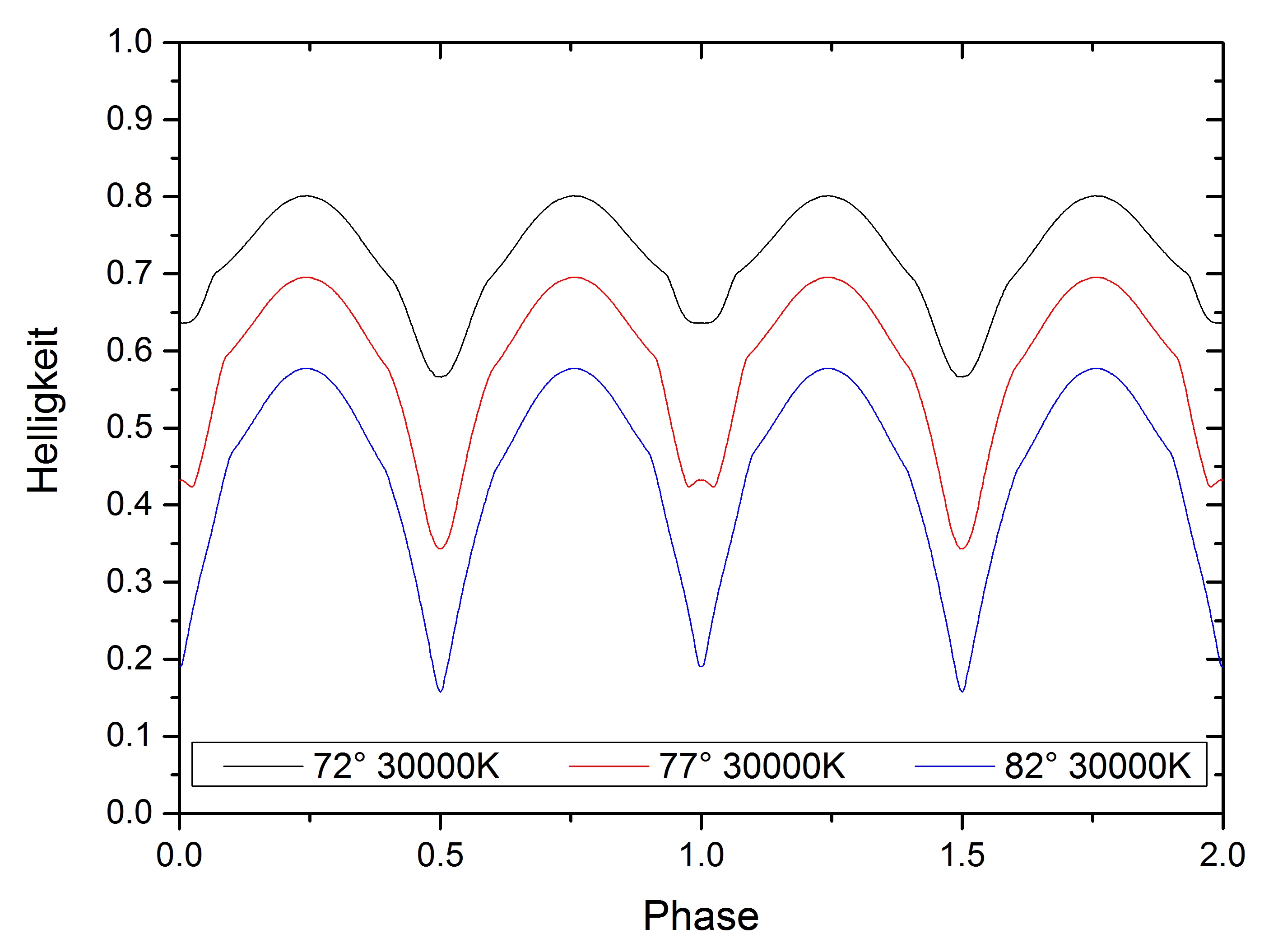}}
\caption{Simulierte Lichtkurven mit variabler Bahnneigung($M_{BH}=6.95 \ M_{\odot}$, $M_S=3.7 \ M_{\odot}$, $T_S=15500K$)}
\end{figure}

Der Parameter der Akkretionsscheibe, die ja eine durchaus komplizierte Geometrie aufweisen kann (Robinson 2014), welcher sich am stärksten auf die Lichtkurve auswirkt, ist ihre Temperatur. In XRbinary wird jeweils die Temperatur am Scheibenaußenrand benutzt um über eine $T \propto R^{-\frac{3}{4}}$ Abhängigkeit (mit dem Scheibenradius R) das Temperaturprofil der Scheibe zu berechnen. In den folgenden Abbildungen ist die Helligkeit einheitslos gewählt, da die absolute Helligkeit keine Rolle spielt.\\
Wenn bei einer festen Scheibentemperatur von 10000K am Außenrand der Scheibe die Bahnneigung vergrößert wird (vgl. Abbildung \ref{fig:NWVar}), werden die Amplituden wesentlich größer. Vor allem am tiefen Minimum findet eine starke Veränderung statt. Bei großen Winkeln bildet sich dort ein kleiner Knick aus, welcher durch die Bedeckung der Scheibe durch den Stern verursacht wird.
Bei einer Erhöhung der Scheibentemperatur (vgl. Abbildung \ref{fig:TDVar} $\longrightarrow$ es sind relative Magnituden mit einem willkürlichen Offset dargestellt) geht natürlich der Fluss nach oben. Die Amplitude verringert sich dabei in allen Bändern gleichermaßen mit der Temperatur der Scheibe.
Wegen der Bedeckung der Scheibe durch den Stern ist die Amplitude des tieferen Minimums größer im Vergleich dazu, wenn keine Scheibe angenommen wird. Bei Scheibentemperaturen, die größer sind als die Sterntemperatur, tritt ebenfalls eine Veränderung des seichten Minimums auf.
Dies hängt mit dem Winkel zusammen und kann am besten mit den Abbildungen \ref{fig:TD10NWVar} und \ref{fig:TD30NWVar} erklärt werden. Bei einer Scheibentemperatur von 10000K wird hauptsächlich das tiefe Minimum ausgeprägter, bei einer hohen Scheibentemperatur erkennt man eben dann einen leichten Buckel im seichten Minimum. Dies hängt damit zusammen, dass bei einem Winkel von circa 77° die Scheibe vom Stern halb bedeckt wird. Wenn man nun von der anderen Seite in genau diesem Winkel das System betrachtet, kann man durch den Innenrand der Disk hindurch schauen und den Stern leicht sehen.

\subsubsection{Hot Spot}
\begin{figure}[h]
\begin{center}
\includegraphics[width=8cm]{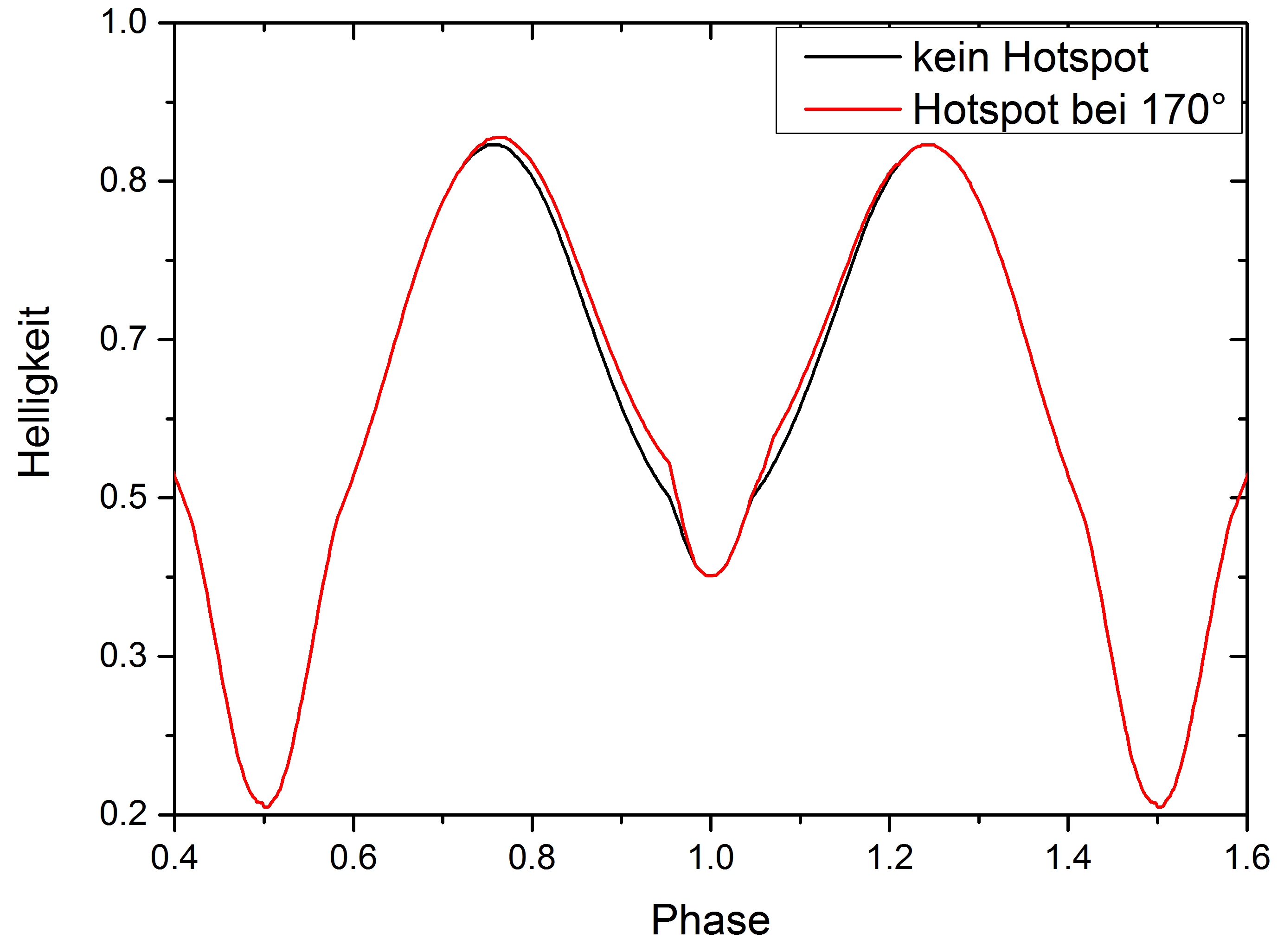}
\caption{Simulierte Lichtkurve ohne und mit Hot Spot bei einem Winkel von 170° ($M_P=6.95 \ M_{\odot}$, $M_S=3.7 \ M_{\odot}$, $T_S=15500K$), $T_D=26000K$, $T_{HS}=50000K$, $i=69.3°$}
\label{fig:HSStudie}
\end{center}
\end{figure}

Beim Hot Spot trifft Materie aus dem Massenstrom vom Begleitstern auf die Akkretionsscheibe. Durch die Reibung entsteht an diesem Punkt eine kleine Stelle, welche heißer ist als der Außenrand der Scheibe. In der simulierten Lichtkurve in Abbildung \ref{fig:HSStudie} ist ein Hot Spot bei 170° zu sehen, wobei der Winkel auf der Scheibe vom Punkt aus gemessen wird, der gegenüber dem Begleitstern liegt. Man erkennt, dass das Maximum, welches näher am Hot Spot liegt leicht erhöht und zur Position des Hot Spots hin verschoben wird. Viel auffälliger ist aber der Knick, der in der Lichtkurve entsteht und dadurch zu erklären ist, dass der Hot Spot vom Begleitstern bei 170° vorauseilt und vor dem Minimum durch die Scheibe bedeckt wird.  Dadurch entsteht eine Asymmetrie des Minimums.\\
In der späteren Modellierung wird der Hot Spot allerdings weggelassen, da die Abdeckung mit Datenpunkten zu gering ist, um eine solche Asymmetrie zu erkennen, welche dazu noch signifikant ist.

\section{Datenauswertung}

\subsection{Verwendete Daten}
Die Daten, auf denen diese Arbeit aufbaut, wurden im Zeitraum vom 20. Oktober 2007 bis zum 13. Dezember 2013 immer wieder vereinzelt mit dem GROND Imager (siehe Kapitel \ref{sec:GROND}) am MPI/ESO 2.2m Teleskop in La Silla (Chile) aufgenommen. Die Integrationszeit variierte zwischen vier über acht und zehn bis zwanzig Minuten.

\subsection{Vorbereitung der Roh-Daten}
In diesem Abschnitt wird kurz beschrieben, wie mit den GROND-Bildern (vlg. Kapitel \ref{sec:GROND}) weiter verfahren wird, um diese einer quantitativen Analyse zu unterziehen.
\subsubsection{Reduktion der Daten}
\begin{figure}[h]
\center
\subfloat[\label{fig:ds9OPT}Aufnahme im g-Band]
{\includegraphics[width=4cm]{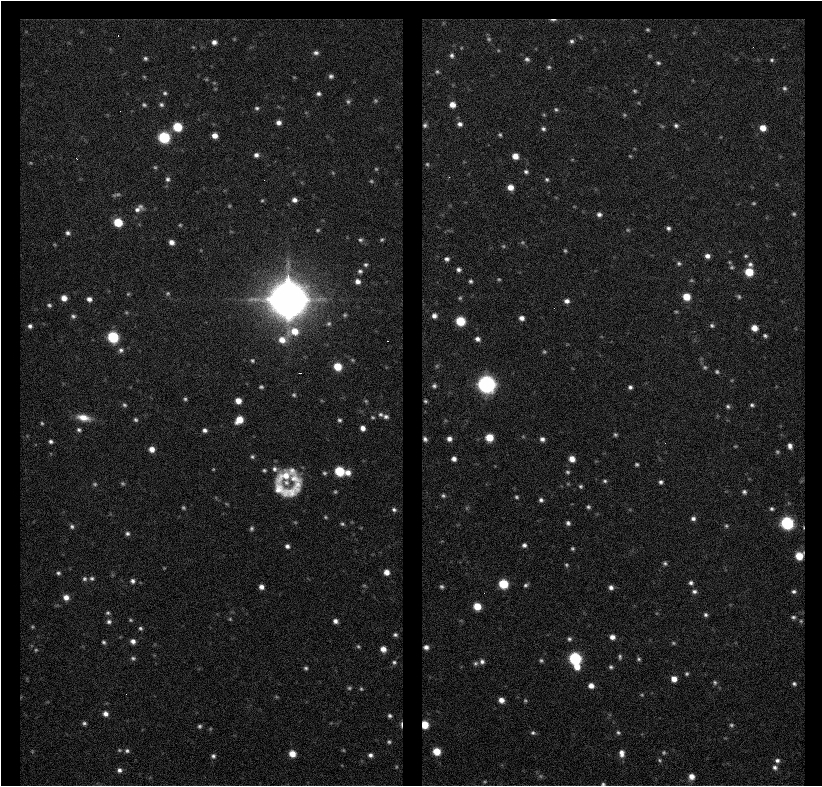}}
\quad
\subfloat[\label{fig:ds9NIR}Aufnahme der Infrarotbänder J, H, K]
{\includegraphics[width=11.5cm]{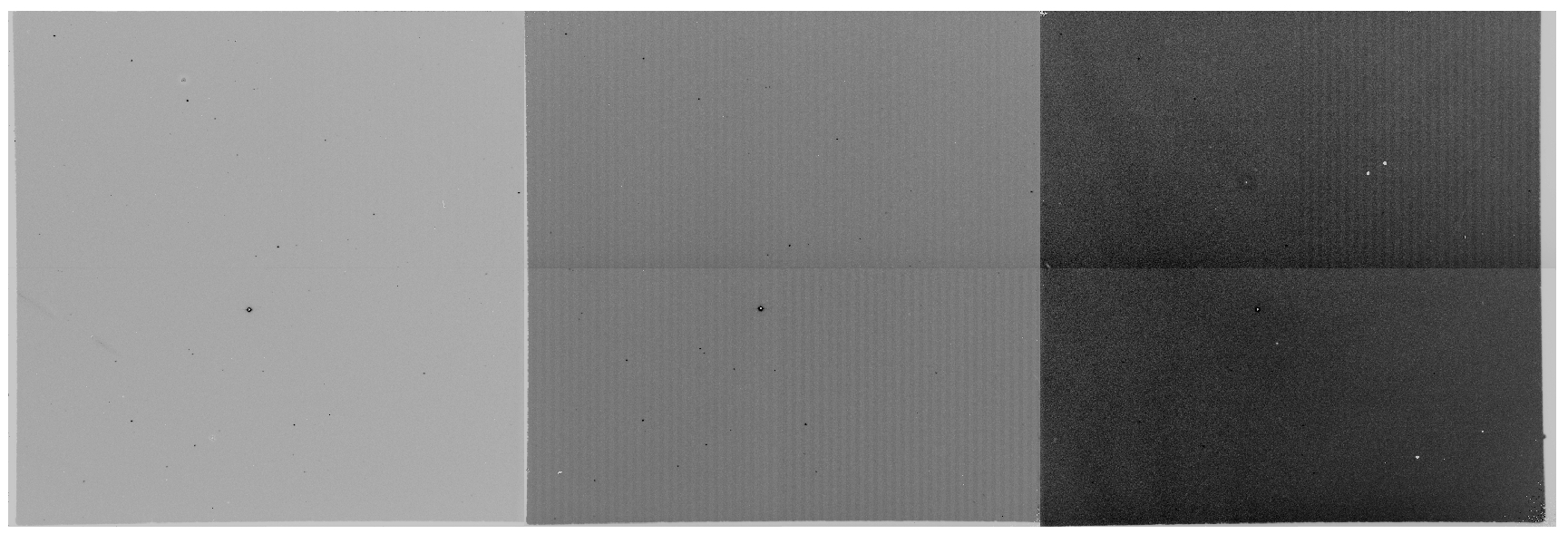}}
\caption{GROND-Aufnahmen vor der Reduktion der Daten}
\end{figure}

Das Ziel der Datenreduktion ist es aus den mit GROND aufgenommenen Daten pro Observation Block pro Band ein Bild zu bekommen welches nach Konvention so ausgerichtet ist, dass Norden oben und Osten links ist. Dazu werden die Aufnahmen gespiegelt, rotiert, geschnitten, aneinander angeglichen und schließlich pro Band zusammen kombiniert. Danach werden sie mit Hilfe der Daten aus den Kalibrationsdateien (siehe nächster Absatz) kalibriert, was mit dem Programm \glqq gr\_redcomb.py\grqq \ geschieht. (Für nähere Information zum Reduktionsprozess siehe: F. Knust 2013) In Abbildung \ref{fig:ds9OPT} ist eine GROND-Aufnahme des g-Bandes zu sehen. Das Bild ist in zwei Hälften geteilt, weil der CCD-Sensor von zwei Seiten ausgelesen wird. Abbildung \ref{fig:ds9NIR} zeigt die GROND-Aufnahme der Bänder im nahen Infrarotbereich, wobei alle drei Bänder (J, H, K) in einem Bild nebeneinandergereiht sind. Man kann auf den noch nicht reduzierten Bildern im Infraroten deshalb so wenig erkennen, weil die Belichtungszeit mit 10 Sekunden kurz, und zusätzlich der Himmel im Infraroten hell ist. Erst durch die Reduktion ergibt sich ein verwertbares Bild. 
\\
Jeden Tag, bzw. vor jeder Observation mit GROND werden sogenannte \glqq bias\grqq \ und \glqq darks\grqq \ am Teleskop erstellt. Diese Datensätze beinhalten das Rauschen der Elektronik, die sogenannten \glqq Bad Pixels\grqq, also Pixel auf dem Sensor, die kein Signal mehr detektieren, und das thermische Rauschen. Zusätzlich werden, wenn es nicht bewölkt ist in der Dämmerung sogenannte \glqq skyflats\grqq \ aufgenommen. Diese dienen dazu, globale Merkmale, die intrinsisch die Sensoren beeinflussen zu detektieren. Die Datensätze aus \glqq bias\grqq \ und \glqq darks\grqq \ werden mit dem Programm \glqq gr\_bida.py\grqq \ und \glqq skyflats\grqq \ mit\glqq gr\_flat.py\grqq \ reduziert (vgl. GRONDwiki $\rightarrow$ How to produce master calibration frames).\\
Für die Reduktion werden nun die bias und darks Belichtungen zu einem Bild für jedes Band kombiniert, welches den Durchschnitt der einzelnen Bilder darstellt. Diese \glqq average bias\grqq \ und \glqq average darks\grqq \ werden nun von den Bildern von LMC X-3 subtrahiert. Des Weiteren werden die skyflats ebenfalls zu einem Durchschnittsbild für jedes Band kombiniert und daraufhin normiert. Die Aufnahme von LMC X-3 wird dann durch das normierte skyflat Bild dividiert. Nach diesen Schritten sind die Bilder befreit vom Rauschen durch das Auslesen der Sensoren, vom Dunkelstrom und von Variationen in der Pixelsensitivität. Wenn nun vom gleichen Objekt für jedes Band mehrere Aufnahmen gemacht wurden pro OB, dann müssen diese noch angeglichen und kombiniert werden, was zum Resultat hat, dass die Kosmische Hintergrundstrahlung größtenteils entfernt wird.

\subsubsection{Astrometrie und Photometrie}
\begin{figure}[h]
\vspace*{-2cm}
\center
\subfloat[\label{fig:OPTRegion}im OPT]
{\includegraphics[width=8cm]{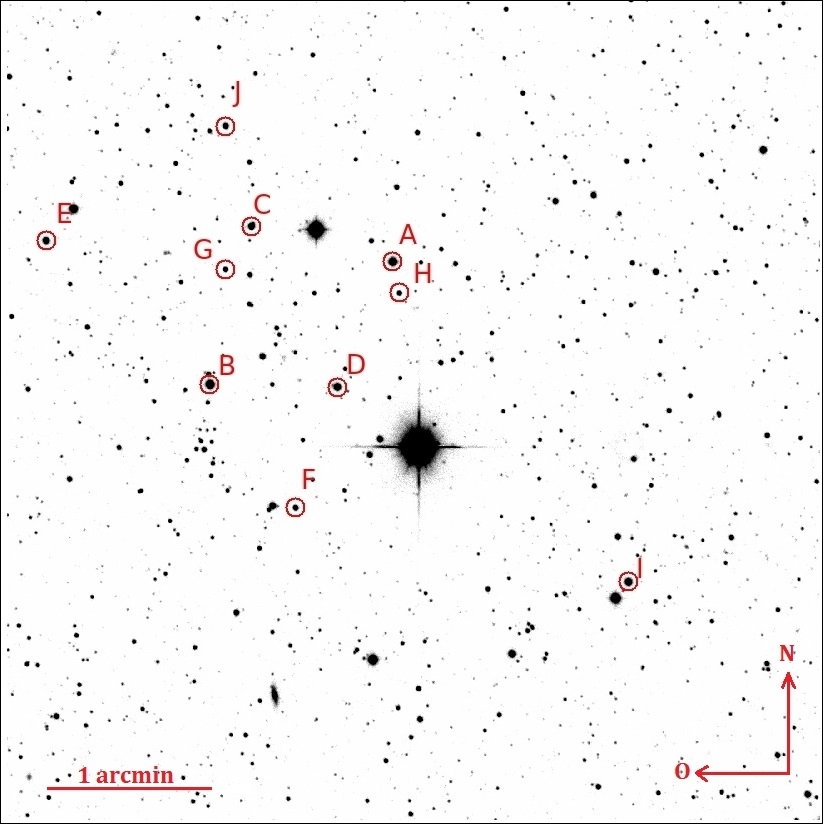}}
\subfloat[\label{fig:NIRRegion}im NIR]
{\includegraphics[width=8cm]{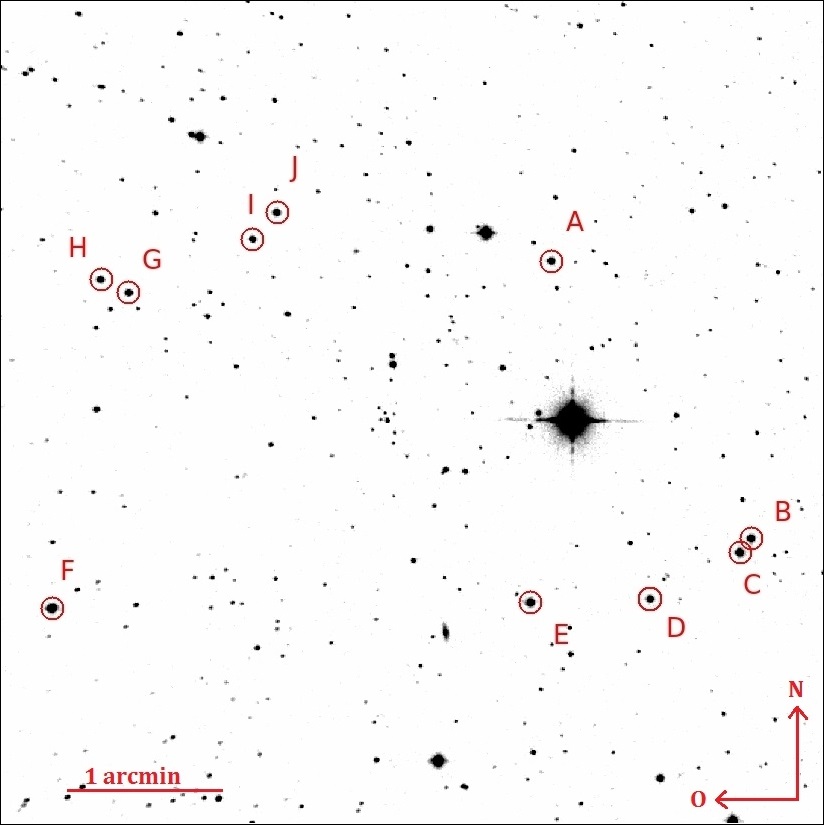}}
\caption{Bild der Kalibrationssterne}
\end{figure}

Durch die Reduktion der Daten hat man nun mehr oder minder schön aussehende Aufnahmen des Objekts, dem unser Interesse gilt, und mitunter relativ vieler anderer Sterne in dessen naher Umgebung. Nun ist der nächste Schritt die Helligkeit der jeweiligen Objekte auf dem Bild zu ermitteln. Die einzige Information, die man vom Sensor erhält ist der integrierte Fluss der Photonen pro jeweiligem Pixel, also eine Zählrate.
\begin{itemize}
\item \textbf{Astrometrie:} Hierzu benutzt man aus den Bildern helle Sterne mit einer guten Form, also kreisförmig und vergleicht diese mit einem Katalog, in dem die scheinbaren Helligkeiten und die Koordinaten einiger Sterne gespeichert sind. Im optischen Bereich ist dies der \glqq Sloan Digital Sky Survey\grqq \ (SDSS) im AB-System (A.S. Szalay et al. 2002) sowie der \glqq United States Naval Observatory B1.0\grqq \ (USNO-B1.0) (Monet et al. 2003), und im infraroten der \glqq Two Micron All Sky Survey\grqq \ (2MASS) im Vega-System (M.F. Skrutskie et al. 2006). Der SDSS Katalog ist genauer als der USNO Katalog, deckt aber den Südhimmel nur teils ab, weshalb der USNO Katalog im Optischen verwendet wird. Somit kann ein Koordinatensystem über das Bild gelegt werden, bzw. jedem Stern seine Koordinaten zugeordnet werden. Darauf erhält man Bilder wie in Abbildung \ref{fig:ds9g}, welches eine Aufnahme im g-Band ist und Abbildung \ref{fig:ds9K}, welches im K-Band aufgenommen wurde. Das generell weniger Sterne im K-Band zu erkennen sind und die Helligkeit einiger Sterne geringer ist als im g-Band, liegt an der Skalierung, welche in diesen Bildern so gewählt wird, dass das Hintergrundrauschen unterdrückt wird.
\begin{figure}[h]
\center
\subfloat[\label{fig:ds9g}g-Band]
{\includegraphics[width=6cm]{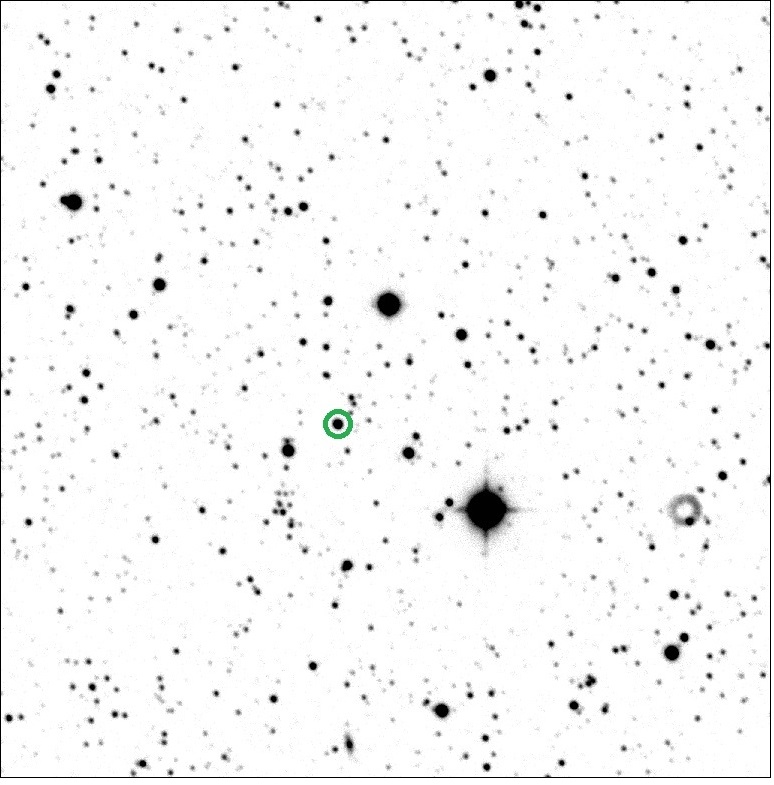}}
\quad
\subfloat[\label{fig:ds9K}K-Band]
{\includegraphics[width=6cm]{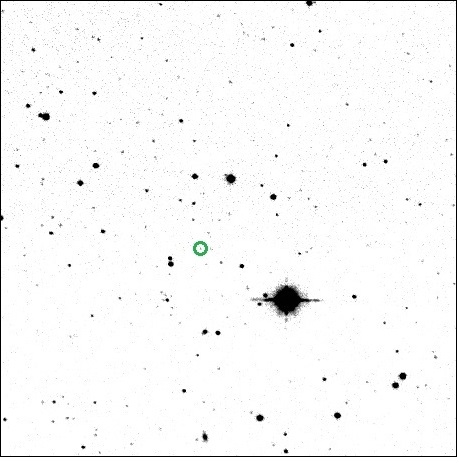}}
\caption{Aufnahmen nach der Astrometrie}
\end{figure}
\vspace{-0.6cm}
\item \textbf{Photometrie:} Wenn eine bestimmte Region auf den Bildern einen signifikant höheren Photonenfluss aufweist, als der Hintergrund, wird von dieser Region, also diesem Stern der Fluss berechnet. Auf Grund von atmosphärischen Abbildungsverzerrungen ist der Stern nicht punktförmig, sondern hat die Form einer zweidimensionalen Gaußverteilung. Um nun den Fluss des Sterns zu messen, wird der Fluss jedes Pixels innerhalb eines Kreises um den Stern, welcher mit der Halbwertsbreite skaliert, gemessen. Um nun eine Verfälschung durch überlappende Sterne, bzw. deren Abbild auf dem Sensor zu vermeiden, wird die sogenannte Point-Spread-Funktion (PSF)-Photometrie angewandt, welche eine zweidimensionale Gaußfunktion benutzt, die an die Counts der einzelnen Pixel gefittet wird. Somit ist es möglich die Ränder der Sterne zu korrigieren und dem jeweiligen Stern zuzuordnen.\\ 
\end{itemize}
Nun ist jedem detektierten Stern eine Magnitude zugeordnet, die auf einem Vergleich mit den Katalogsternen beruht. Für die NIR-Bänder ist dies ausreichend, wenn die Magnitude der Sterne relativ mit hellen Sternen des Katalogs im Blickfeld verglichen wird. Der SDSS Katalog hingegen deckt den Südhimmel nur teilweise ab, weshalb USNO verwendet wird, welcher aber nicht die Genauigkeit von SDSS aufweißt. Deshalb muss im optischen relative Photometrie (wie auch im Infraroten) und schließlich darauf basierend absolute Photometrie angewandt werden.
\begin{itemize}

\item \textbf{absolute Photometrie:}
Alle paar Wochen, wenn die Witterungsbedingungen gut sind, wird am Teleskop eine Aufnahme eines Himmelsausschnittes gemacht, in dem einige Standardsterne liegen. Standardsterne sind stabile, das heißt im Fluss nicht variierende, Fixsterne, welche über sehr lange Zeiträume genau vermessen sind. Die Standardsternaufnahme die für diese Thesis verwendet wird, wurde am 26. Februar 2012 um 2.01 Uhr UTC aufgenommen. Es wird genau diese Standardsternaufnahme verwendet, da noch in der gleichen Nacht Aufnahmen von LMC-X3 gemacht wurden, was sehr wichtig ist, denn die Zeit zwischen der Standardsternaufnahme und der Beobachtung von LMC-X3 sollte minimal sein, um Effekte, die auf die zeitliche Variation der Atmosphäre zurückgehen zu minimieren. Aus der Aufnahme der Standardsterne wird der sogenannte \glqq Zero Point\grqq ermittelt, welcher die untere Helligkeitsgrenze angibt, bei der ein Objekt noch detektierbar ist, es also 1 Count pro Sekunde verursacht. Mit genau diesem Zeropoint (bei Airmass 1) wird der zeitnäheste OB (am 26. Februar 2012 um 2.10 Uhr UTC OB5-1) der Aufnahme von LMC-X3 kalibriert. Nun wird mit Hilfe dieses Kalibrations-OBs OB5\_1 die Helligkeitskalibration aller anderen OBs mit der relativen Photometrie durchgeführt.
\item \textbf{relative Photometrie:} 
Diese basiert darauf den Fluss eines Sterns relativ einem Satz anderer Sterne zu berechnen. Diese Vergleichsterne müssen stabil in ihrer Helligkeit sein und ihre scheinbare Helligkeit sollte größer oder vergleichbar groß mit der des Objekts, welches es zu vermessen gilt, sein. Damit kann die statistische Unsicherheit bei der Flussbestimmung minimiert werden, da der Fehler mit $\sqrt{N}$ skaliert, wobei N die Counts pro Quelle sind und schwache Sterne mit großen statistischem Fehler nicht berücksichtigt werden. Der systematische Fehler setzt sich dann aus dem rms der Vergleichsterne zusammen. Die Daten der Vergleichsterne, die hierbei verwendet werden, werden aus dem Kalibrations-OB OB5\_1 entnommen und sind im Anhang \ref{tab:KalibrationssterneOPT} und \ref{tab:KalibrationssterneNIR} aufgelistet. Hierbei werden für den optischen Bereich teils andere Sterne verwendet, als für den nahen Infrarotbereich (vgl. Abbildung \ref{fig:OPTRegion} und \ref{fig:NIRRegion} im Anhang). In Abbildung \ref{fig:GRONDeog} sind die mit GROND vermessenen Magnituden einiger relativ hellen Sterne über den Katalogmagnituden aufgetragen. Man erkennt hierbei, dass die Streuung und der statistische Fehler bei größeren Magnituden steigen. Somit ist es sinnvoll nur sehr helle Sterne für die relative Photometrie zu nehmen, welche sich in diesem Diagramm hauptsächlich in der linken Hälfte befinden.
\begin{figure}[h]
\begin{center}
\includegraphics[width=10cm]{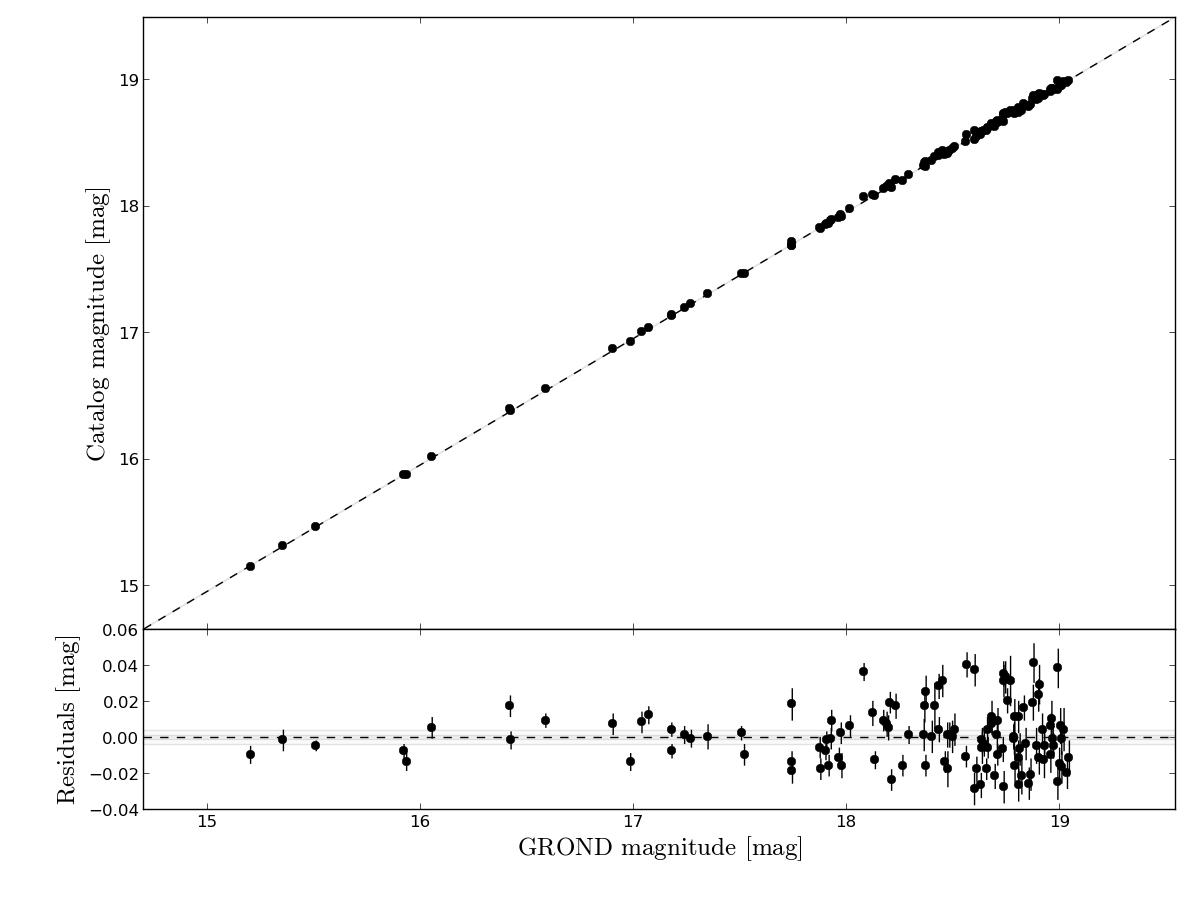}
\caption{Magnituden im r-Band des SDSS verglichen mit den GROND-Messdaten}
\label{fig:GRONDeog}
\end{center}
\end{figure}
Da die absolute Magnitude dieser Sterne aus OB5\_1, die nun zur relativen Photometrie verwendet werden, über den Zero Point bekannt ist, erhält man nun die korrekte scheinbare Helligkeit für alle Objekte der jeweiligen Aufnahmen, welche über die Magnitudenabhängigkeit berechnet wird:

\begin{equation}
m - m_0 = -2.5 \cdot log_{10}\left(\frac{f_0}{f}\right)
\end{equation}
wobei m und f die Magnitude und der Fluss des Observationsobjekts sind und $m_0$ sowie $f_0$ jeweils dem Vergleichstern zugeordnet sind.
\end{itemize}

\vspace{2cm}

\subsubsection{Vega-System und Extinktionskoeffizienten}
Die nun erhaltenen Helligkeiten der Observationsobjekte sind in zwei verschiedenen Magnituden-Systemen. Die optischen Bänder im AB-System, in dem wir weiter arbeiten werden, die NIR-Bänder im Vega System. Die Bänder im nahen Infrarotbereich müssen also umgerechnet werden. Dies geschieht durch eine additive Konstante, die zur Vega-Magnitude dazugezählt wird, sodass die Helligkeit der Infrarot-Bänder abnimmt. Der Summand lautet (vgl. GRONDwiki):\\
\ \ J-Band:  0.929 mag \ \ \ \ \ \ \ H-Band: 1.394 mag \ \ \ \ \ \ \ K-Band: 1.859 mag\\
Verursacht durch das Interstellare Medium, welches sich zwischen der Lichtquelle und dem Beobachter befindet und hauptsächlich aus Staub und Gas besteht, erscheinen die Lichtquellen je nach Position leicht gerötet auf Grund von Streuung und Absorption von Photonen. Dies verursacht die sogenannte \glqq Galaktische Vordergrundextinktion\grqq, weswegen die Magnituden nochmals um einen bandabhängigen Subtrahenden korrigiert werden müssen, wobei das Modell von Cardelli- Clayton- Mathis (CCM) (vgl. GRONDwiki) verwendet wird. Der Extinktionskoeffizient in der Region von LMC-X3 in der Großen Magellanschen Wolke liegt für das V-Band inklusive der galaktischen Vordergrundextinktion bei $(A_V=0.223\pm0.030)$ mag (Orosz et al. 2014). Die Korrekturwerte für die GROND Bänder lauten nach dem CMM-Modell wie folgt:
\begin{itemize}
\item $A_g=1.273 \cdot A_V = (0.284\pm0.038)$ mag vor März 2008
\item $A_g=1.255 \cdot A_V = (0.280\pm0.038)$ mag nach März 2008
\item $A_r=0.866 \cdot A_V = (0.193\pm0.026)$ mag
\item $A_i=0.648 \cdot A_V = (0.145\pm0.020)$ mag
\item $A_z=0.482 \cdot A_V = (0.107\pm0.014)$ mag
\item $A_J=0.283 \cdot A_V = (0.063\pm0.008)$ mag
\item $A_H=0.181 \cdot A_V = (0.044\pm0.006)$ mag
\item $A_K=0.116 \cdot A_V = (0.026\pm0.003)$ mag
\end{itemize}

\subsection{Bestätigung der Periode von LMC X-3}
\begin{figure}[h!]
\begin{center}
\includegraphics[width=16cm]{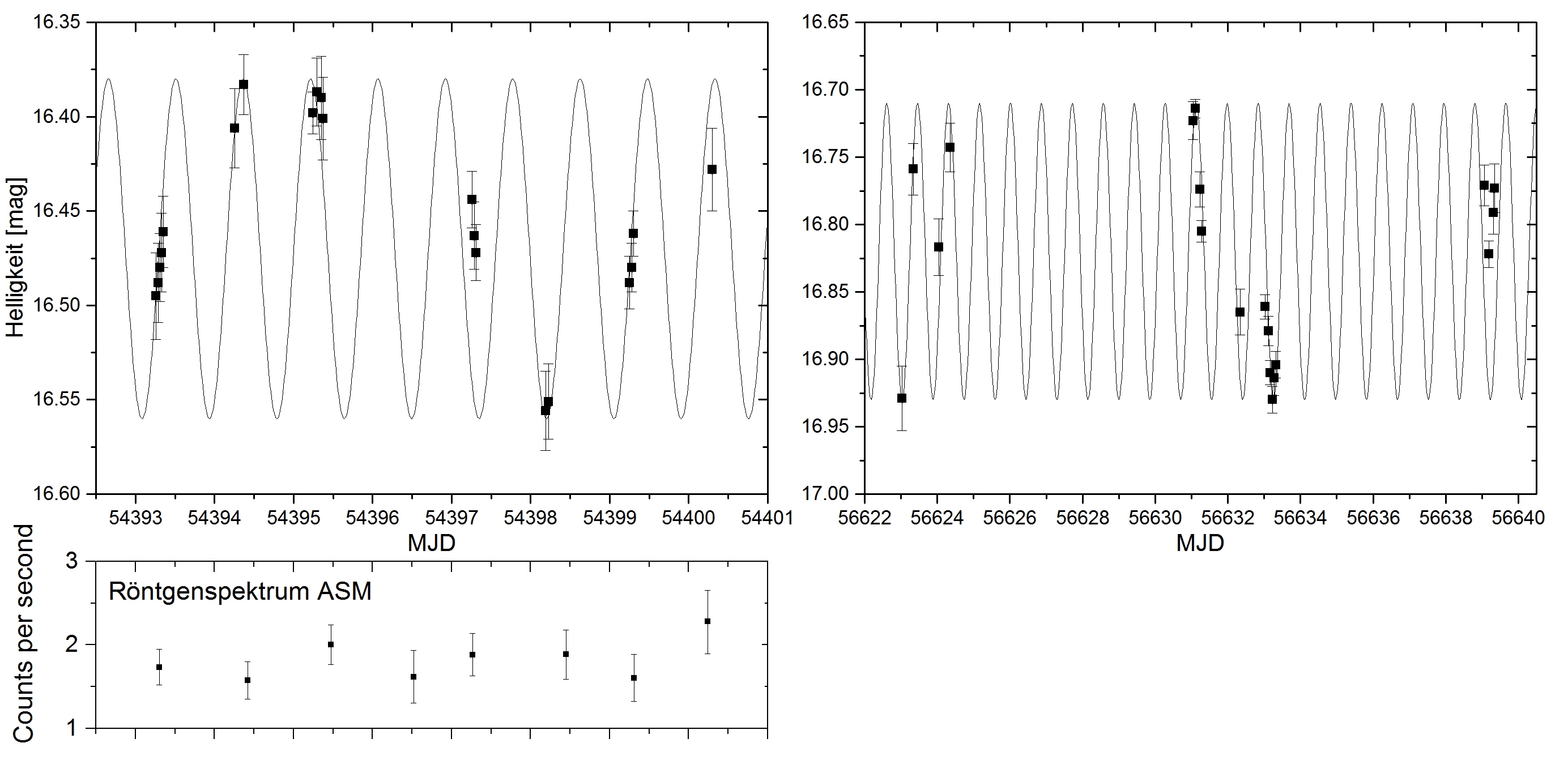}
\caption{Sinus mit Frequenz $2 \times 1,70481$ Tage über das g-Band der Epoche 2007 (links) und Epoche 2013 (rechts)}
\label{fig:Periodencheck}
\end{center}
\end{figure}

Da die Observation Blocks von LMC X-3 zeitlich sehr unregelmäßig weit auseinander liegen und meist nur wenige OBs in einer Nacht aufgenommen wurden, welche einen zeitlichen Abstand haben, der groß genug ist um die Helligkeitsschwankung aufzulösen, ist es sehr schwer mit den vorhandenen GROND Daten eine Periode des Doppelsternsystems zu bestimmen. Ebenfalls kommt erschwerlich hinzu, dass die Helligkeit des Systems zwischen 2007 und 2013 um circa 0.3 mag abgenommen hat, was bedeutet, dass die verschiedenen Epochen der Aufnahmen von LMC-X3 für eine Periodenbestimmung einzeln betrachtet werden müssen, da die Amplitude der Helligkeitsschwankung in der Größenordnung der absoluten Helligkeitsschwankung ist. Im Wesentlichen gibt es also drei Epochen: 18.-27. Oktober 2007, 22.-27. Februar 2012 und 26. November bis 12. Dezember 2013, wobei die Daten in 2012 komplett aus drei direkten Abfolgen von 4 Minuten OBs bestehen und so gut wie keine Helligkeitsänderung zu sehen ist (vgl. Kapitel \ref{sec:LK} Abbildung \ref{fig:LKr}), was darauf schließen lässt, dass die Periode (mit der der Begleitstern und das schwarze Loch um ihren gemeinsamen Schwerpunkt kreisen) in LMC-X3 lange im Vergleich zu den meisten anderen Doppelsternsystemen sein muss.\\
Auf Grund der wenigen und zeitlich nicht gleichmäßig verteilten Datenpunkte der einzelnen Epochen liefert die Lomb-Scargle-Methode zum Auffinden der Periode keine signifikanten Ergebnisse. Daher wird eine Periode von P=1,70481 Tage (J.A. Orosz et al. 2014) angenommen, welche eine Präzision von circa 100 Millisekunden aufweist. Dies wird dadurch versucht zu bestätigen, dass über die ungefalteten Lichtkurven der Epochen 2007 und 2013 eine Sinusfunktion gelegt wird und die Übereinstimmung der Periodizität der Lichtkurve mit der Sinuskurve optisch überprüft wird (siehe Abb. \ref{fig:Periodencheck}).\\
\vspace{-0.5cm}
Aufgetragen ist das g-Band der beiden Epochen. Der Sinus hat die doppelte Periodizität, wie die eigentliche Lichtkurve, da zur groben Abschätzung das tiefe und seichtere Minimum als identisch angenommen werden können. Wie in der Abbildung zu sehen ist, geht der Sinus durch die meisten Punkte hindurch. Jedoch ist es sehr vage eine Aussage über die Richtigkeit der Periode von P=1.70481 Tagen zu treffen. Im Vergleich mit den Röntgendaten lässt sich auch kein Zusammenhang finden zwischen den, vom Sinus abweichenden Punkten, und einem vermuteten Anstieg im Röntgenspektrum. Die mit ASM aufgenommenen Datenpunkte im Röntgenbereich sind im Bereich der Unsicherheit identisch. Ein vermuteter Anstieg im Röntgenbereich bei MJD 54397, wo die Punkte über dem Sinus liegen ist nicht zu finden. Im Weiteren wird die Periode von P=1.70481 Tagen als richtig angenommen und weiter verwendet.

\subsection{Lichtkurven von LMC X-3}
\subsubsection{Faltung}
Zur Erstellung der Lichtkurven von LMC-X3 wird die Helligkeit des BHBs über der Periode des Systems aufgetragen. Hierfür wird die absolute Zeitachse in MJD über die Periode gefaltet. Somit lässt sich auf Grund der durchaus großen zeitlichen Abstände der OBs die Periodizität der Helligkeit von LMC-X3 um einiges besser erkennen. In die Faltung der Lichtkurve über die Periode geht zum einen die Annahme ein, dass die Kreisfrequenz des Systems über den Zeitraum zwischen 2007 und 2013, also über sechs Jahre annähernd konstant ist. Diese Annahme scheint durchaus berechtigt, da der Massenstrom vom Begleitstern auf die Akkretionsscheibe und schließlich ins Schwarze Loch nur einen winzigen Anteil an der Masse des Systems ausmacht und somit wegen der Drehimpulserhaltung die Periode über einen, astronomisch gesehen, so kurzen Zeitraum konstant ist. Zum anderen geht die Annahme ein, dass das System auf der kurzen Zeitskala von wenigen Jahren keine Präzessionsbewegung ausführt und somit die große Halbachse des Doppelsternsystems konstant in Raum und Zeit ist.\\
In die Beobachtungszeit ist die Auswirkung des Baryzentrums im System Sonne-Erde nicht miteingerechnet. Auf Grund der Endlichkeit der Lichtgeschwindigkeit ist die Laufzeit des Lichts und damit auch die Ankunftszeit auf der Erde abhängig von der Position der Erde relativ zur Sonne. Somit kann die Entfernung der Erde maximal, falls das Beobachtungsobjekt sich in der Ekliptik befindet, um zwei Astronomische Einheiten, also 16 Lichtminuten je nach Jahreszeit schwanken. Da LMC X-3 aber unter einem Winkel von circa 64° zur Ekliptik steht ist die Laufzeitverlängerung des Lichts durch diesen Effekt nur bei maximal $sin(26°) \cdot 16 \ min = 7 \ min$. Da die Daten von 2007, wo man diesen Effekt möglicherweise beobachten könnte, weniger als ein viertel Jahr in der Phasenlage der Erdbewegung um die Sonne von den Daten aus 2013 entfernt sind, liegt die Laufzeitverlängerung bei 3.5 min. Dies ist also ein sehr kleiner Effekt, welcher auf die Lichtkurve so gut wie keine Auswirkungen hat. Die zeitliche Verschiebung der Messpunkte durch diesen Effekt liegt innerhalb der Zeitunsicherheit der Messungen.\\

\subsubsection{Variation zwischen den Epochen}\label{sec:LK}
\begin{figure}[h!]
\begin{center}
\includegraphics[width=13cm]{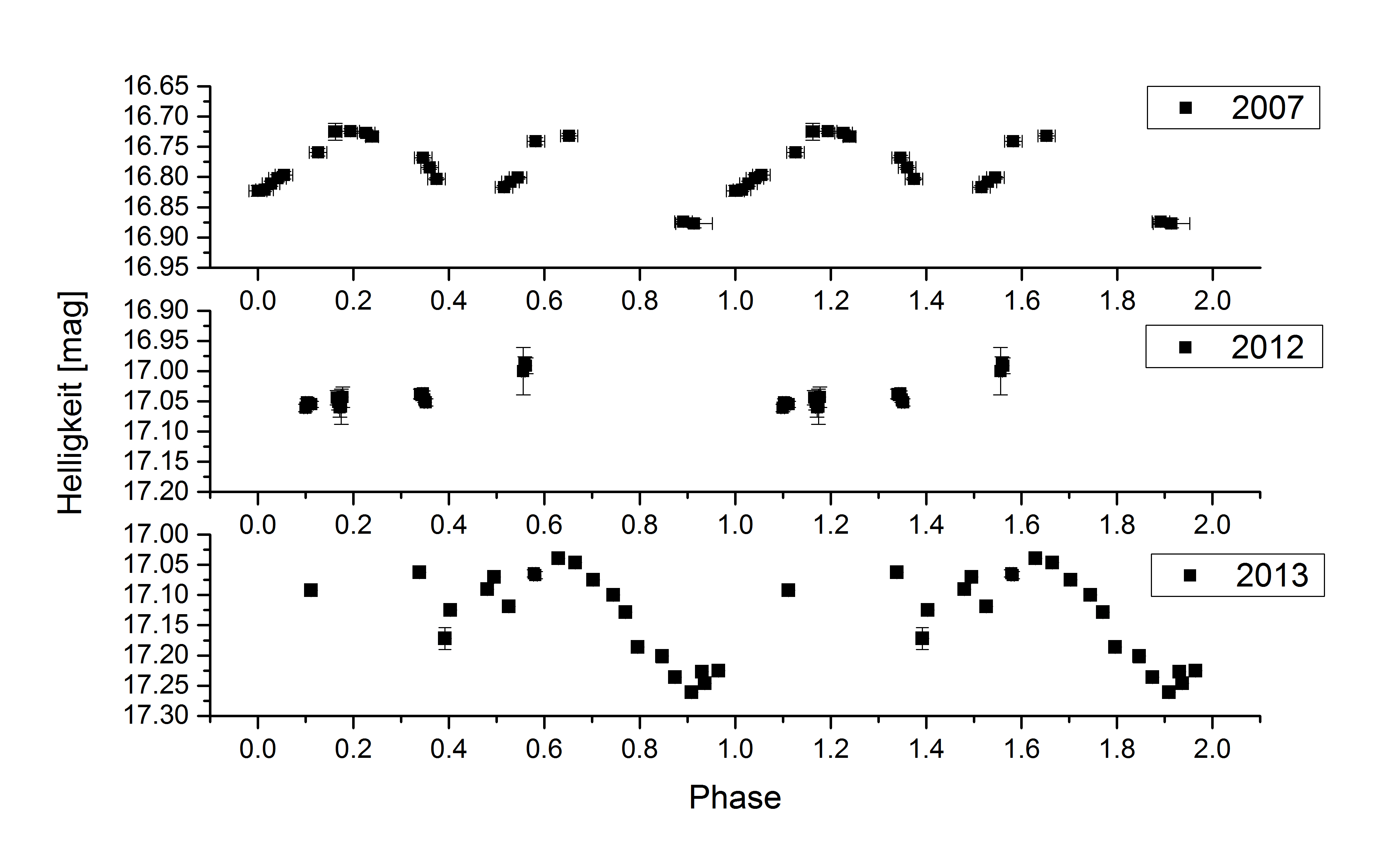}
\caption{Periodengefaltete Lichtkurven im r-Band: Jahre 2007, 2012 und 2013 separat}
\label{fig:LKr}
\end{center}
\end{figure}

Die Datenpunkte für die Lichtkurven sind im Anhang in Tabelle \ref{tab:Daten} zu finden. Als Beispiel ist hier die Lichtkurve für das r-Band dargestellt. In Abbildung \ref{fig:LKr} sind die drei Epochen, mit Ausnahme der vier Datenpunkte von 2008 aufgezeichnet. Zunächst fällt es auf, dass die Helligkeiten im Zeitraum von 2007 zu 2013 um circa 0.25 mag schwanken, beziehungsweise dass die Helligkeit von LMC X-3 stetig abgenommen hat in diesem Zeitraum, was es nötig macht die einzelnen Zeiträume auch separat zu betrachten, da die Helligkeitsänderung auf eine Änderung der Leuchtkraft im Doppelsternsystem zurückzuführen sein muss. Weiter bemerkt man sofort, dass aus den Daten von 2012 wenig Aussagen getroffen werden können, weil die Messungen viel zu eng liegen, bzw. zu stark gehäuft sind. Deshalb werden für die weitere Betrachtung der Lichtkurven diese nicht mehr herangezogen.
\begin{figure}[h!]
\begin{center}
\includegraphics[width=13cm]{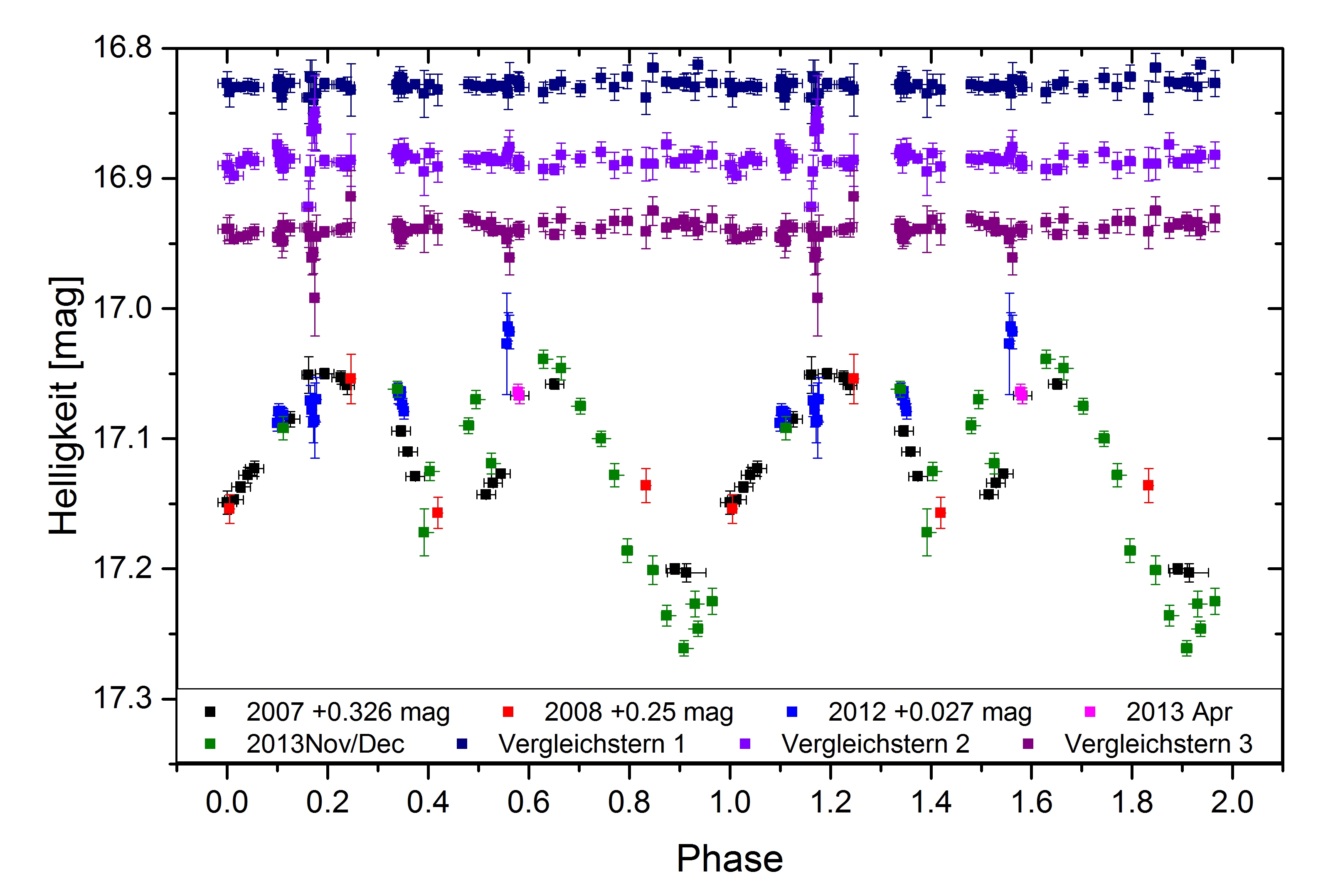}
\caption{Lichtkurven im r-Band: um den Magnitudenoffset der versch. Jahre verschoben; zusätzlich drei Vergleichsterne}
\label{fig:LKrverschoben}
\end{center}
\end{figure}
In Abbildung Abb. \ref{fig:LKrverschoben} sind die Lichtkurven der einzelnen Jahre für das r-Band so verschoben, dass sie übereinanderliegen. Man kann erkennen, dass die Amplituden der einzelnen Epochen, gerade zwischen 2007 und 2013 unterschiedlich sind. Dies kann einen physikalische Effekt haben, und zum Beispiel eine Auswirkung der Akkretionsscheibe sein, oder es ist auf die Abdeckung mit Datenpunkten zurückzuführen, die, wie man sehen kann, nicht optimal ist.\\

\subsubsection{Kalibration}
\begin{figure}[h!]
\begin{center}
\includegraphics[width=10cm]{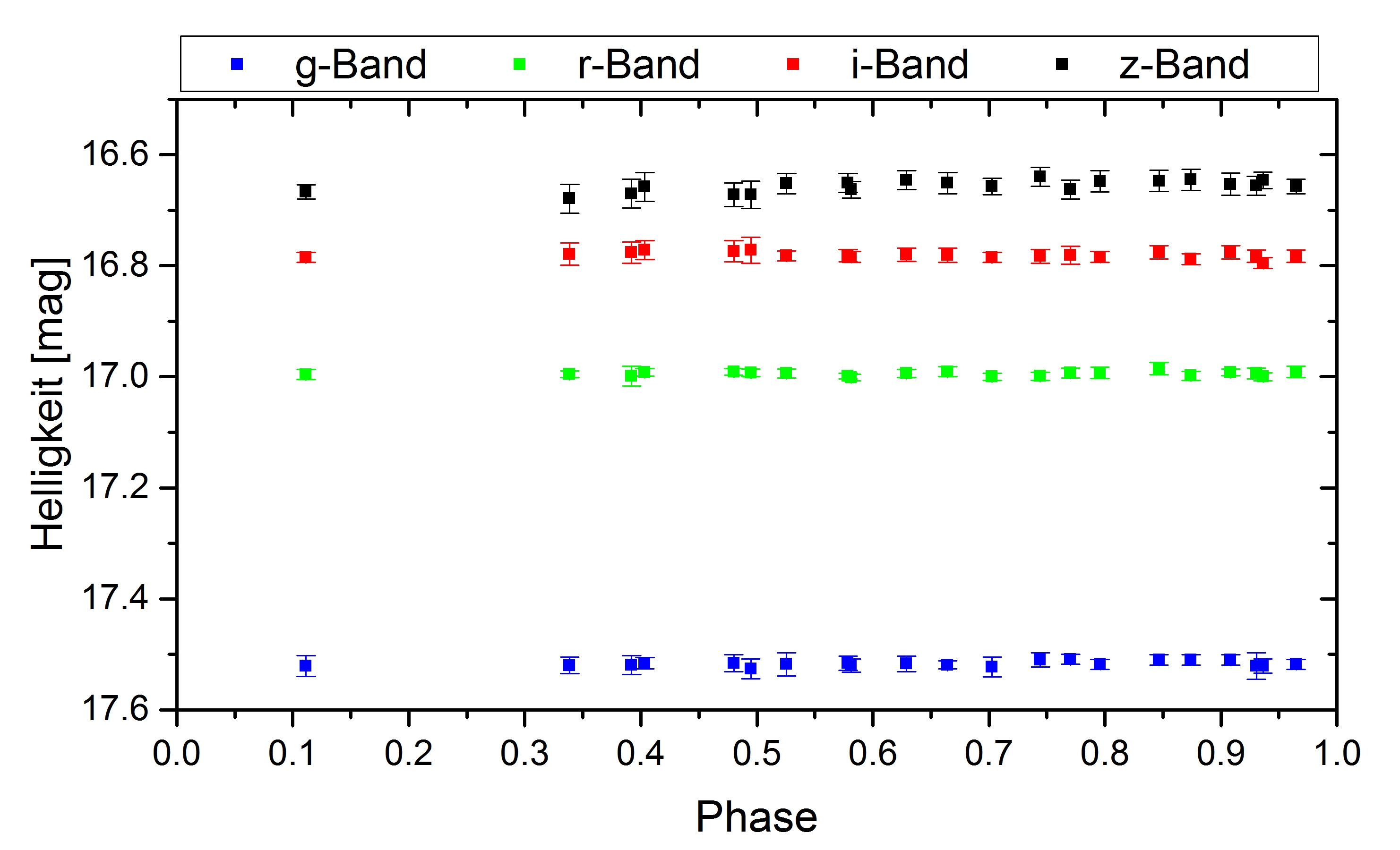}
\caption{Lichtkurve eines Vergleichsterns in den optischen Bändern mit den Daten aus 2013}
\label{fig:VglStern}
\end{center}
\end{figure}

Um Fehler in der Lichtkurve auszuschließen, die durch die Kalibration entstehen, sind in Abbildung \ref{fig:LKrverschoben} zusätzlich über der Lichtkurve von LMC X-3 die Lichtkurven von drei Vergleichsternen, welche zur Kalibration benutzt wurden aufgetragen. Man erkennt, dass die Lichtkurven im Rahmen der Unsicherheit einen konstanten Wert annehmen. Daraus kann geschlossen werden, dass mit stabilen unveränderlichen Lichtquellen kalibriert wurde und daher keine Kalibrationseffekte in der Lichtkurve von LMC X-3 zu sehen sind. Die Koordinaten dieser Sterne sind:\\ Vergleichstern 1: 84.74431 -64.07585;\ \ Vergleichstern 2: 84.70564 -64.07820;\ \ Vergleichstern 3: 84.74421 -64.06192.\\
Zusätzlich als Vergleich sind die Lichtkurven aus dem Jahr 2013 in allen optischen Bändern von Vergleichstern 3 in Abbildung \ref{fig:VglStern} aufgetragen. Man erkennt, dass die Helligkeit nur innerhalb der Unsicherheit schwankt, also als konstant angenommen werden kann.

\subsubsection{Relative Amplituden}
Um zu sehen, welche Komponenten Auswirkung auf die Lichtkurve haben, ist es interessant, die Amplitude der Lichtkurve in den einzelnen Bändern zu untersuchen. In Tabelle \ref{tab:Amplituden} sind die Amplituden, die sich aus der Differenz der Magnituden von OB0\_0, welches sich an einem Maximum befindet und OB17\_4, welches sich im tiefen Minimum befindet eingetragen. Dabei ist das K-Band weggelassen, das die Unsicherheiten zu groß sind um eine Aussage zu treffen und man in der Lichtkurve kein eindeutiges Maximum und Minimum erkennen kann.
\begin{table}[h]
\begin{center}
\begin{tabular}{|c|c|} \hline
\textbf{Band} & \textbf{Magnitudendifferenz}  \\ \hline
$g$	& $0,216\pm0,017$ \\
$r$	& $0,222\pm0,011$ \\
$i$	& $0,225\pm0,024$ \\
$z$	& $0,226\pm0,038$ \\
$J$	& $0,301\pm0,075$ \\
$H$	& $0,354\pm0,151$ \\
\hline
\end{tabular}
\caption{Amplituden der Lichtkurven von 2013 in den verschiedenen Bändern (ohne K-Band)}
\label{tab:Amplituden}
\end{center}
\end{table}
Die ermittelten Amplituden stimmen alle im Rahmen der Unsicherheit überein. Es ist aber eine Tendenz zu erkennen, dass die Amplitude mit steigender Wellenlänge größer wird. Dies könnte auf eine konstante zusätzliche blaue Komponente, welche heißer als der Begleitstern ist, zurückzuführen sein.\\
\vspace{-0.5cm}
\subsubsection{Mögliche Asymmetrie}
Über eine mögliche Asymmetrie der Maxima, welche auf einen Hot Spot hinweisen könnte, lässt sich bei einzelner Betrachtung der Epochen schwer eine Aussage treffen, was die Betrachtung eines Hot Spots bei Regionen um 90° und 270° ausschließt. Ebenso lässt sich in den Minima keine Asymmetrie und spezielle Form erkennen, die signifikant ist, wie man sie bei einem Hot Spot in der Nähe von 0° und 180° erwarten würde (vgl. Kapitel \ref{sec:Parameterstudie}). Aus diesen Gründen, die speziell auf die Datenabdeckung zurückzuführen sind, wird bei der Simulation und Modellierung der Lichtkurven eine Hot Spot-Komponente weggelassen.\\
In Kapitel \ref{sec:LKModell} sind die Lichtkurven von LMC X-3 von 2007 und 2013 in allen Bändern zusammen mit ihren simulierten Lichtkurven, welche am besten auf die experimentellen Daten passen, gezeigt.

\subsubsection{Analyse der Aktivität von LMC X-3 im Röntgenbereich}\label{sec:Roentgen}
\begin{figure}[h]
\center
\subfloat[\label{fig:XTE}Summenspektrum von LMC X-3 aufgenommen mit dem Detektor ASM am XTE]
{\includegraphics[width=11cm]{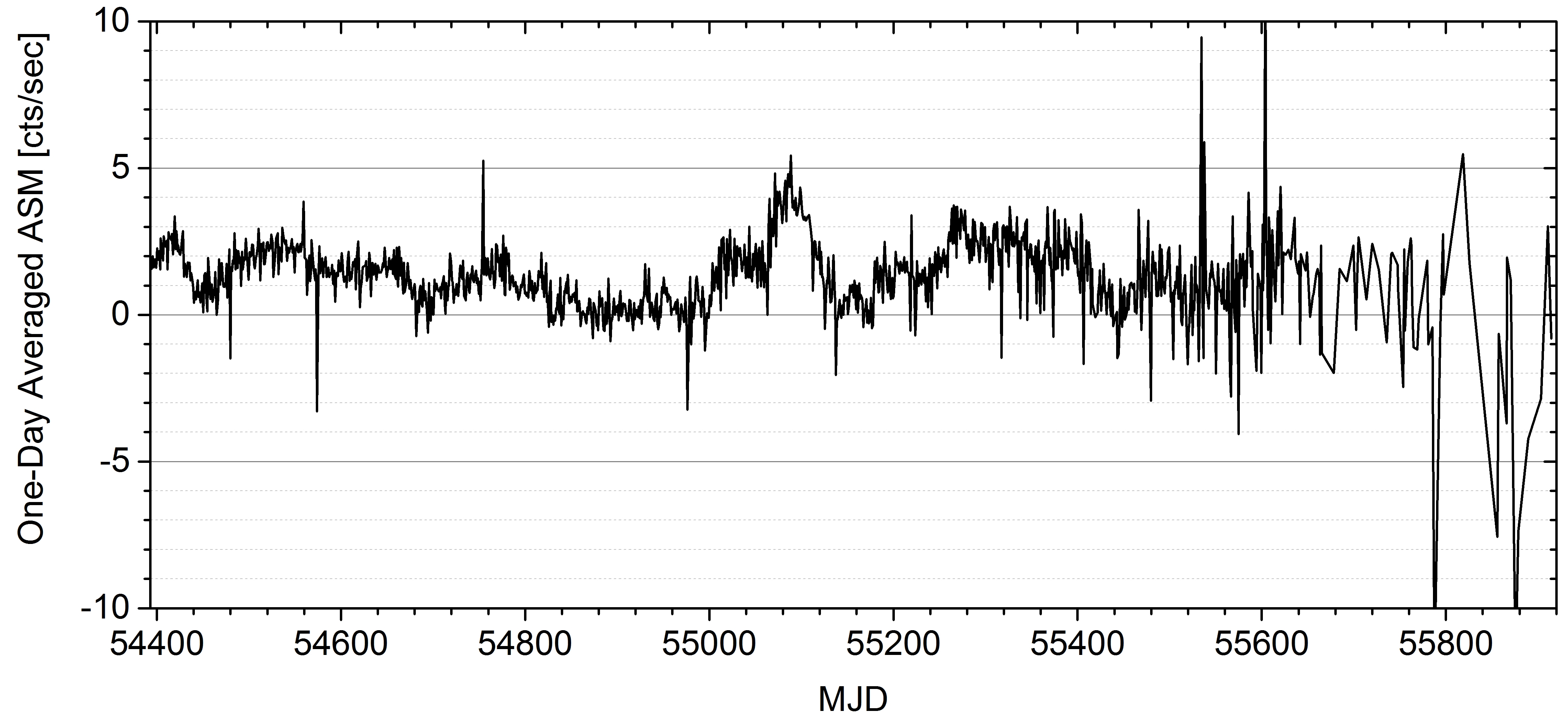}}
\quad
\subfloat[\label{fig:MAXI}Summenspektrum von LMC X-3 aufgenommen mit MAXI]
{\includegraphics[width=12cm]{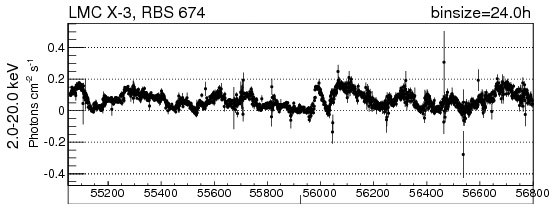}}
\caption{Röntgenspektrum von LMC X-3 vom 01.01.2007 (XTE) bis zum 23.05.2014 (MAXI)}
\end{figure}

Um herauszufinden, weshalb die Messpunkte von 2007 im Vergleich zu denen von 2013 ungefähr 0.25 mag heller sind wird das Röntgenverhalten von LMC X-3 überprüft. Dies geschieht mit den Daten von ASM am RXTE-Satelliten und denen von MAXI auf der ISS. Es werden dafür zwei verschiedene Messdatensätze gebraucht, weil RXTE am 5. Januar 2012 deaktiviert wurde und MAXI erst am 7. August 2009 in Betrieb ging. Für diesen Zeitraum überlappen sich also die Lichtkurven beider Detektoren. In Abbildung \ref{fig:XTE} und \ref{fig:MAXI} sind die Lichtkurven der beiden Detektoren ab 2007 aufgezeichnet, wobei die ASM Daten in Energiebereich zwischen 2-10 keV liegen. 
Es sind in der Beobachtungszeit keine signifikanten Röntgenausbrüche zu beobachten, wie es in Kapitel \ref{sec:XRAY} der Fall war. Es gilt also den Röntgenfluss genauer zu untersuchen: Die Daten von MAXI könnten anhand des Überlappungszeitraumes auf die von ASM normiert werden, wobei die Daten von MAXI einfach mit dem Faktor 25 multipliziert werden müssen. Die Unterschiede in den Zählraten von ASM und MAXI hängen mit dem Detektordesign zusammen. Es sind unkalibrierte Zählraten, die von der Empfindlichkeit der jeweiligen Sensoren abhängen. Die Einheit der Zählrate bei ASM ist Counts per Second, während die von MAXI noch auf die Fläche von einem Quadratzentimeter normiert ist. Als Resultat erhält man ungefähr folgende Zählraten für die jeweiligen Beobachtungszeiträume:
\begin{itemize}
\item $\approx$ 2 cts/s		für 2007
\item $\approx$ 1.5 cts/s	für 2008
\item $\approx$ 1.25 cts/s für 2013 Apr
\item $\approx$ 2.5 cts/s 	für 2013 Nov/Dez
\end{itemize}
Man erkennt also, dass die Röntgenaktivität nicht über die ganze Zeit abnimmt, wie eigentlich erwartet. Daraus kann man schließen, dass kein erkennbarer direkter Zusammenhang der sinkenden Magnitude im optischen und nahen Infrarotbereich über die sechs Jahre Beobachtungszeitraum mit der Röntgenaktivität von LMC X-3 besteht. Es folgt daraus, dass die Helligkeitsabnahme wahrscheinlich mit der Leuchtkraft der Disk zusammenhängt, da diese in kurzen Zeiträumen sich stark verändern kann.\\
Im Jahr 2001 gab es zwei Veröffentlichungen, welche die Röntgenaktivität von LMC X-3 genauer untersuchten. Sie fanden eine signifikante orbitale Phasenmodulation im 2-10 keV Röntgenfluss (Boyd et al. 2001), welche sie auf einen Hot Spot, oder eine Region kühlerer absorbierender Materie, oder beides zurückführen. Die andere Veröffentlichung (Brocksopp et al. 2001) untersuchte die Langzeitvariabilität von LMC X-3 und fanden heraus, dass die periodischen Schwankungen im Röntgenbereich nicht besonders deutlich im Optischen zu sehen sind. Weiter fanden sie aber, dass die Röntgenkurven den optischen um 5-10 Tage zeitlich versetzt auftreten. Daraus schlossen sie eine variable Akkretionsrate mit einem Massenstrom, der diese Zeit braucht sich vom Außenrand zum Innenrand der Scheibe zu bewegen. Eine Erklärung für den variablen Massenstrom kann eine variierende Begleitsterngröße sein, welche temporär den Roche-Lobe überschreitet.\\
Dies alles lässt darauf schließen, dass sich LMC X-3 nicht in einem wirklichen ruhigen Zustand befindet. Dies kann dadurch erklärt werden, dass der Massenüberfluss des massiven Begleitsterns sehr groß ist und im Vergleich zu anderen BHBs auf viel kürzeren Zeitskalen schwankt. Dies kann so interpretiert werden, dass sich LMC X-3 ständig in einem Röntgen-Ausbrauchstadium befindet (Orosz et al. 2014).

\subsection{Spektrale Energieverteilung (SED) von LMC X-3}
\begin{figure}[h!]
\begin{center}
\includegraphics[width=16.5cm]{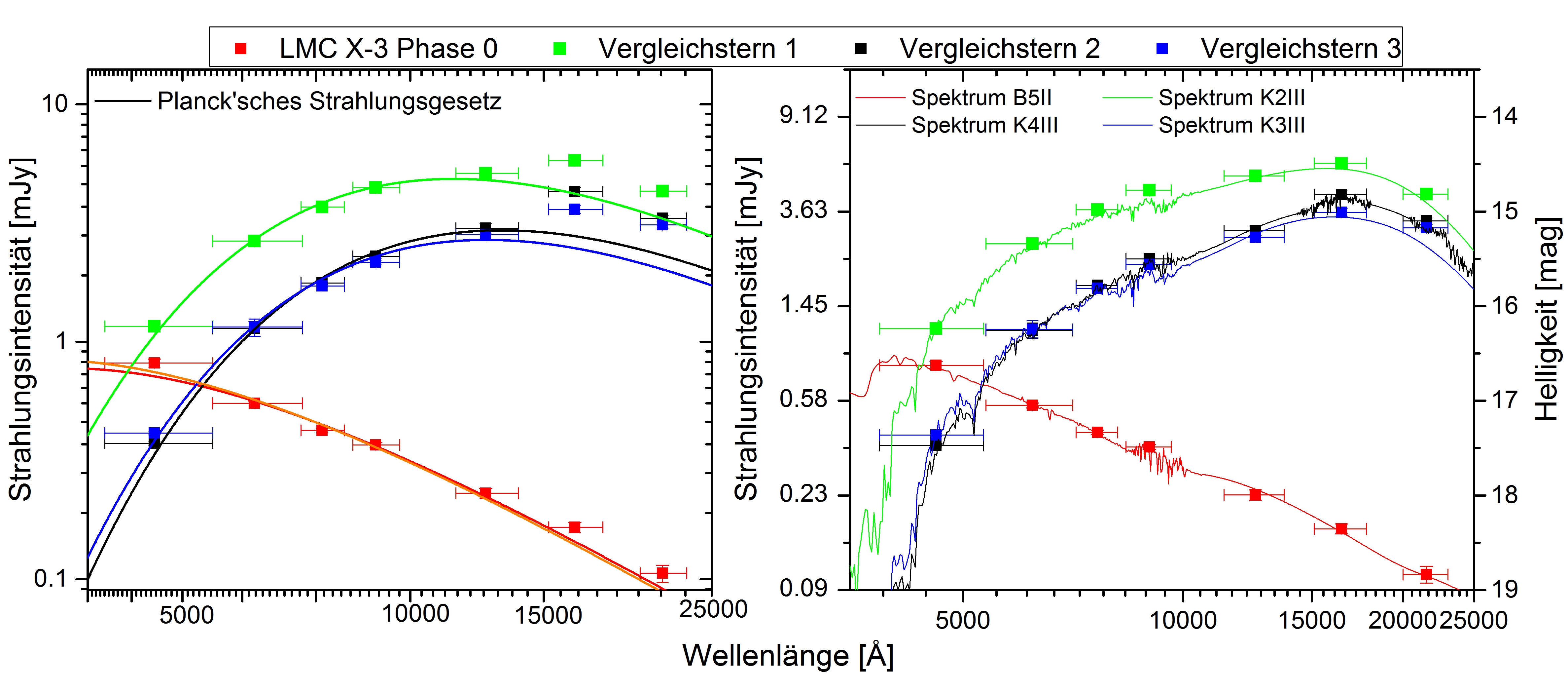}
\caption{SED von LMC X-3 und drei Vergleichsternen: links mit Planck'schem Strahlungsgesetz und rechts mit Sternspektren gefittet}
\label{fig:Spektrum2}
\end{center}
\end{figure}

\begin{figure}[h!]
\begin{center}
\includegraphics[width=13cm]{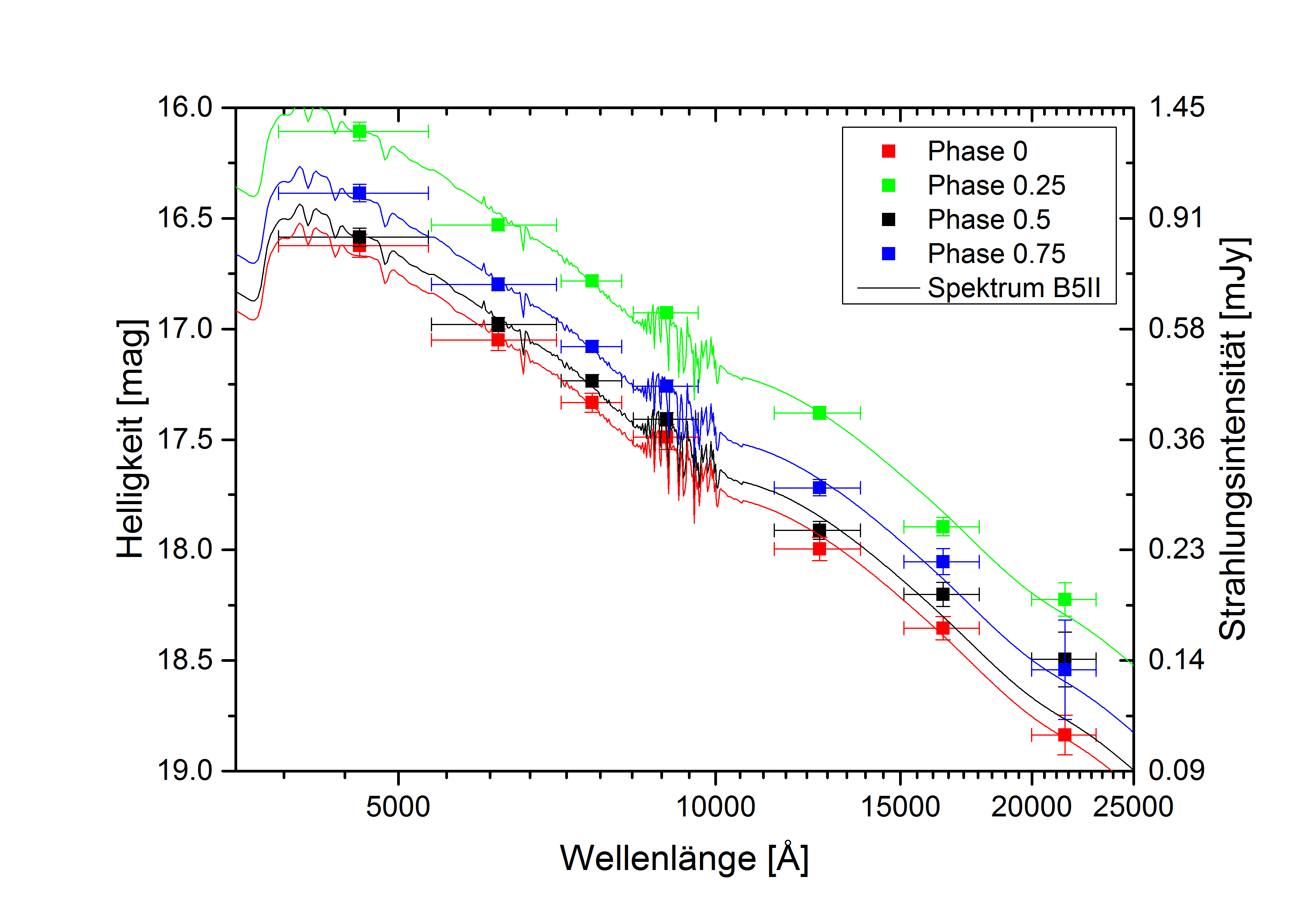}
\caption{SED der vier Phasen von LMC X-3 mit dem Spektrum eines B5II-Sterns gefittet}
\label{fig:Spektrum1}
\end{center}
\end{figure}

Die Daten von GROND bieten den großen Vorteil, dass in sieben Energiebändern gleichzeitig gemessen werden kann. Dies lässt sich nutzen, um im Wellenlängenbereich von circa 4000\AA \ bis 21000\AA \ eine spektrale Energieverteilung von LMC-X3 zu erstellen. Da die Quelle in ihrer Helligkeit nicht konstant ist und die verschiedenen Bänder unterschiedlich große Amplituden in ihren Helligkeitsschwankungen haben, macht es Sinn für jede viertel Phase des Systems eine eigene SED zu erstellen, da durch die verschiedenen Betrachtungswinkel das Spektrum sich unterscheiden wird. Hierzu wurden für die jeweiligen Bänder die Daten folgender OBs verwendet:
\begin{itemize}
\item Phase 0:		7. Dezember 2013 OB17\_4
\item Phase 0.25	6. Dezember 2013 OB16\_1
\item Phase 0.5	24. Februar 2012 OB5\_3
\item Phase 0.75	22. Oktober 2007 OB3\_2
\end{itemize}
Durch die Form der Spektren kann man Rückschlüsse ziehen, welche Temperatur der Begleitstern besitzt und ob die Akkretionsscheibe eine wichtige oder eher untergeordnete Rolle im Spektrum spielt. Folglich kann man daraus schließen, welche Komponente wie stark in die einzelnen Bänder eingeht, da das Strahlungsverhalten des Sterns als Schwarzkörperspektrum angenommen wird und das Spektrum der Disk als Überlagerung mehrerer Schwarzkörperspektren, die über die Temperatur der Scheibe am inneren und äußeren Rand begrenzt werden. Diese Schwarzkörperspektren folgen dem Planck'schen Strahlungsgesetz:
\begin{equation}
w_{\nu}(\nu)=\frac{8\pi h \cdot \nu^3}{c^3}\cdot \frac{1}{e^{\frac{h \nu}{k_B T}}-1}
\end{equation}
was der Planck'schen Strahlungsverteilung in Frequenzdarstellung entspricht (W. Demtröder 2012). Das Akkretionsscheibenspektrum hingegen ist komplizierter, da es mit $\nu^2$ anfänglich nach dem Rayleigh-Jeans-Gesetzt, dann mit $\nu^{\frac{1}{3}}$ steigt und schließlich mit $\nu^3 \cdot e^{\frac{- h \nu}{k_B T}}$, dem Wien'schen Strahlungsgesetz folgend abfällt.\\
Beim Untersuchen der SED wird ein Planck'sches Strahlungsgesetz gefittet. Es ergeben sich folgende Temperaturen für die jeweiligen Phasen:
\begin{itemize}
\item Phase 0   : $T=(14900\pm1400)K$
\item Phase 0.25: $T=(13800\pm1500)K$
\item Phase 0.5 : $T=(14600\pm1600)K$
\item Phase 0.75: $T=(13200\pm1200)K$
\end{itemize}
Die Temperaturen für die jeweiligen Phasen stimmen im Rahmen von $1\sigma$ miteinander überein. Es ist also schwer eine Aussage darüber zu machen, ob die Akkretionsscheibe heißer, oder kälter als der Stern ist. Trotzdem fällt auf, dass tendenziell die Temperaturfits für die Phasen 0,25 und 0,75, also bei seitlicher Betrachtung kleinere Temperaturen ergeben. Dagegen wird die Temperatur bei Blick auf die Scheibe mit Begleitstern dahinter tendenziell niedriger gefittet. Zur genaueren Bestimmung der Temperatur des Begleitsterns, und um eine Aussage über die Scheibentemperatur machen zu können, werden die Bilder der OBs, welche sich im tiefen Minimum bei Phase 0 befinden (OB10\_1, OB17\_3 bis OB17\_6) übereinandergelegt und auf diese Summe erneut die Photometrie anwendet. Man erwartet also ein stark vom Stern dominiertes Spektrum des gegebenenfalls eine Scheibenkomponente beinhaltet. Um zu sehen, ob die Kalibration korrekt ist, wird ebenfalls eine SED von drei Vergleichsternen erstellt, welche im g-Band sehr hell sind und folgende Koordinaten besitzen:
\begin{itemize}
\item Vergleichstern 1: 84.70717	-64.07516
\item Vergleichstern 2: 84.73859	-64.07173
\item Vergleichstern 3: 84.78426	-64.07311
\end{itemize}
Da die Vergleichsterne am hellsten im r-Band sind haben sie eine Oberflächentemperatur von etwa 4500K. Dadurch ist ihre Leuchtkraft wesentlich schwächer als die von LMC X-3. Daher kann man davon ausgehen, dass sie sich nicht in der Großen Magellanschen Wolke befinden, sondern noch in der Milchstraße. Folglich muss ein anderer Extinktionsfaktor für die Vordergrundextinktion für die Vergleichsterne verwendet werden. Er beträgt in die Richtung von LMC X-3: E(B-V)=0,05 mag (NASA/IPAC Extragalactic Database). Bim Fit eines Schwarzkörperspektrums erkennt man gerade bei den Vergleichsternen eine starke Abweichung im J- und K-Band (vgl. Abbildung \ref{fig:Spektrum2} links). Ebenso ist bei LMC X-3 bei der orangen Linie, welche einen Fit bei 15500K darstellt, eine geringe Abweichung im K-Band festzustellen. Die rote Linie ist die beste angepasste Kurve ($T=(14500\pm1600)K$ und zeigt eher eine Abweichung im g-Band.\\
Um festzustellen ob ein Kalibrationsfehler vorliegt werden nun reale Sternspektren (Pickles 1998) über die Vergleichsterne und über LMC X-3 bei Phase 0 gelegt (vgl. Abbildung \ref{fig:Spektrum2} rechts). Darin erkennt man, dass die Sternspektren relativ gut mit den Datenpunkten übereinstimmen. Nur beim Vergleichstern 3 ist eine sehr leichte Abweichung im K-Band zu beobachten. Die Spektralklassen der Spektren sind im Bild abzulesen. Für LMC X-3 wurde ein B5II-Stern gefittet. Dies stimmt nicht ganz mit der von Orosz et al. 2014 verwendeten Spektralklasse von B3V (Cowley et al. 1983) überein, ist jedoch berechtigt, da neue Ergebnisse eine Schwankung zischen B3V und B5V aufweisen (Val-Baker et al. 2007) und die Helligkeitsklasse durch den Amplitudenfit keine Rolle spielt. Ebenfalls wurde von Soria et al. (2001) auf der Basis von XMM Optical Monitor Daten postuliert, dass eine Spektralklasse von B5IV, also die eines Unterriesen passender ist als B3V. Ebenfalls bestätigte Orosz et al. (2014) auch, dass die Oberflächengravitationsmessungen eher einem Riesen (Klasse III) anstelle eines Hauptreihensterns (Klasse V) entsprechen. Val-Baker et al. (2007) bestimmten den Spektraltyp auf B3V während der Phase bei der das X-Ray heating dominant ist, und B5V bei schwachem X-Ray heating; Orosz et al. (2014) hingegen schließen auf Grund ihrer weit größeren Datenmenge einen mäßigen Effekt des X-Ray heatings.
Wenn man nun das Spektrum des B5II-Sterns über alle vier Phasen von LMC X-3 legt, ergibt sich Abbildung \ref{fig:Spektrum1}.
Die Anpassung der Sternspektren wurde durch Arne Rau durchgeführt. Es lässt sich darin erkennen, dass für alle Phasen das Spektrum an die Datenpunkte fittet, außer bei Phase 0.5 im K-Band, welches einen höheren Fluss aufweist, als das Spektrum. Da dies die Position ist, bei der man direkt auf die Akkretionsscheibe blickt und der Begleitstern sich dahinter befindet ist dies ein starker Hinweis darauf, dass die Temperatur der Scheibe kälter ist, als die des Sterns.

\subsection{Farbveränderungen von LMC X-3}
\begin{figure}[h!]
\begin{center}
\includegraphics[width=13.5cm]{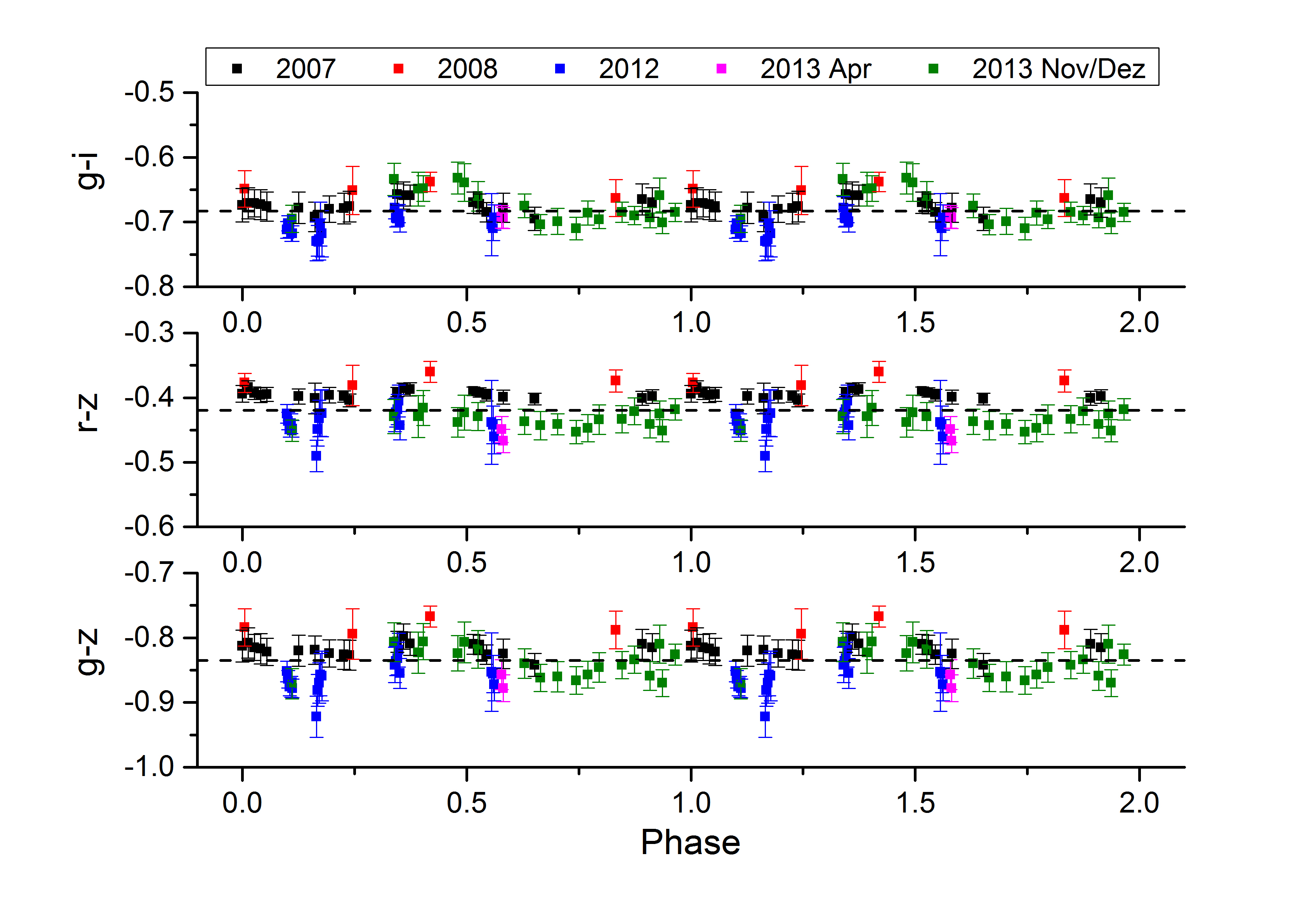}
\caption{Differenz der Helligkeiten diverser Bänder über der Periode: Oben: g-i, Mitte: r-z, Unten: g-z}
\label{fig:Farben}
\end{center}
\end{figure}

Für eine Auftragung der Farbveränderung wird in Abbildung \ref{fig:Farben} die Differenz aus den verhältnismäßig blaueren mit den verhältnismäßig röteren Bändern aufgetragen (g-i, r-z, g-z).
Zusätzlich zu den Datenpunkten ist jeweils der Mittelwerte der einzelnen Datenpunkte als gestrichelte Linie aufgetragen. Die Mittelwerte lauten:\\
\ \ für g-i: 0.683 mag \ \ \ \ \ \ für r-z: 0.419 mag \ \ \ \ \ \ für g-z: 0.834 mag \\
Es ist zu erkennen, dass relativ zum Mittelwert gesehen, die Datenpunkte des Jahres 2008 nach rot verschoben sind, während die des Jahres 2007 nach blau verschoben sind. Ebenfalls sieht man, dass die Daten von 2013 im Mittel bläulicher sind, als die des Jahres 2007. Es lässt sich also keine einheitliche Tendenz erkennen, wie sich das System farblich gesehen auf längeren Zeitskalen entwickelt, da für die beobachteten Effekte, wie man vor allem in den Daten von 2013 sieht, auch Schwankungen auf einer Skala von Tagen oder Wochen verantwortlich sind.\\
Was allerdings deutlich zu erkennen ist, gerade bei g-i, ist ein leichtes Maximum bei circa Phase 0.5, also wenn hauptsächlich die Akkretionsscheibe zu sehen ist. Da dies röter ist, kann dadurch auf eine Scheibe geschlossen werden, welche kälter ist als der Begleitstern. Für alle anderen Phasen stimmen die Datenpunkte fast immer im Rahmen der Unsicherheit mit dem Mittelwert überein.

\clearpage

\section{Modellierung}
\subsection{Erweiterung von existierenden Programmen und neu erstellte Programme}
Zur Beschreibung der existierenden Programme, welche für die Modellierung der Lichtkurven verwendet werden können, siehe F. Schiegg 2013. Einige dieser Programme mussten verändert und erweitert werden:
\begin{itemize}
\item \textbf{xrb\_MakeDB.py:} Dies ist ein Programm von Fabian Knust und Florian Schiegg, welches die von XRbinary erzeugten Lichtkurven in die GROND-Bänder umrechnet und generell XRbinary Parameter übergibt. Es wurde insofern geändert, dass nun zusätzlich über die Periode des Doppelsternsystems, weiter über die Masse und die Temperatur des Begleitsterns unabhängig iteriert werden kann. Das war nötig, weil die verwendete Formel zur Berechnung der Masse des Begleitsterns über seine Temperatur für Sterne der Spektralklasse B sehr schlechte Werte für die Masse liefert, da sie auf kleinere Sterne optimiert ist.\\
Des Weiteren wurde die Benennung der Ausgabedatei so verändert, dass alle veränderlichen Parameter im Dateinamen direkt genannt werden, und nicht mehr einer Massencodierung folgen, was schließlich sehr beim weiteren Benutzen dieser Dateien nützt, da auf den ersten Blick alle Parameter erkennbar sind. Zusätzlich wurde der Header der Ausgabedateien um die neuen Parameter ergänzt.
\item \textbf{fs\_ExpLK.py:} Dies ist ein Programm um aus den Gesamtdaten die einzelnen Bänder in eigene Dateien zu schreiben. Es wurde so modifiziert, dass es nun auch mit unvollständigen Datensätzen (für ein paar OBs war aus diversen Gründen keine Magnitude für LMC X-3 vorhanden) arbeitet, insofern für die Magnitude OBs im fehlenden Band der Wert \glqq -99\grqq \ eingetragen wird.
\item \textbf{fs\_CalcPhaseOffset.py:} Dieses Programm dient dazu den Phasenoffset der experimentellen Lichtkurve gegenüber einer beliebigen theoretischen Lichtkurve zu bestimmen. Hierzu wird nach der $\chi^2$-Methode vorgegangen und der Phasenoffset mit dem kleinsten $\chi^2$ ausgegeben. Es wurde ebenfalls so erweitert, dass unvollständige Datensätze eingelesen werden können.
\item \textbf{fs\_CalcChi.py:} Diesem Programm wird ein Verzeichnis mit mehreren simulierten Lichtkurven übergeben, daraus werden die $\chi^2$ und der Magnitudenoffset berechnet und in eine Datei geschrieben. Ebenfalls wurde dieses Programm erweitert, dass unvollständige Datensätze eingelesen werden können. Im Header stehen nun alle in xrb\_MakeDB.py veränderlichen Parameter.
\item \textbf{fs\_errLKComp.py:} In diesem Programm werden die simulierten Lichtkurven über die experimentellen gelegt. Dies wird auf Basis des berechneten Phasen- und Magnitudenoffsets bewerkstelligt. Es wurde insofern geändert, dass nun immer zwei komplette Phasen dargestellt werden, was so Standard ist und der Übersicht dient. Des Weiteren können nun für jeden OB die Unsicherheiten in der Zeit getrennt sowohl für die optischen, als auch für die Bänder im nahen Infrarotbereich dargestellt werden, wobei ebenfalls wie in den oben beschriebenen Programmen nun auch unvollständige Datensätze eingelesen werden können.
\item \textbf{fs\_TheoLKComp.py:} Mit diesem Programm können simulierte Lichtkurven dargestellt werden. Es wurde ebenfalls so erweitert, dass zwei volle Perioden angezeigt werden.
\end{itemize}
\clearpage

Neu erstellt wurden folgende Programme:
\begin{itemize}
\item \textbf{dg\_HSPlot.py, dg\_NW\_HSTemp.py, dg\_MBH\_i.py} In diesen Programmen wird ein zweidimensionales $\chi^2$-Diagramm erstellt, in welchem einmal der Winkel der Position des Hot Spots gegen die Temperatur des Hot Spots aufgetragen wird, einmal die Bahnneigung des Doppelsternsystems über der Hot Spot Temperatur und einmal die Masse des Schwarzen Lochs über der Bahnneigung dargestellt wird.
\item \textbf{dg\_findminima.py} Dieses Programm liest eine Datei, die von \glqq fs\_CalcPhaseOffset.py\grqq \ oder \glqq fs\_CalcChi.py\grqq \ erstellt wurde und die $\chi^2$ enthält ein und hat als Ausgabe eine Liste mit den Phasenoffsets und den zugehörigen $\chi^2$, welche darauf in einem Diagramm dargestellt werden können. Weiter findet das Programm den kleinsten $\chi^2$-Wert und den zugehörigen Phasenoffset.
\end{itemize}

\subsection{Simulation der Lichtkurven}\label{sec:Parameter}
\subsubsection{Parameter}
Bei der Simulation der Lichtkurven mit dem Programm XRbinary welches eine Vielzahl von Parametern enthält, können insbesondere, übergeben durch das Programm xrb\_MakeDB, folgende Parameter zur Erstellung eines Sets an Lichtkurven variiert werden: Masse des Schwarzen Lochs, Masse des Begleitsterns, Temperatur des Begleitsterns, Bahnperiode, Bahnneigung, Temperatur am Außenrand der Akkretionsscheibe, Position des Hot Spots, Temperatur des Hot Spots. Diese Parameter bilden also einen Achtdimensionalen Raum in welchem die am besten fittende Lichtkurve liegen sollte. Auf Grund des immensen Rechenaufwands, der exponentiell mit der Anzahl der Parameter steigt, ist es sinnvoll nur wenige Parameter zu variieren und sich so an ein Minimum heranzutasten. Einige Parameter, wie die Massen und die Bahnperiode ändern nur die Form des Roche-Lobe und haben so, wie in der Parameterstudie beschrieben nur eine sehr geringe, optisch nicht erkennbare Auswirkung auf die Lichtkurve. Die beiden Parameter des Hot Spots werden, wie in Kapitel \ref{sec:Parameterstudie} erwähnt mangels guter Datenabdeckung weggelassen. Bei der Temperatur des Begleitsterns wird ein fester Wert von 15500 K (Orosz et al. 2014) angenommen, der aus spektroskopischen Messungen hervorgeht und hier nicht genauer bestimmt werden kann. Ebenfalls bei der daraus hervorgehenden Masse des Begleitsterns von $3.70\pm0.24$ $M_\odot$ und der bestätigten Periode von 1.70481 Tagen (Orosz et al. 2014). Die Masse des Schwarzen Loches ändert die Form der Lichtkurve auch nur minimalst, weswegen sie hier nicht variiert werden braucht und ebenfalls auf den von Orosz et al. stammenden Wert von $6.95\pm0.33$ $M_\odot$ festgelegt wird. In Kapitel \ref{sec:Winkel} wird dieses Vorgehen nochmal anhand eines $\chi^2$-Diagramms legitimiert. Später wird die Masse des Schwarzen Lochs mit der ermittelten Bahnneigung über die Massenfunktion berechnet. Es bleiben also noch zwei zu variierende Parameter übrig, nämlich die Bahnneigung und die Temperatur der Akkretionsscheibe.

\subsubsection{Phasenoffset zwischen simulierter und experimenteller Lichtkurve}
In den simulierten Lichtkurven, ist, wie in den Abbildungen aus Kapitel \ref{sec:Parameterstudie} zu sehen ist, das tiefere Minimum in der Lichtkurve immer genau bei Phase 0.5 platziert. In den experimentellen Lichtkurven ist die Phase erst einmal willkürlich, wird aber so gewählt, dass das tiefe Minimum in etwa bei Phase 0 liegt, da dies auch so bei den neueren Veröffentlichungen gewählt wurde und die Diskussion erleichtert.\\
Zwischen den Datenpunkten aus 2007 und 2013 ist kein Phasenoffset zu erkennen, wie man Abbildung \ref{fig:LKrverschoben} entnehmen kann, daraus folgt, dass der Phasenoffset zwischen experimenteller und simulierter Lichtkurve für beide Epochen gleich sein muss. Trotzdem ist es notwendig die Daten von 2007 und 2013 für eine Analyse getrennt zu betrachten, da ein Magnitudenoffset vorhanden ist. Mit dem Programm fs\_CalcPhaseOffset.py wird ein $\chi^2$-Test zwischen den theoretischen und experimentellen Lichtkurven, welche um eine gewisse Phase verschoben werden, durchgeführt. Es ergibt sich ein Phasenoffset zwischen experimenteller und theoretischer Lichtkurve von $0.572\pm0.016$ im Jahr 2007 und $0.602\pm0.011$ im Jahr 2013. Diese Offsets stimmen innerhalb einer Unsicherheit von $1\sigma$ nicht komplett überein, sollten aber, wie aus vorangegangener Diskussion hervorgeht, identisch sein. Die Unsicherheit von $1\sigma$ wird dabei berechnet, indem der Toleranzbereich der Abweichung des $\chi^2$ um 1 projiziert auf die Phase angegeben wird (vgl. R. Andrae 2010). Wenn beide Lichtkurven übereinander gelegt werden und der Phasenoffset bestimmt wird, erhält man einen Wert von $0.580\pm0.011$. Die Abweichung kommt daher zustande, dass die Abdeckung der Datenpunkte nicht gleichmäßig ist (vgl. Abbildung \ref{fig:LKr}). Im Folgenden wird der Wert 0,582 verwendet, welcher nach optischer Überprüfung des Phasenoffsets am besten passt.

\subsection{Bestimmung des Neigungswinkels und der Scheibentemperatur}\label{sec:Winkel}
\begin{figure}[h!]
\begin{center}
\includegraphics[width=12cm]{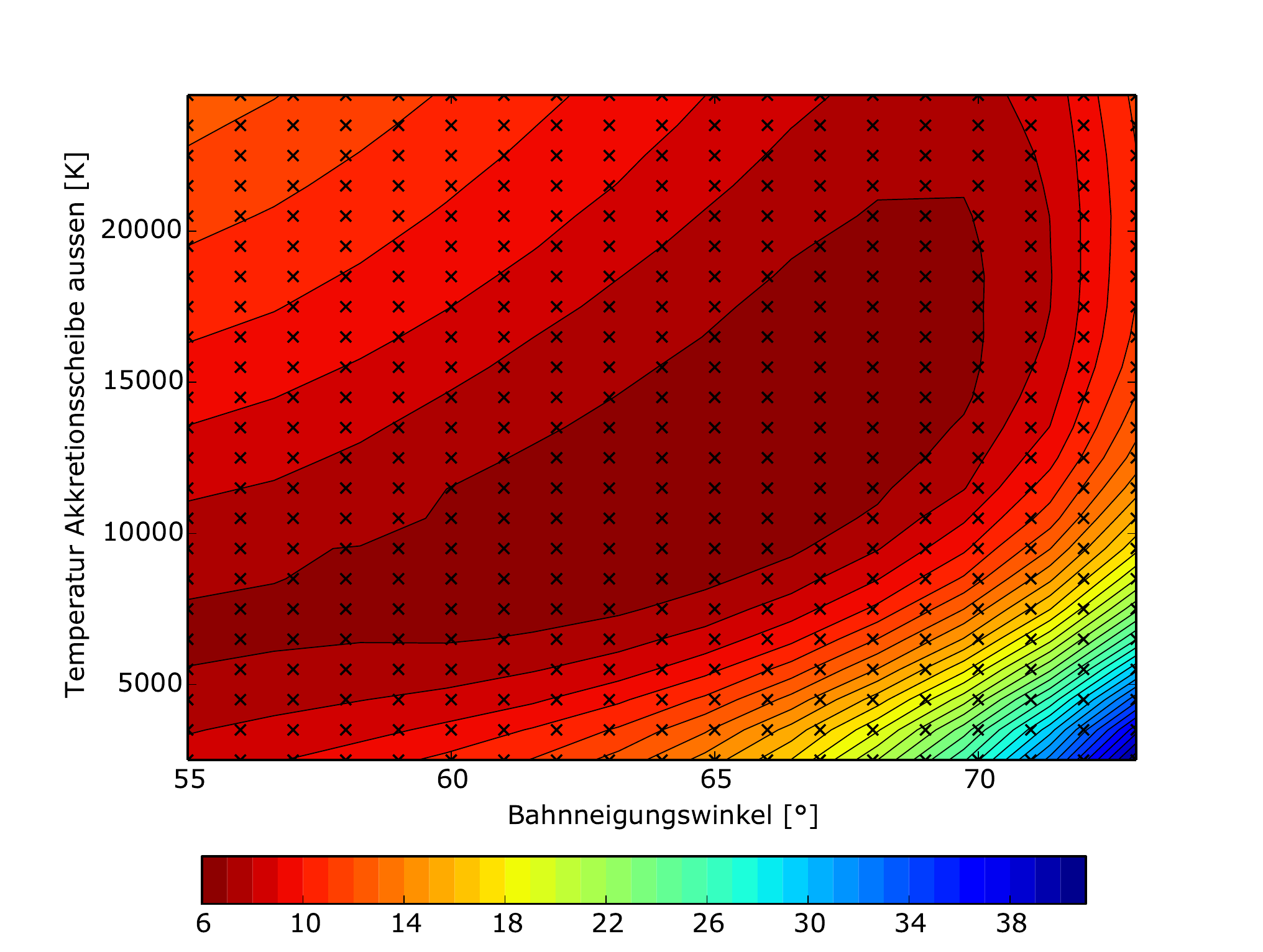}
\caption{$\chi^2$-Verteilung über der Neigungswinkel-Scheibentemperatur-Ebene für die Lichtkurven von 2007}
\label{fig:Chi2007}
\end{center}
\end{figure}

\begin{figure}[h!]
\begin{center}
\includegraphics[width=12cm]{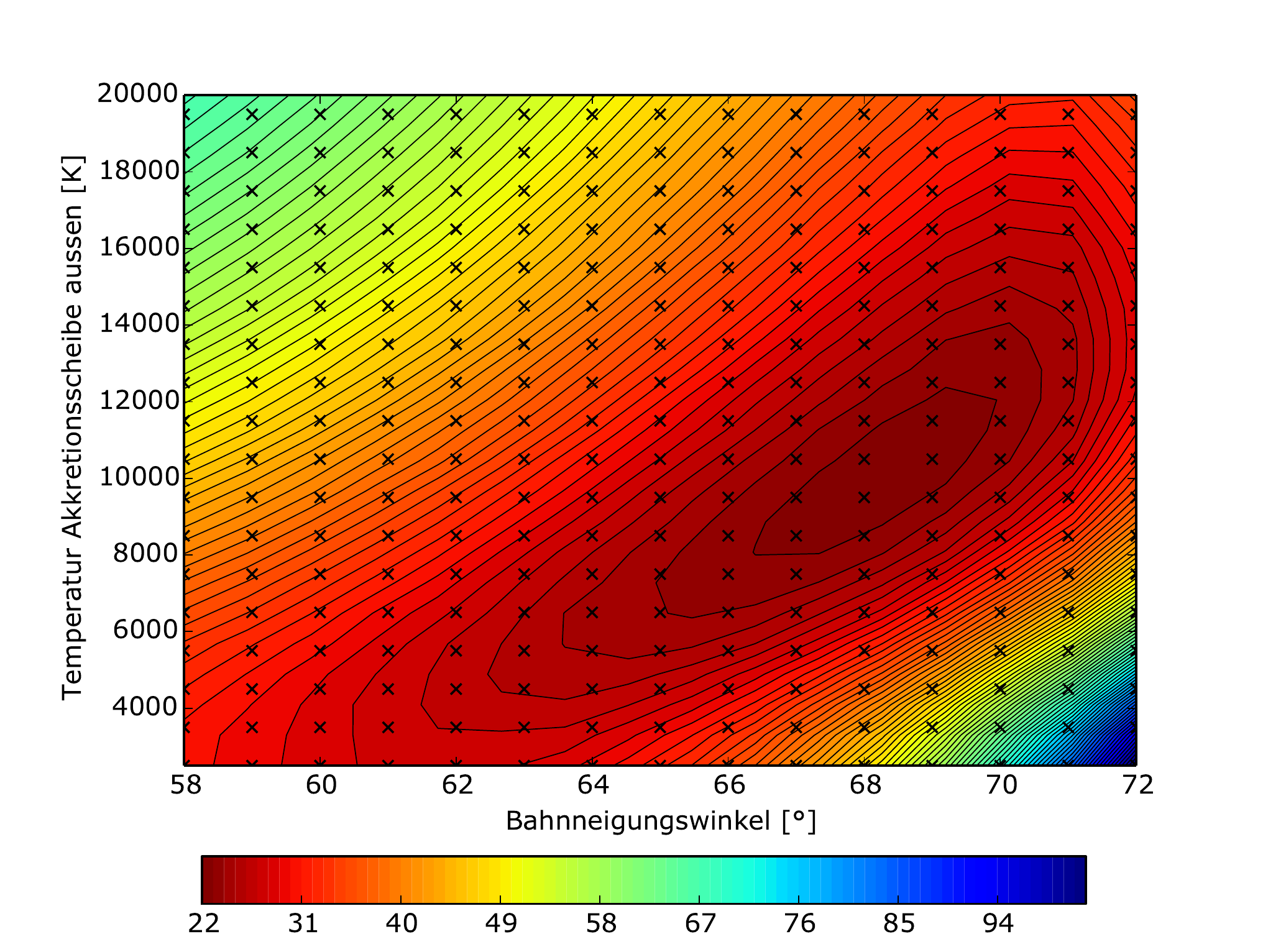}
\caption{$\chi^2$-Verteilung über der Neigungswinkel-Scheibentemperatur-Ebene für die Lichtkurven von 2013}
\label{fig:Chi2013}
\end{center}
\end{figure}

\begin{figure}[h]
\begin{center}
\includegraphics[width=12cm]{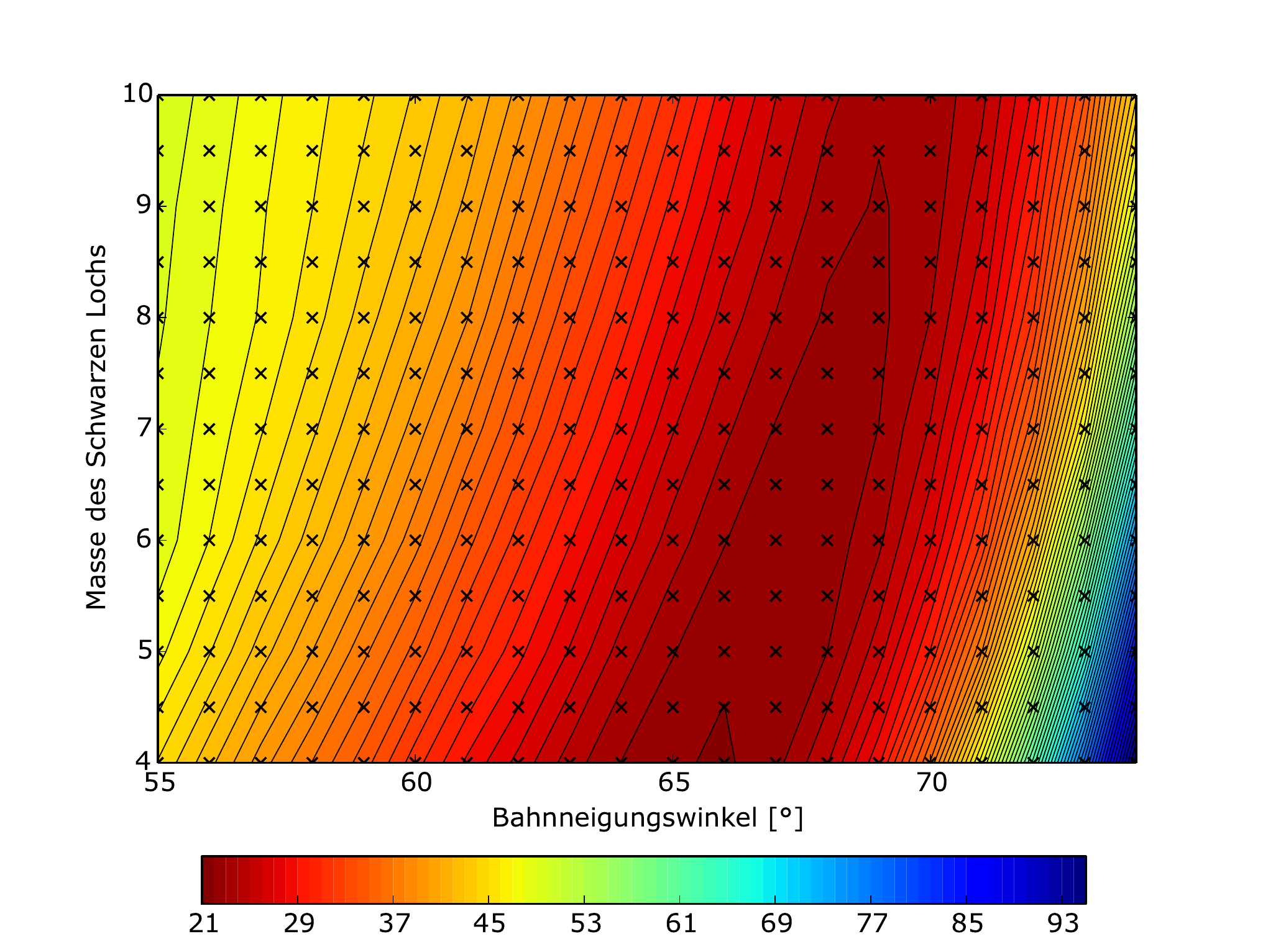}
\caption{$\chi^2$-Verteilung über der Neigungswinkel-Massen-Ebene für die Lichtkurven von 2013}
\label{fig:Massenvariation}
\end{center}
\end{figure}

Für alle simulierten Lichtkurven werden die $\chi^2$-Werte berechnet. Da sowohl die Scheibentemperatur, als auch der Neigungswinkel jeweils starke Auswirkung auf die Form der Lichtkurve haben, werden diese beiden Parameter in einem $\chi^2$ Diagramm aufgetragen, um ein Minimum zu bestimmen. Dies wird getrennt für die Lichtkurven aus den Daten von 2007 bzw. 2013 gemacht und ist in Abbildung \ref{fig:Chi2007} bzw. \ref{fig:Chi2013} zu sehen. Dabei wird für die Berechnung des $\chi^2$ das K-Band auf Grund der großen Streuung weggelassen. Die Kreuze in den Diagrammen geben die Orte an, an welchem das $\chi^2$ exakt berechnet wurde. Für alle anderen Punkte im Raum wurde eine Extrapolation durchgeführt.
Wie aus beiden Abbildungen der $\chi^2$-Verteilungen zu erkennen ist, bilden die Bereiche mit niedrigem $\chi^2$ eine von links unten nach rechts oben aufsteigende Insel. Dies resultiert daraus, dass eine höhere Scheibentemperatur gerade eine größere Bahnneigung kompensiert. Dabei geben die Werte, bei denen das $\chi^2$ um eins abfällt, den Unsicherheitsbereich von $1\sigma$ an.\\ 
Es fällt auf, dass die $\chi^2$ in den Daten von 2007 wesentlich kleiner sind als die von 2013. Dies liegt daran, dass die Unsicherheiten der Datenpunkte im Jahr 2007 größer sind als 2013 und die Streuung geringer ist (vgl. Abbildung \ref{fig:LK2007} und \ref{fig:LK2013}). Ein Resultat daraus ist auch, dass im Plot von 2013 die Verteilung wesentlich schneller abfällt und dadurch die Unsicherheit der Scheibentemperatur und der Bahnneigung wesentlich geringer ist. Der Bereich des kleinsten $\chi^2$ im Jahr 2013 liegt somit vollständig im Bereich von $1\sigma$ aus dem Jahr 2007.\\
Da die Temperatur in den beiden $\chi^2$-Verteilungen die Temperatur des Außenrandes der Akkretionsscheibe darstellt mag es erstaunlich erscheinen, dass die Temperatur des besten Fits um 10000 K liegt, was bedeutet, dass der Innenrand der Scheibe etwa das fünf bis zehnfache beträgt. Dies würde wiederum im Widerspruch zur Annahme einer Akkretionsscheibe stehen, welche als kalt im Vergleich zum Begleitstern aus den SEDs hervorgeht. Da aber keine direkte Aussage über die Helligkeit der Disk getroffen wird, folgt, dass die Scheibe nur eine sehr kleine Leuchtkraft haben kann, da sonst in den SEDs eine Abweichung vom Sternspektrum in den kurzen Wellenlängenbereichen zu sehen sein müsste. Hierauf wird in Kapitel \ref{sec:LKModell} genauer eingegangen.\\
Da keine Präzessionsbewegung des Doppelsternsystems erwartet wird, sollte der reale Bahnneigungswinkel konstant sein. Aus den $\chi^2$-Verteilungen und den Punkten mit dem niedrigsten $\chi^2$ gehen aber zwei verschiedene Neigungswinkel hervor, die aber im Rahmen ihrer Unsicherheit einen Überlapp zeigen. Der, per $\chi^2$-Test bestimmte Winkel für die Daten von 2007 ist $i=66_{-12}^{+4}$, und für 2013 $i=68_{-3}^{+2}$. Es ergibt sich also, dass der Winkelbereich von 2013 im Bereich von 2007 liegt und deshalb in den Graphen zur Massenbestimmung verwendet wird.\\
In Grafik \ref{fig:Massenvariation} sind die $\chi^2$ über der Ebene, welche durch die Masse des Schwarzen Lochs und die Bahnneigung aufgespannt wird, aufgetragen. Hierzu wurden die Datenpunkte von 2013 benutzt. Man erkennt, dass der Neigungswinkel nur schwach mit der Masse des Schwarzen Lochs korreliert, also die Masse des Schwarzen Lochs die Lichtkurve nur schwach beeinflusst. Der $1\sigma$-Bereich der Bahnneigung erstreckt sich von 64° bis 69°, was bis auf 1° auch dem Konfidenzbereich der zuvor bestimmten Bahnneigung entspricht. Es ist also gerechtfertigt, die Bahnneigung als konstant bezüglich Variationen der Masse des Schwarzen Lochs anzunehmen.\\
Im Vergleich zu der vermuteten Bahnneigung von $i=68_{-3}^{+2}$ liegen die Werte von Orosz et al. (2014) für den Fall, dass kein X-Ray heating in ihrem Modell verwendet wird zwischen 64.5° und 71.3° und mit X-Ray heating zwischen 62.6° und 72.1° wobei die Modelle eine Häufung bei 69° zeigen, übereinstimmend mit dem Wert, welcher in dieser Arbeit ermittelt wurde.

\subsection{Modellierte Lichtkurven}\label{sec:LKModell}
\begin{figure}[h]
\center
\subfloat[\label{fig:LK2007}Daten aus 2007]
{\includegraphics[width=8cm]{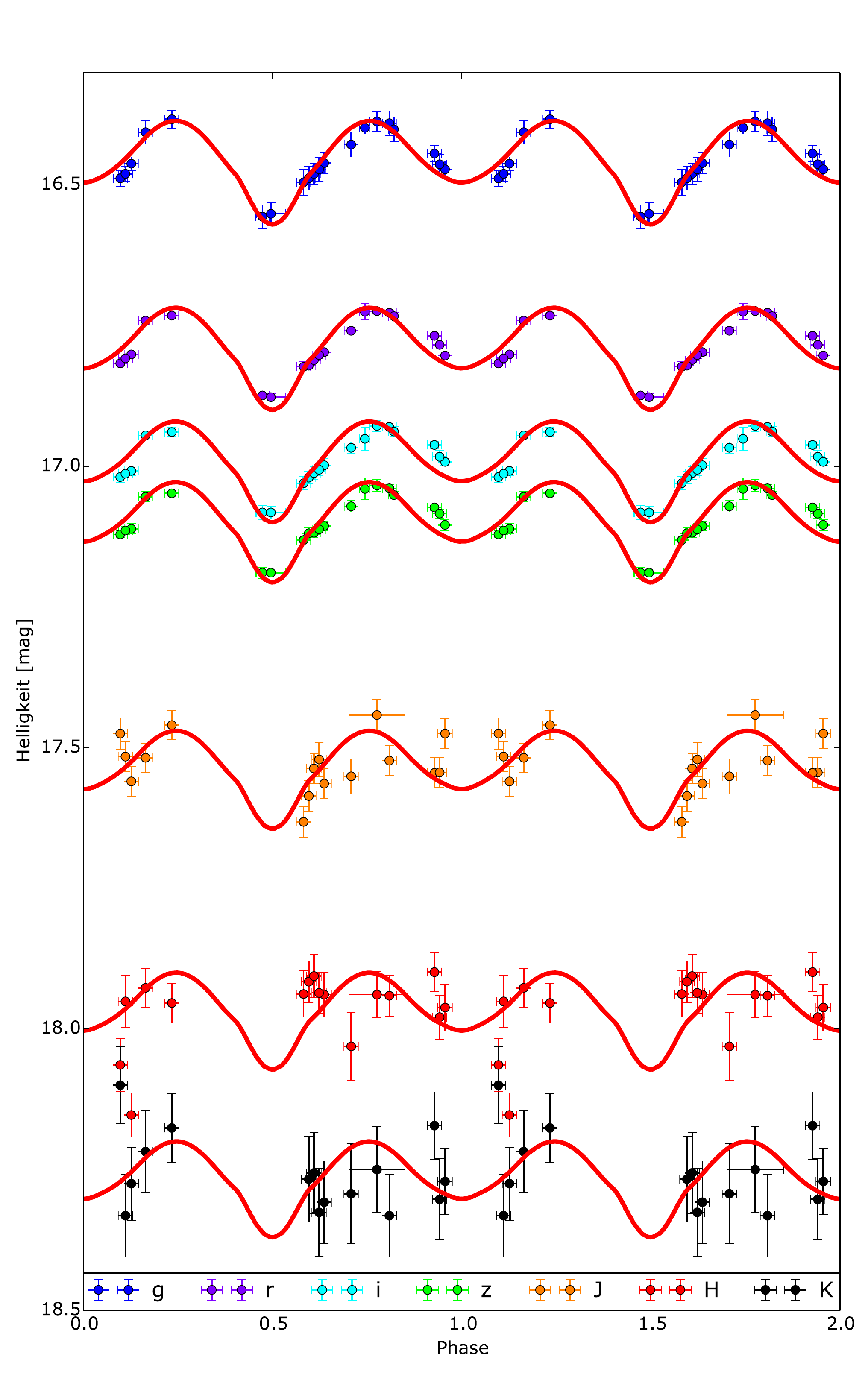}}
\subfloat[\label{fig:LK2013}Daten aus 2013]
{\includegraphics[width=8cm]{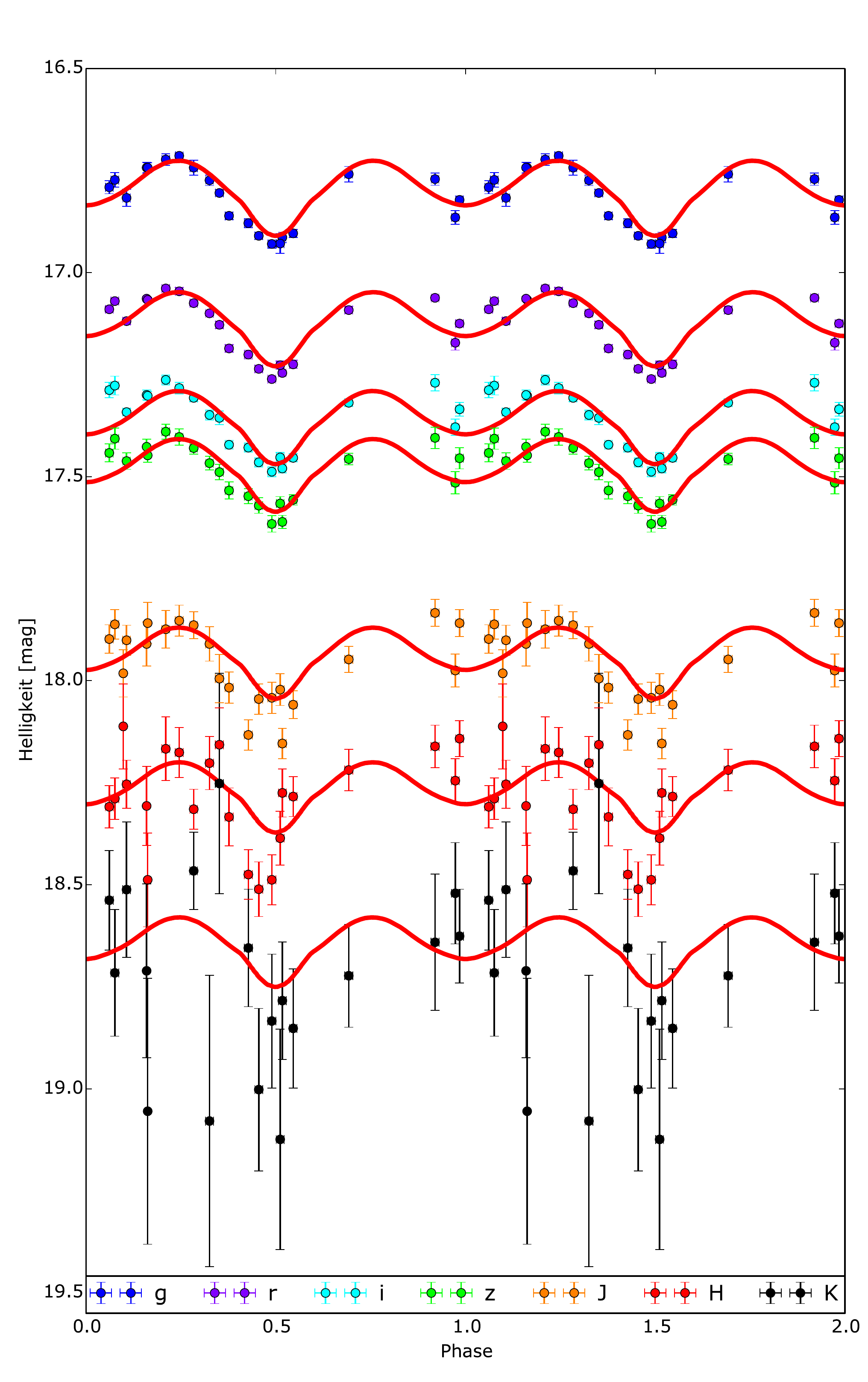}}
\caption{Lichtkurven der Daten von 2007 und 2013 in allen sieben Bändern zusammen mit den jeweils besten Modellen der simulierten Lichtkurven}
\end{figure}

\begin{figure}[h]
\center
\subfloat[\label{fig:2007ohne}Daten aus 2007]
{\includegraphics[width=8cm]{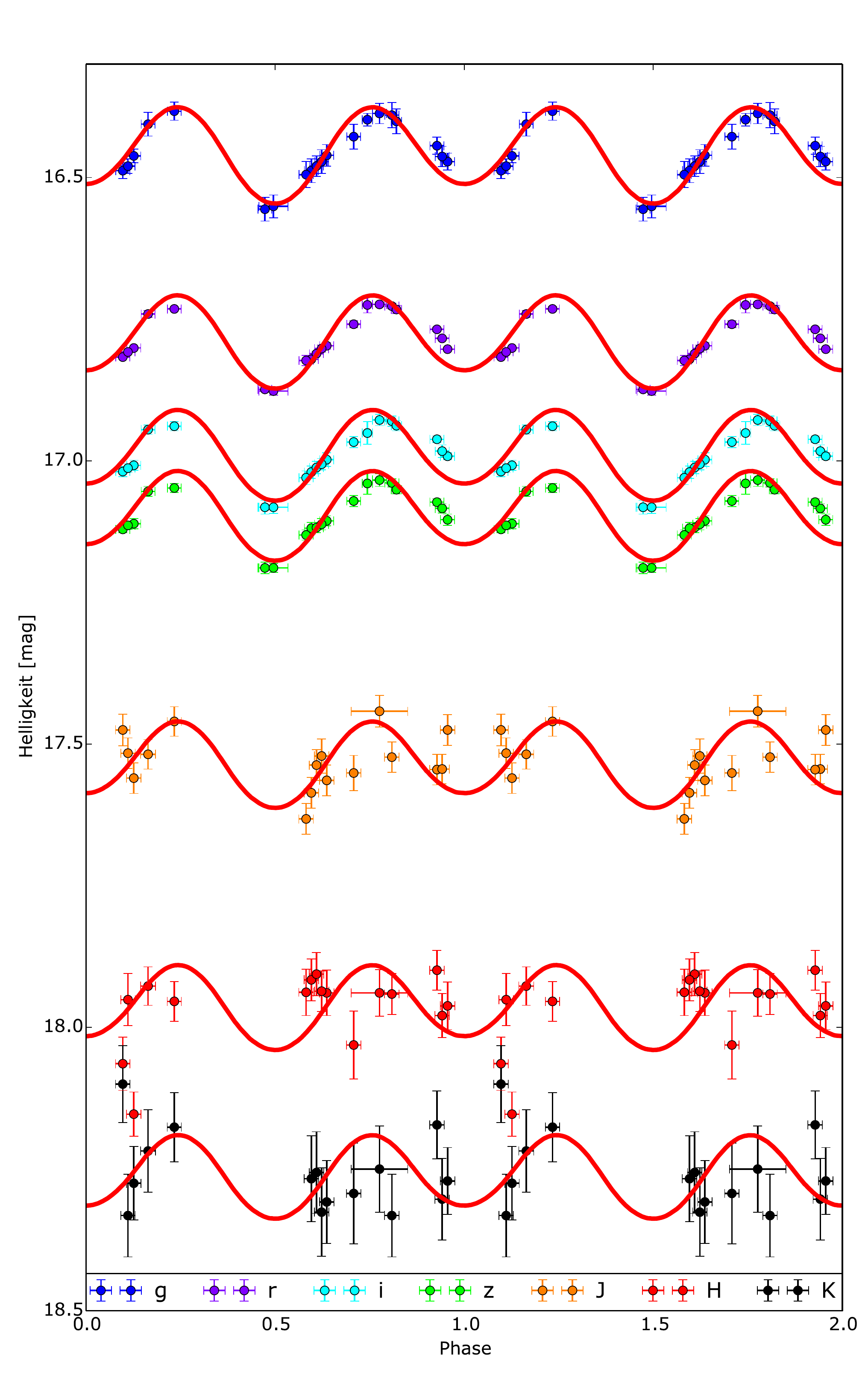}}
\subfloat[\label{fig:2013ohne}Daten aus 2013]
{\includegraphics[width=8cm]{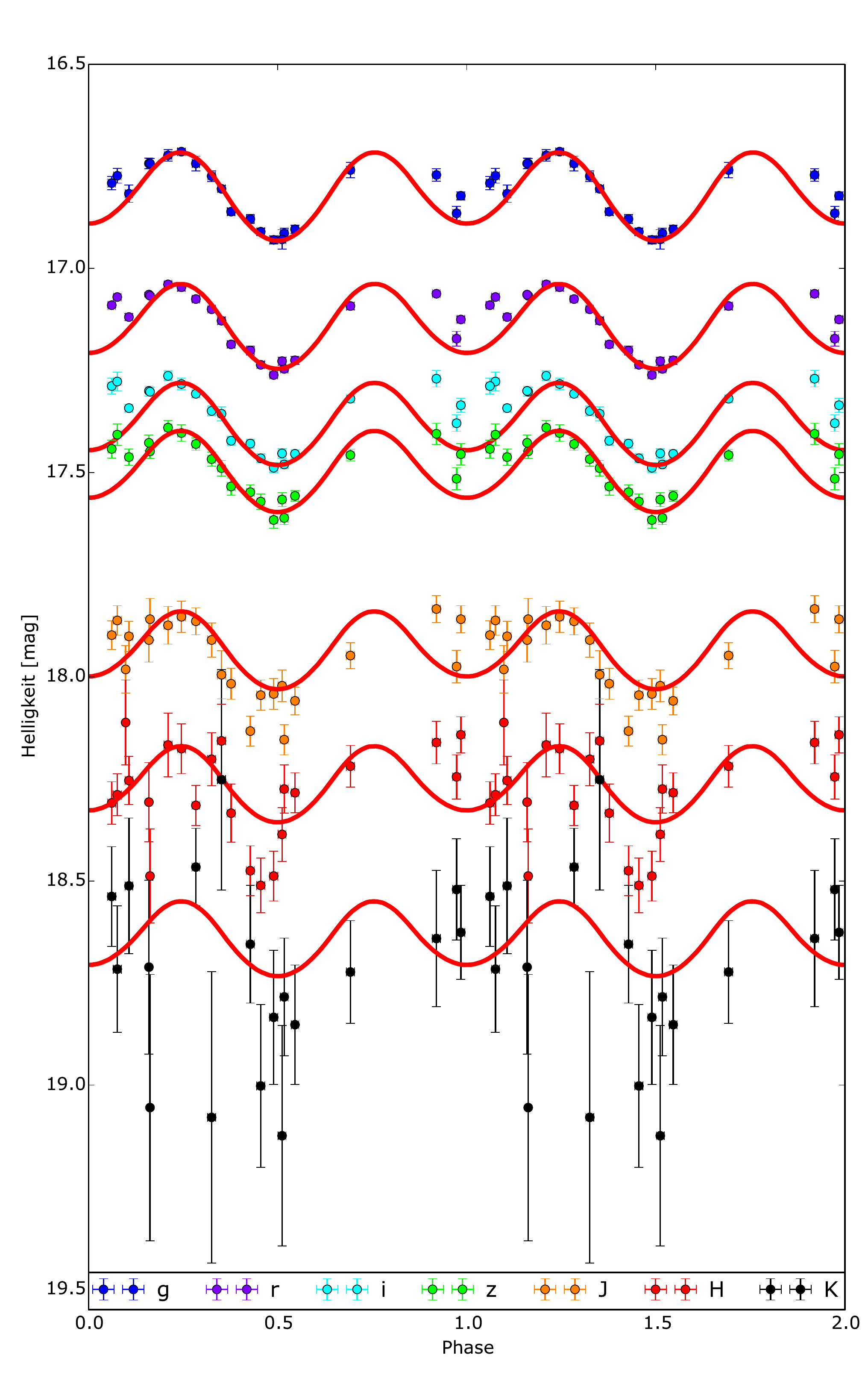}}
\caption{Lichtkurven der Daten von 2007 und 2013 in allen sieben Bändern mit simulierten Lichtkurven ohne Akkretionsscheibe}
\end{figure}

\begin{figure}[h]
\center
\subfloat[\label{fig:LKgNWvar2007}Daten aus 2007]
{\includegraphics[width=8cm]{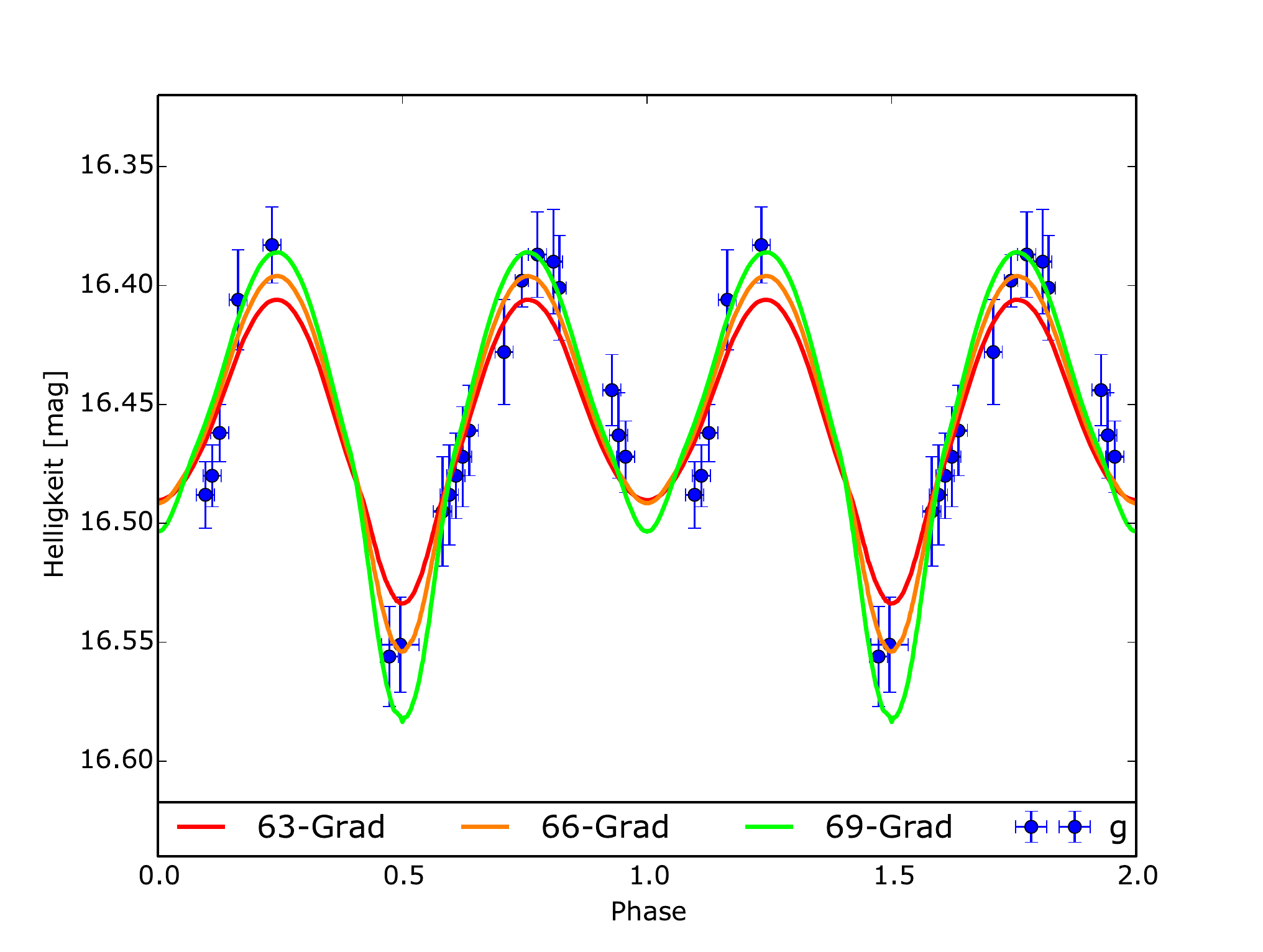}}
\subfloat[\label{fig:LKgNWvar2013}Daten aus 2013]
{\includegraphics[width=8cm]{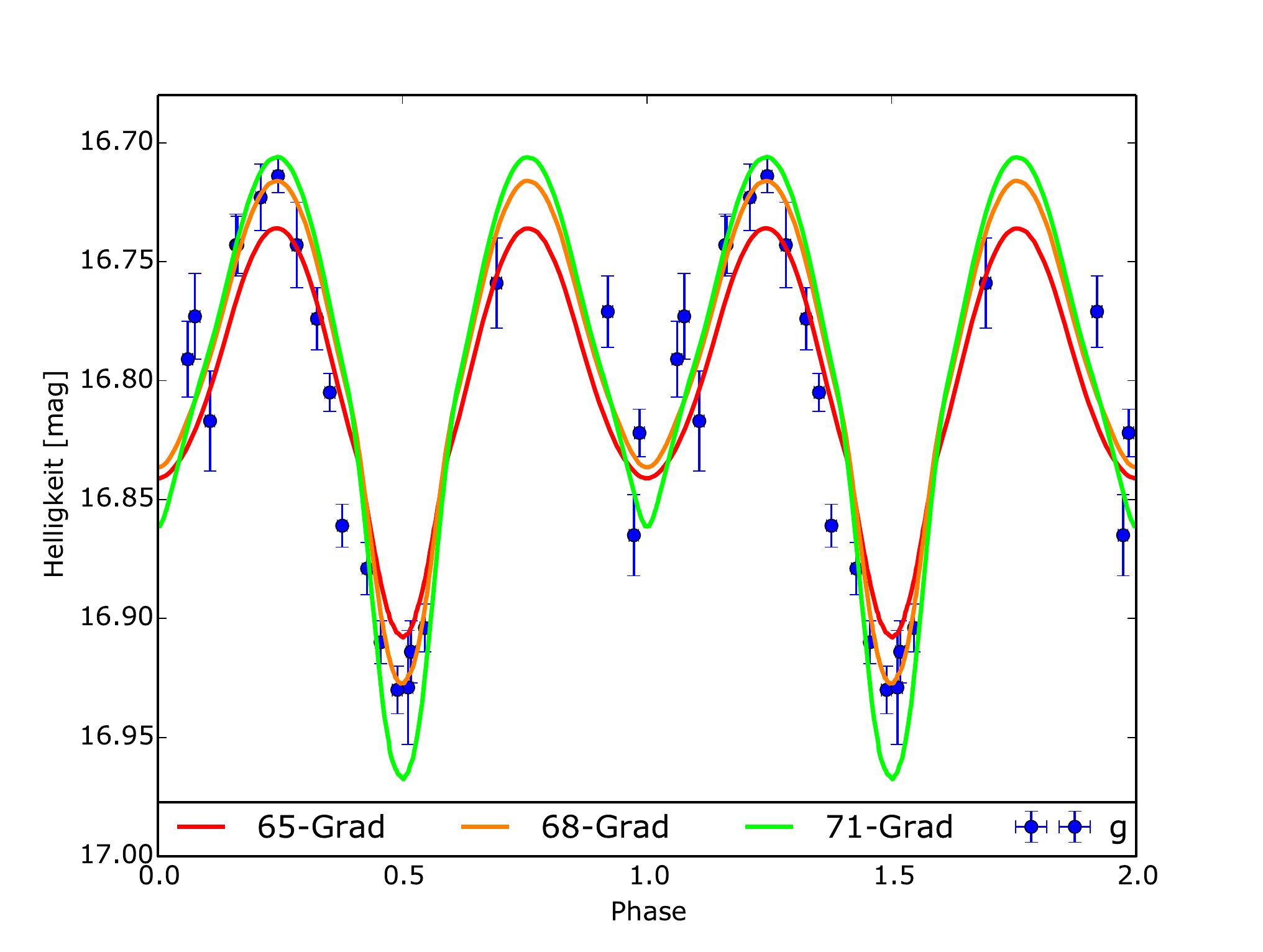}}
\caption{Lichtkurven der Daten von 2007 und 2013 im g-Band zusammen simulierten Lichtkurven}
\end{figure}

Nachdem aus dem Minimum der $\chi^2$-Verteilungen die Parameter für die am besten auf die jeweiligen Daten passende Lichtkurve bekannt sind, können die fertig modellierten Lichtkurven über den experimentellen Datenpunkten aufgetragen werden (siehe Abbildung \ref{fig:LK2007} für die Simulationen an die Daten von 2007 und Abbildung \ref{fig:LK2013} für die Simulationen an die Daten von 2013). Für dargestellten simulierten Lichtkurven aus 2007 beträgt die Bahnneigung 66° und der Außenrand der Scheibe hat eine Temperatur von 11500 K, während für 2013 die Bahnneigung 68° und der Außenrand der Akkretionsscheibe eine Temperatur von 9500 K aufweist.
Beim genauen Betrachten weisen die simulierten Lichtkurven von 2013 gegenüber denen von 2007 ein minimal helleres seichteres Minimum bei Phase 0 auf. Der anfängliche Abfall der Lichtkurve ins tiefe Minimum findet bei den Daten von 2007 dafür leicht steiler statt. Allgemein kann man erkennen, dass die simulierten Lichtkurven nicht optimal auf die Datenpunkte passen. Dies liegt teils an der Streuung der Datenpunkte, wie sie vor allem im seichten Minimum 2013 zu erkennen ist und teils an der fehlenden Abdeckung mit Datenpunkte sowohl 2007 als auch 2013. Dazu kommt noch, dass ein möglicher Hot Spot, der nicht modelliert wurde, einen zusätzlichen kleinen Einfluss auf die Lichtkurve hat. In den Datenpunkten im H und K-Band fallen neben der großen Streuung sofort die großen Unsicherheiten auf, die eine Aussage über eine Variation in der Lichtkurve unmöglich machen.\\
Trotz alledem wurde über den $\chi^2$-Test herausgefunden, dass die Simulationen mit einer nicht stellaren Lichtkomponente, also sprich mit einer Akkretionsscheibe wesentlich besser an die Datenpunkte passen als die Simulationen ohne diese Komponente, was vielleicht nicht sofort aus den Lichtkurven ersichtlich ist.\\
Interessant ist es den besten Fit der beiden Datensätze mit einem Modell zu vergleichen, bei dem die Akkretionsscheibe nicht berücksichtigt wird. Wenn dabei die sonstigen Parameter festgehalten werden kann man den relativen Fluss der Akkretionsscheibe im Vergleich zum Begleitstern berechnen. Daraus ergibt sich eine phasenabhängig erhöhte Strahlungsintensität im Prozentbereich, welche in Tabelle \ref{tab:Strahlung} beispielsweise für das g und K-Band aufgetragen ist, wobei sich die Phasenangaben auf die Abbildungen mit den modellierten Lichtkurven beziehen.
\begin{table}[h!]
\vspace{1cm}
\begin{center}
\begin{tabular}{|c|c|c|c|c|} \hline
\textbf{Phase} & \textbf{g-Band 2013} & \textbf{g-Band 2007} & \textbf{K-Band 2013} & \textbf{K-Band 2007}\\ \hline
$0$	   & $4.8\%$ & $2.9\%$	& $3.6\%$ & $2.4\%$\\
$0.25$	& $1.6\%$ & $0.8\%$	& $1.1\%$ & $0.8\%$\\
$0.5$  	& $3.8\%$ & $1.0\%$	& $2.1\%$ & $0.5\%$\\
$0.75$	& $1.6\%$ & $0.8\%$	& $1.1\%$ & $0.8\%$\\
\hline
\end{tabular}
\caption{Strahlungsintensität der Akkretionsscheibe verglichen mit dem Begleitstern für das g und K-Band und die Epochen 2007 und 2013}
\label{tab:Strahlung}
\end{center}
\end{table}\\
Über alle Phasen ist der Fluss der Scheibe im einstelligen Prozentbereich, trägt also allgemein nur einen kleinen Teil zur Strahlungsintensität bei. Man erkennt, das bandunabhängig der Fluss der Scheibe relativ am größten ist, wenn der Begleitstern sich dahinter befindet. Von den beiden Seiten aus betrachtet ist der Anteil der Scheibe natürlich gleich und jeweils am niedrigsten, da der Stern dort am hellsten strahlt. Auffällig ist auch, dass der Anteil der Scheibe im Jahr 2013 stärker ausgeprägt ist, als 2007, was durch Veränderungen in der Scheibenstruktur, die sich auf der Zeitskala von Tagen abspielen können zurückzuführen ist. Dies ist konsistent mit der Beobachtung, dass die Datenpunkte von 2007 heller sind als die von 2013. Bei Phase 0.5 ist zu sehen, dass im K-Band der Fluss im Gegensatz zum g-Band tiefer als bei 0.25 ist, was darauf zurückzuführen ist, dass der Begleitstern noch die kalten Außenbereiche der Akkretionsscheibe bedeckt.\\
Zum Vergleich der Lichtkurven mit Akkretionsscheibe sind in Abbildung \ref{fig:2007ohne} und \ref{fig:2013ohne} die Lichtkurven ohne Scheibe für eine Bahnneigung von 54° bzw. 68° aufgetragen.
In den Abbildungen \ref{fig:LKgNWvar2007} und \ref{fig:LKgNWvar2013} ist die Lichtkurve des g-Bandes jeweils für 2007 und 2013 gezeigt. Dabei ist die simulierten Lichtkurven aufgetragen, welche am besten fitten und jeweils eine Lichtkurve bei der die Bahnneigung im 3° variiert wurde. Der Parameter der Akkretionsscheibentemperatur am Außenrand wurde bei 2007 auf 11500K und bei 2013 auf 9500K eingestellt, was den am besten fittenden Temperaturen entspricht.
Insgesamt lässt sich sagen, dass die Modelle den vorhandenen Datensatz nicht optimal beschreiben.

\vspace{5cm}

\subsection{Bestimmung der Masse des Schwarzen Lochs}\label{sec:Masse}
\begin{figure}[h]
\center
\subfloat[\label{fig:Massenbestimmung}Ebene der Bahnneigung über der Masse des Schwarzen Lochs. Der überlappende Bereich ist die Region, in der sich die Masse des Schwarzen Lochs befinden muss.]
{\includegraphics[width=7.5cm]{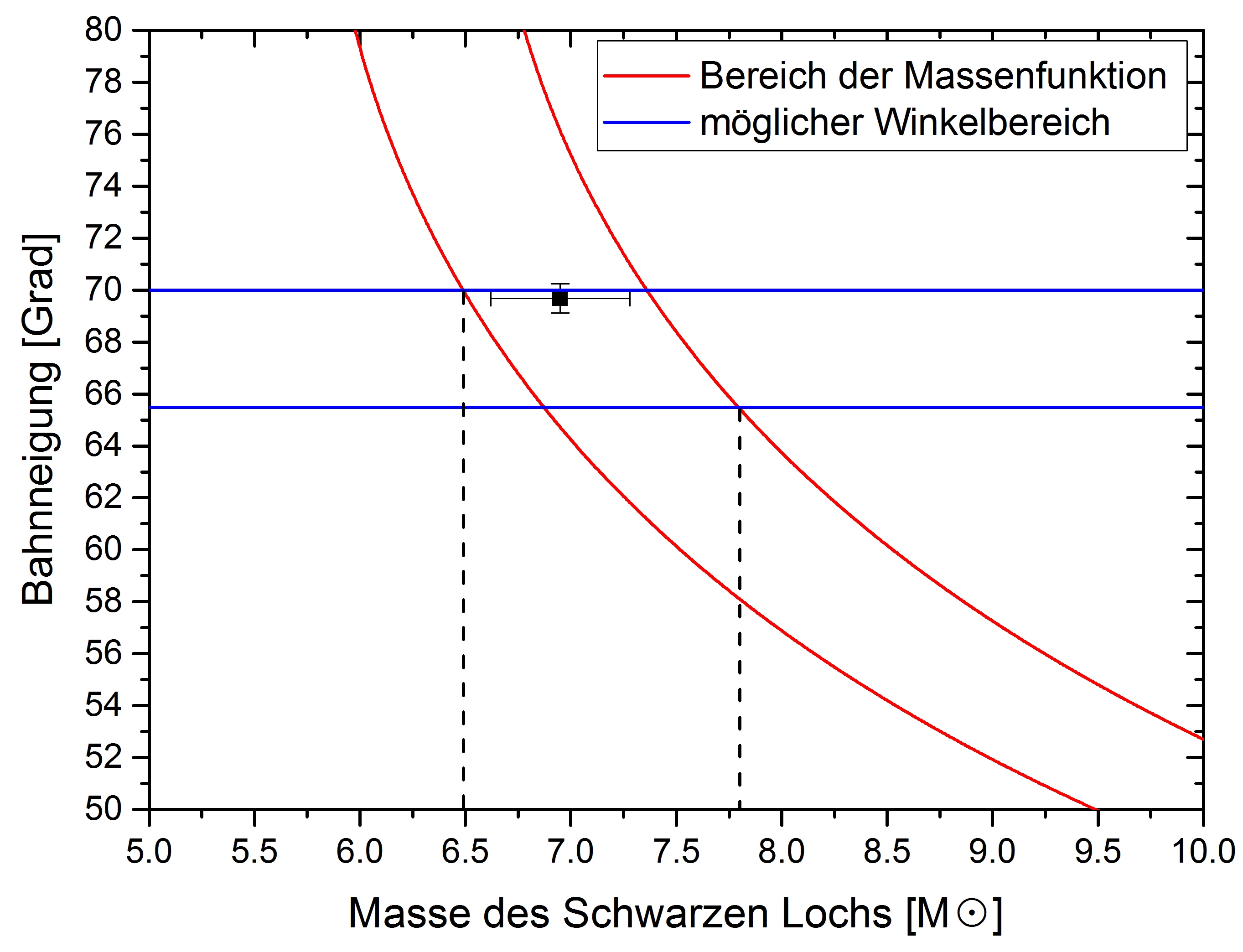}}
\quad
\subfloat[\label{fig:Massenverhaeltnis}Masse des Begleitsterns über Masse des Schwarzen Lochs mit möglichem Winkelbereich, wobei die grünen Linien den Fehlerbereich der Masse des Begleitsterns prozentual auf $6.5\%$ einbeziehen.]
{\includegraphics[width=7.5cm]{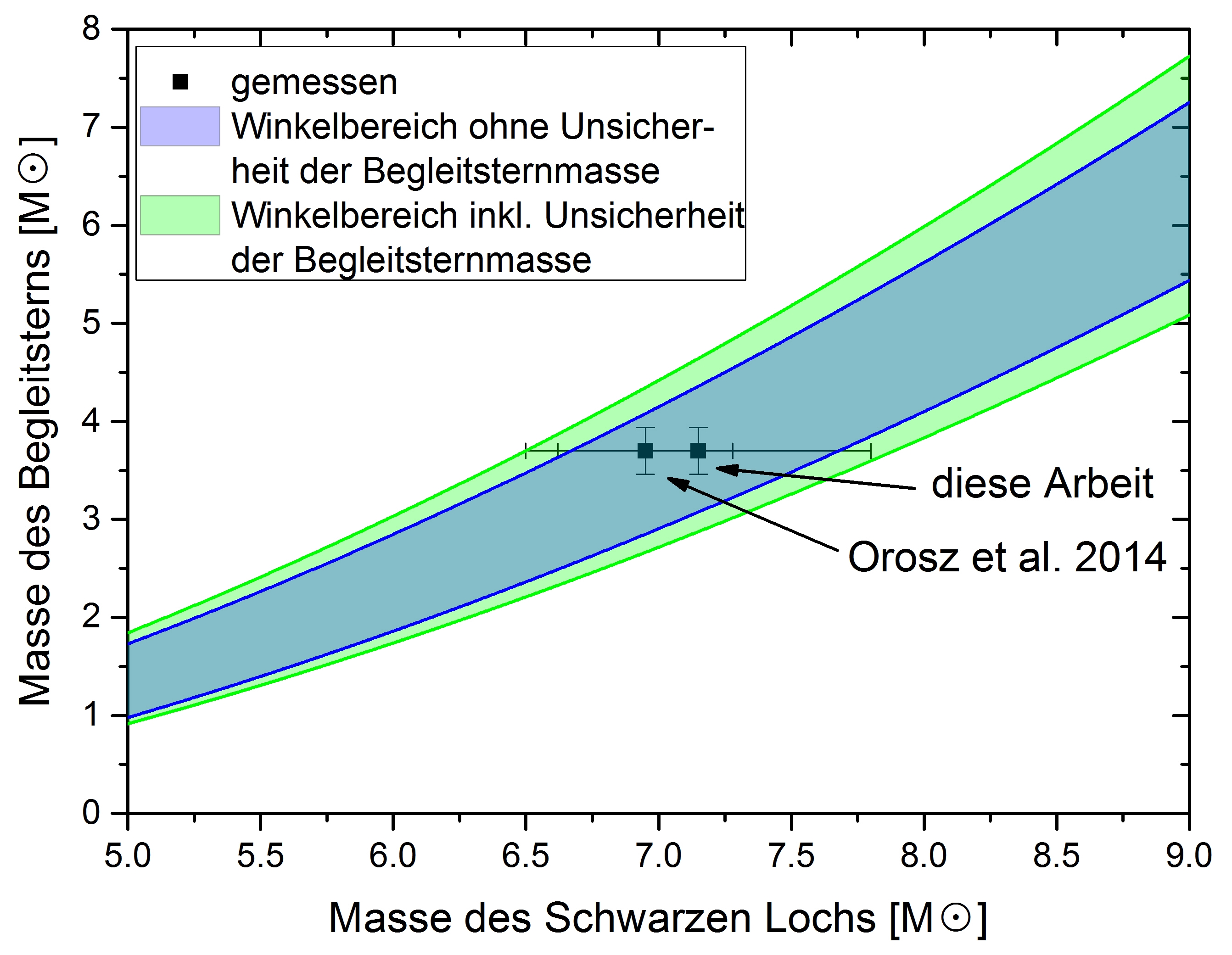}}
\caption{Einordnung der Masse des Schwarzen Lochs in LMC X-3}
\end{figure}

Nachdem der Neigungswinkel vom LMC X-3 bestimmt ist kann über die Massenfunktion die Masse des schwarzen Lochs berechnet werden. Hierzu wird für den Begleitstern eine Bahngeschwindigkeit von $K_S=239.85\pm4.82$ km/s, sowie eine Masse von $3.70\pm0.24$ $M_{\odot}$ angenommen, wobei die Masse aus Oberflächengravitationsmessungen bestimmt wurde, welche genauer sein sollten, als spektrale Messungen bei einem Stern, welcher einen nicht zu unterschlagenden Anteil seiner Masse verloren hat (siehe. Orosz et al. 2014). Da nun die Massenfunktion bis auf zwei Parameter bestimmt ist, kann diese umgestellt werden zu:
\begin{equation}
i=arcsin\left(\frac{P \ K_S^3}{2\pi \ G}\left(1+\frac{M_S}{M_{BH}}\right)^2\frac{1}{M_{BH}}\right)^\frac{1}{3}
\end{equation}
Damit kann nun die Bahnneigung über der Masse des Schwarzen Lochs aufgetragen werden. Dies ist in Abbildung \ref{fig:Massenbestimmung} zu sehen, wo die obige Funktion jeweils für die obere und untere Grenze des Unsicherheitsbereichs aufgetragen ist. Der Bereich zwischen den beiden roten Linien ist der mögliche Bereich für das Tupel Bahnneigung und Masse des Schwarzen Lochs. Da aber aus Kapitel \ref{sec:Winkel} der in Frage kommende Winkelbereich bekannt ist für beide Epochen, kann dieser Winkelbereich ebenfalls in das Diagramm eingezeichnet werden. Dabei kann der Winkel als konstant gegenüber der Masse des Schwarzen Lochs angenommen werden, da diese, wie in der Parameterstudie beschrieben nur sehr geringe Auswirkungen auf die Form der Lichtkurve hat, da durch ihre Variation die Form des Roche-Lobe nur minimal verändert wird. Aus diesem Grund muss die Bahnneigung nicht für jede mögliche Masse des Schwarzen Lochs berechnet werden, sondern kann als konstant angenommen werden (vgl. auch $\chi^2$-Plot in Kapitel \ref{sec:Winkel}).\\
Für die beiden Epochen gibt es laut Kapitel \ref{sec:Winkel} für die Bahnneigung verschiedene Unsicherheitsbereiche. Da aber die Bahnneigung als konstant angenommen werden kann (Vernachlässigung der Präzessionsbewegung) bleibt als relevanter Bereich nur die Region, in der sich beide Winkelbereiche überschneiden. Die Region der aufgetragenen Massenfunktionen (rote Linien) mit diesem relevanten Bahnneigungsbereich gibt projiziert auf die Achse der Masse des Schwarzen Lochs den Bereich an, in der sich die Masse des Schwarzen Lochs befinden muss. Es ergibt sich ein Bereich zwischen 6.5 and 7.8 $M_{\odot}$. Es ergibt sich also im Mittel für die Masse des Schwarzen Lochs $M_{BH}=7.15\pm0.65 \ M_{\odot}$\\
Zum Vergleich ist der Wert der letzten Veröffentlichung aufgetragen (Orosz et al. 2014).\\
Des Weiteren wird nun die Masse des Begleitsterns über der Masse des Schwarzen Lochs aufgetragen (vgl. Abbildung \ref{fig:Massenverhaeltnis}), was schließlich das Massenverhältnis der beiden Hauptkomponenten darstellt. Hierzu wurde die Massenfunktion umgestellt nach:
\begin{equation}
M_S=\left(\frac{2\pi \ G sin^\frac{1}{3} \ i \ M_{BH}^3}{P \ K_S^3}\right)^\frac{1}{2}-M_{BH}
\end{equation}
Es wird jeweils wieder die Massenfunktion für die obere und untere Grenze des Unsicherheitsbereichs für des Winkelbereichs von 2013 aufgetragen und wie zuvor der beste Fit mit dem Wert aus der Veröffentlichung von Orosz et al. 2014.
Da bei der Auftragung der Masse des Begleitsterns über der Masse des Schwarzen Lochs durch die Massenfunktion jeweils einem exakten Wert des Begleitsterns ein Massenbereich für das Schwarze Loch zugeordnet wird, die Masse des Begleitsterns jedoch auch fehlerbehaftet ist, wird ein Prozentualer Fehler von $6.5\%$ in der Masse des Begleitsterns durch einen erweiterten Unsicherheitsbereich dargestellt, welcher auch in der obigen Veröffentlichung benutzt wird.\\
Soria et al. (2001) setzten eine untere Grenze der Masse des Schwarzen Lochs von $(5.8\pm0.6) \ M_{\odot}$ über spektrale Messungen und Bahngeschwindigkeit des Begleitsterns. Bei einer Annahme von $5.9 \ M_{\odot}$ für den B5V Begleitstern, welcher von Val-Baker et al. (2007) angenommen wurde, bestimmten diese einen Massenbereich für das Schwarze Loch in LMC X-3 zwischen $9.5$ und $13.6 \ M_{\odot}$ was mit dem Wert, der in dieser Arbeit ermittelt wurde nicht übereinstimmt. Dazu ist zu sagen, dass zwei verschiedene Methoden zu Grunde liegen, da Val-Baker keinen Neigungswinkel des Doppelsternsystems bestimmt hat. Hingegen verglichen mit der Masse des Schwarzen Lochs in LMC X-3 von $6.95\pm0.33 \ M_{\odot}$ (Orosz et al. 2014) stimmt der hier ermittelte Wert im Rahmen der Unsicherheit überein.

\clearpage

\section{Zusammenfassung und Ausblick}
\begin{figure}[t]
\includegraphics[width=15cm]{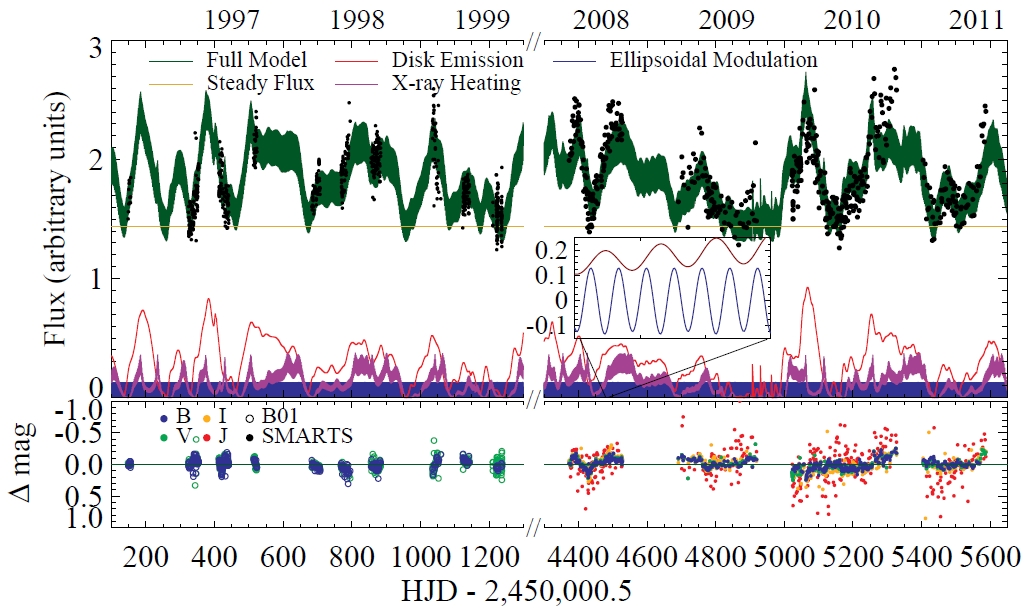}
\caption{Modell von Steiner et al. (2014) bei welchem die Lichtkurve durch drei Komponenten erklärt wird im V-Band. Aufgetragen wird der Fluss über dem Heliozentrischen Julianischen Datum (korrigierte Lichtlaufzeit). Das zusammengesetzte Modell ist in dunkelgrün gezeigt; die Emission der Akkretionsscheibe, welche durch Verwendung der Röntgendaten abgeleitet wurde in rot; die ellipsoidale Veränderliche in blau und die Komponente des X-Ray heatings in lila. Die kleine Box im Bild zeigt eine Vergrößerung. Im unteren Bereich der Abbildung sind die Residuen des Fits aufgetragen.}
\label{fig:Steiner}
\end{figure}

Durch die Anwendung der in dieser Arbeit beschriebenen Methoden ist es möglich, anhand der GROND-Daten die Masse des Schwarzen Lochs in LMC X-3 auf $M_{BH}=(7.15\pm0.65) \ M_{\odot}$ und die Bahnneigung des Röntgen-Doppelsternsystems auf $i=(68_{-3}^{+2})°$ zu bestimmen.\\
Dieses Ergebnis gliedert LMC X-3 im unteren Ende der schmalen Verteilung von Schwarzen Löchern in LMXBs ein, welche ihr Maximum bei $7.8\pm1.2 \ M_{\odot}$ hat (Özel et al. 2010) und einen steilen Abfall beidseitig aufweist, welcher wahrscheinlich mit der speziellen Entwicklung von LMXBs zusammenhängt. Diese Eingliederung in die Klasse von transienten LMXBs ist gerade deshalb interessant, da der Begleitstern von LMC X-3 ein B-Stern ist und Röntgen-Doppelsternsysteme dieser Art vermehrt zu den HMXBs gezählt werden, welche aber überwiegend ununterbrochen eine helle Röntgenintensität aufweisen, deren Begleitsternmasse weitaus größer ist als die von LMC X-3 und welche im Gegensatz zu LMC X-3 nicht durch den Materiefluss über den Roche-Lobe, sondern über stellare Winde akkretieren.\\
Zur Verbesserung der hier dargelegten Studie würde im großen Maße eine gleichmäßige Abdeckung der Lichtkurve beitragen. Somit könnten schon aus der Form der Lichtkurve Aussagen über die Lage eines möglichen Hot Spots gemacht werden. Ebenso liegt eine Verbesserungsmöglichkeit bei der Bestimmung der Temperatur der Akkretionsscheibe und ihrem daraus resultierenden Fluss im Vergleich zum Begleitstern. Nötig hierfür jedoch wäre, um den variablen Massenfluss einzukalkulieren, eine Untersuchung über einen Zeitraum von mehreren Perioden, welche aber gut abgedeckt sein sollten. Bei einer Bahnperiode von 1.7 Tagen, welche größer ist als die Länge einer Nacht auf der Erde wäre dies nur mit nicht-erdgebundenen Teleskopen, Teleskopen, welche jenseits der Polarkreise stehen (für geeignete Himmelskoordinaten), oder mehreren auf der Südhalbkugel verteilten Teleskopen möglich. Eine andere Möglichkeit exaktere Ergebnisse für die Temperatur der Akkretionsscheibe und eines möglichen Hot Spots zu erhalten, wären spektroskopische Messungen in verschiedenen Phasenlagen des Röntgen-Doppelsternsystems.\\
Ebenfalls bleibt zu sagen, dass die angewendeten Modelle nicht optimal die experimentellen Daten wiedergeben. Der $\chi^2$-Test bewirkt im Fall einer Datenabdeckung, wie sie hier vorhanden ist, dass Ausreißer in den Datenpunkten den Fit stark beeinflussen. Es ist gut möglich, das LMC X-3 auf Grund der geringen Leuchtkraft der Akkretionsscheibe auch ohne diese modellieren lässt, oder bei einer sehr geringen Akkretionsscheibentemperatur doch ein Hot Spot teilweise die Datenpunkte besser beschreibt, was zu überprüfen bleibt.\\
Hierbei würde eine Methode helfen, die von Steiner et al. (2014) bei LMC X-3, motiviert durch den ungefähr zwei Wochen betragenden Zeitversatz zwischen den korrelierten Lichtkurven im optischen und Röntgenbereich, durchgeführt wurde. Dabei wird die Emission im Optischen und Infraroten deterministisch als eine Kombination aus einer stellaren Komponente, einer Akkretionsscheibe mit verschiedenen Temperaturbereichen und einer zeitversetzten Leuchtkraft von Stern und Scheibe, die durch das X-Ray heating verursacht wird. Genau hierfür werden die Lichtkurven im Röntgenbereich benutzt, die es ermöglicht die genuine optische Leuchtkraft der Akkretionsscheibe, verursacht durch freigesetzte Gravitationsenergie, vom X-Ray heating zu unterscheiden (vgl. Abbildung \ref{fig:Steiner}).\\

\clearpage

\addcontentsline{toc}{section}{Anhang}
\section*{Anhang}

\subsection*{A Verwendete Sterne für die relative Photometrie}
\begin{table}[h]
\scriptsize
\centering
\subfloat[im OPT; Magnituden im AB-System ohne die systematische Unsicherheit (g-Band: 0.031 mag; r-Band: 0.039 mag; i-Band: 0.037 mag; z-Band: 0.052)]{
\begin{tabular}{c c c c c c c}
 & \textbf{Ra [deg]} & \textbf{Dec [deg]}  & \textbf{g-Band} & \textbf{r-Band} & \textbf{i-Band} & \textbf{z-Band}\\ \hline
A & 84.70721	&	-64.07514	& $	16.426	\pm	0.004	$ & $	15.466	\pm	0.003	$ & $	15.085	\pm	0.004	$ & $	14.856	\pm	0.005	$ \\
B & 84.74778	&	-64.08706	& $	15.791	\pm	0.003	$ & $	15.153	\pm	0.005	$ & $	14.953	\pm	0.006	$ & $	14.822	\pm	0.005	$ \\
C & 84.73855	&	-64.07169	& $	17.671	\pm	0.009	$ & $	16.399	\pm	0.006	$ & $	15.894	\pm	0.006	$ & $	15.596	\pm	0.006	$ \\
D & 84.71953	&	-64.08732	& $	16.260	\pm	0.003	$ & $	15.881	\pm	0.005	$ & $	15.743	\pm	0.006	$ & $	15.652	\pm	0.007	$ \\
E & 84.78411	&	-64.07304	& $	17.549	\pm	0.006	$ & $	16.383	\pm	0.005	$ & $	15.911	\pm	0.007	$ & $	15.651	\pm	0.007	$ \\
F & 84.72884	&	-64.09906	& $	18.912	\pm	0.010	$ & $	17.470	\pm	0.006	$ & $	16.727	\pm	0.006	$ & $	16.377	\pm	0.009	$ \\
G & 84.74434	&	-64.07587	& $	18.445	\pm	0.006	$ & $	17.829	\pm	0.006	$ & $	17.609	\pm	0.010	$ & $	17.481	\pm	0.012	$ \\
H & 84.70577	&	-64.07822	& $	18.497	\pm	0.007	$ & $	17.854	\pm	0.005	$ & $	17.633	\pm	0.007	$ & $	17.478	\pm	0.010	$ \\
I & 84.65493	&	-64.10630	& $	17.374	\pm	0.005	$ & $	15.875	\pm	0.004	$ & $	15.188	\pm	0.004	$ & $	14.848	\pm	0.010	$ \\
J & 84.74423	&	-64.06192	& $	17.505	\pm	0.004	$ & $	17.007	\pm	0.006	$ & $	16.792	\pm	0.005	$ & $	16.678	\pm	0.006	$ \\
\end{tabular}
\label{tab:KalibrationssterneOPT}
}
\qquad

\subfloat[im NIR; Magnituden im Vega-System ohne die systematische Unsicherheit (J-Band: 0.001 mag;  H-Band: 0.001 mag]{
\begin{tabular}{c c c c c c} 
 & \textbf{Ra [deg]} & \textbf{Dec [deg]} & \textbf{J-Band} & \textbf{H-Band} & \textbf{K-Band} \\ \hline 
A & 84.70704	&	-64.07511	& $	13.773	\pm	0.009	$ & $	13.213	\pm	0.010	$ & $	13.088	\pm	0.007	$ \\
B & 84.65464	&	-64.10613	& $	13.582	\pm	0.009	$ & $	12.681	\pm	0.010	$ & $	12.607	\pm	0.004	$ \\
C & 84.65749	&	-64.10772	& $	13.222	\pm	0.009	$ & $	12.790	\pm	0.010	$ & $	12.820	\pm	0.004	$ \\
D & 84.68060	&	-64.11309	& $	13.767	\pm	0.009	$ & $	12.912	\pm	0.010	$ & $	12.864	\pm	0.004	$ \\
E & 84.71151	&	-64.11364	& $	13.222	\pm	0.009	$ & $	12.656	\pm	0.010	$ & $	12.651	\pm	0.004	$ \\
F & 84.83499	&	-64.11481	& $	13.023	\pm	0.009	$ & $	12.540	\pm	0.010	$ & $	12.304	\pm	0.004	$ \\
G & 84.81604	&	-64.07915	& $	14.150	\pm	0.009	$ & $	13.424	\pm	0.010	$ & $	13.194	\pm	0.007	$ \\
H & 84.82324	&	-64.07771	& $	14.298	\pm	0.009	$ & $	13.510	\pm	0.010	$ & $	13.253	\pm	0.007	$ \\
I & 84.78411	&	-64.07300	& $	14.397	\pm	0.009	$ & $	13.655	\pm	0.010	$ & $	13.446	\pm	0.009	$ \\
J & 84.77796	&	-64.06995	& $	13.856	\pm	0.009	$ & $	13.395	\pm	0.010	$ & $	13.289	\pm	0.007	$ \\
\end{tabular}
\label{tab:KalibrationssterneNIR}
}
\caption{Kalibrationssterne für die relative Photometrie}
\end{table}
\vspace*{4cm}

\clearpage

\begin{table}[h]
\subsection*{B Daten der Lichtkurven}
\scriptsize
\hspace*{-1cm}
\begin{tabular}{c c c c c c c c} 
\textbf{MJD} & \textbf{g-Band} & \textbf{r-Band} & \textbf{i-Band} & \textbf{z-Band} & \textbf{J-Band} & \textbf{H-Band} & \textbf{K-Band} \\ \hline 
54393.26137	&	$	16.495	\pm	0.023	$ & $	16.823	\pm	0.009	$ & $	17.030	\pm	0.012	$ & $	17.131	\pm	0.009	$ & $	17.632	\pm	0.027	$ & $	17.938	\pm	0.041	$ & $	18.577	\pm	0.108	$ \\
54393.28479	&	$	16.488	\pm	0.021	$ & $	16.821	\pm	0.003	$ & $	17.020	\pm	0.011	$ & $	17.119	\pm	0.010	$ & $	17.586	\pm	0.027	$ & $	17.916	\pm	0.037	$ & $	18.267	\pm	0.076	$ \\
54393.30806	&	$	16.480	\pm	0.018	$ & $	16.811	\pm	0.004	$ & $	17.012	\pm	0.011	$ & $	17.118	\pm	0.009	$ & $	17.537	\pm	0.027	$ & $	17.906	\pm	0.038	$ & $	18.256	\pm	0.072	$ \\
54393.33124	&	$	16.472	\pm	0.021	$ & $	16.802	\pm	0.006	$ & $	17.006	\pm	0.010	$ & $	17.112	\pm	0.010	$ & $	17.521	\pm	0.030	$ & $	17.936	\pm	0.036	$ & $	18.326	\pm	0.078	$ \\
54393.35426	&	$	16.461	\pm	0.019	$ & $	16.797	\pm	0.006	$ & $	16.998	\pm	0.012	$ & $	17.106	\pm	0.009	$ & $	17.564	\pm	0.027	$ & $	17.939	\pm	0.040	$ & $	18.308	\pm	0.073	$ \\
54394.25376	&	$	16.406	\pm	0.021	$ & $	16.741	\pm	0.006	$ & $	16.945	\pm	0.008	$ & $	17.054	\pm	0.009	$ & $	17.518	\pm	0.026	$ & $	17.927	\pm	0.034	$ & $	18.218	\pm	0.073	$ \\
54394.37204	&	$	16.383	\pm	0.016	$ & $	16.732	\pm	0.004	$ & $	16.939	\pm	0.008	$ & $	17.048	\pm	0.008	$ & $	17.460	\pm	0.026	$ & $	17.954	\pm	0.035	$ & $	18.176	\pm	0.061	$ \\
54395.24235	&	$	16.398	\pm	0.011	$ & $	16.725	\pm	0.014	$ & $	16.951	\pm	0.020	$ & $	17.040	\pm	0.019	$ & $	--	--	--	$ & $	--	--	--	$ & $	--	--	--	$ \\
54395.29702	&	$	16.387	\pm	0.018	$ & $	16.724	\pm	0.004	$ & $	16.928	\pm	0.010	$ & $	17.034	\pm	0.011	$ & $	17.442	\pm	0.028	$ & $	17.939	\pm	0.041	$ & $	18.250	\pm	0.076	$ \\
54395.35258	&	$	16.390	\pm	0.022	$ & $	16.727	\pm	0.005	$ & $	16.930	\pm	0.009	$ & $	17.039	\pm	0.007	$ & $	17.523	\pm	0.027	$ & $	17.941	\pm	0.036	$ & $	18.332	\pm	0.073	$ \\
54395.37328	&	$	16.401	\pm	0.022	$ & $	16.733	\pm	0.007	$ & $	16.938	\pm	0.009	$ & $	17.051	\pm	0.007	$ & $	--	--	--	$ & $	--	--	--	$ & $	--	--	--	$ \\
54397.26078	&	$	16.444	\pm	0.015	$ & $	16.768	\pm	0.004	$ & $	16.962	\pm	0.006	$ & $	17.073	\pm	0.006	$ & $	17.545	\pm	0.027	$ & $	17.899	\pm	0.035	$ & $	18.172	\pm	0.060	$ \\
54397.28420	&	$	16.463	\pm	0.018	$ & $	16.784	\pm	0.003	$ & $	16.983	\pm	0.011	$ & $	17.084	\pm	0.008	$ & $	17.544	\pm	0.026	$ & $	17.979	\pm	0.039	$ & $	18.303	\pm	0.072	$ \\
54397.30866	&	$	16.472	\pm	0.015	$ & $	16.803	\pm	0.002	$ & $	16.992	\pm	0.006	$ & $	17.104	\pm	0.010	$ & $	17.475	\pm	0.027	$ & $	17.962	\pm	0.042	$ & $	18.271	\pm	0.059	$ \\
54398.19009	&	$	16.556	\pm	0.021	$ & $	16.874	\pm	0.004	$ & $	17.082	\pm	0.012	$ & $	17.189	\pm	0.010	$ & $	--	--	--	$ & $	--	--	--	$ & $	--	--	--	$ \\
54398.22854	&	$	16.551	\pm	0.020	$ & $	16.877	\pm	0.007	$ & $	17.082	\pm	0.011	$ & $	17.189	\pm	0.008	$ & $	--	--	--	$ & $	--	--	--	$ & $	--	--	--	$ \\
54399.25418	&	$	16.488	\pm	0.014	$ & $	16.817	\pm	0.003	$ & $	17.019	\pm	0.009	$ & $	17.121	\pm	0.006	$ & $	17.475	\pm	0.028	$ & $	18.064	\pm	0.047	$ & $	18.100	\pm	0.068	$ \\
54399.27759	&	$	16.480	\pm	0.013	$ & $	16.808	\pm	0.002	$ & $	17.013	\pm	0.008	$ & $	17.114	\pm	0.007	$ & $	17.516	\pm	0.027	$ & $	17.951	\pm	0.046	$ & $	18.332	\pm	0.073	$ \\
54399.30380	&	$	16.462	\pm	0.012	$ & $	16.801	\pm	0.002	$ & $	17.008	\pm	0.009	$ & $	17.111	\pm	0.009	$ & $	17.560	\pm	0.027	$ & $	18.153	\pm	0.039	$ & $	18.275	\pm	0.065	$ \\
54400.29506	&	$	16.428	\pm	0.022	$ & $	16.759	\pm	0.006	$ & $	16.967	\pm	0.010	$ & $	17.071	\pm	0.010	$ & $	17.551	\pm	0.031	$ & $	18.031	\pm	0.060	$ & $	18.293	\pm	0.089	$ \\
54484.03602	&	$	16.482	\pm	0.030	$ & $	16.804	\pm	0.019	$ & $	16.994	\pm	0.022	$ & $	17.099	\pm	0.025	$ & $	17.571	\pm	0.034	$ & $	18.002	\pm	0.066	$ & $	18.301	\pm	0.142	$ \\
54485.03634	&	$	16.563	\pm	0.027	$ & $	16.886	\pm	0.013	$ & $	17.087	\pm	0.009	$ & $	17.174	\pm	0.011	$ & $	17.613	\pm	0.034	$ & $	18.084	\pm	0.066	$ & $	18.545	\pm	0.173	$ \\
54486.03581	&	$	16.591	\pm	0.012	$ & $	16.907	\pm	0.012	$ & $	17.090	\pm	0.009	$ & $	17.181	\pm	0.011	$ & $	17.580	\pm	0.035	$ & $	18.060	\pm	0.076	$ & $	19.167	\pm	0.309	$ \\
54487.03463	&	$	16.588	\pm	0.027	$ & $	16.904	\pm	0.011	$ & $	17.098	\pm	0.008	$ & $	17.195	\pm	0.010	$ & $	17.640	\pm	0.034	$ & $	18.235	\pm	0.060	$ & $	18.716	\pm	0.174	$ \\
55979.01565	&	$	16.699	\pm	0.024	$ & $	17.044	\pm	0.012	$ & $	17.294	\pm	0.018	$ & $	17.448	\pm	0.021	$ & $	17.726	\pm	0.051	$ & $	18.053	\pm	0.091	$ & $	18.525	\pm	0.536	$ \\
55979.02090	&	$	16.706	\pm	0.011	$ & $	17.051	\pm	0.013	$ & $	17.300	\pm	0.021	$ & $	17.414	\pm	0.022	$ & $	17.780	\pm	0.052	$ & $	18.052	\pm	0.083	$ & $	18.288	\pm	0.273	$ \\
55979.02612	&	$	16.709	\pm	0.021	$ & $	17.060	\pm	0.016	$ & $	17.301	\pm	0.024	$ & $	17.406	\pm	0.023	$ & $	17.767	\pm	0.052	$ & $	17.974	\pm	0.078	$ & $	19.231	\pm	0.622	$ \\
55979.03117	&	$	16.713	\pm	0.025	$ & $	17.059	\pm	0.029	$ & $	17.281	\pm	0.023	$ & $	17.397	\pm	0.021	$ & $	17.690	\pm	0.051	$ & $	18.064	\pm	0.085	$ & $	18.811	\pm	0.506	$ \\
55979.03619	&	$	16.695	\pm	0.021	$ & $	17.043	\pm	0.017	$ & $	17.278	\pm	0.029	$ & $	17.381	\pm	0.032	$ & $	17.742	\pm	0.046	$ & $	18.274	\pm	0.094	$ & $	18.519	\pm	0.266	$ \\
55981.01832	&	$	16.710	\pm	0.013	$ & $	17.038	\pm	0.008	$ & $	17.253	\pm	0.011	$ & $	17.379	\pm	0.024	$ & $	17.804	\pm	0.048	$ & $	18.020	\pm	0.080	$ & $	18.516	\pm	0.277	$ \\
55981.02333	&	$	16.714	\pm	0.008	$ & $	17.040	\pm	0.007	$ & $	17.273	\pm	0.011	$ & $	17.377	\pm	0.021	$ & $	17.789	\pm	0.048	$ & $	18.017	\pm	0.079	$ & $	18.518	\pm	0.241	$ \\
55981.02833	&	$	16.710	\pm	0.011	$ & $	17.037	\pm	0.004	$ & $	17.266	\pm	0.011	$ & $	17.369	\pm	0.023	$ & $	17.835	\pm	0.050	$ & $	18.151	\pm	0.088	$ & $	18.359	\pm	0.201	$ \\
55981.03334	&	$	16.720	\pm	0.010	$ & $	17.047	\pm	0.005	$ & $	17.272	\pm	0.010	$ & $	17.366	\pm	0.024	$ & $	17.917	\pm	0.055	$ & $	17.920	\pm	0.075	$ & $	18.521	\pm	0.243	$ \\
55981.03832	&	$	16.727	\pm	0.010	$ & $	17.052	\pm	0.006	$ & $	17.291	\pm	0.013	$ & $	17.409	\pm	0.021	$ & $	17.893	\pm	0.051	$ & $	18.057	\pm	0.080	$ & $	18.403	\pm	0.232	$ \\
55983.09088	&	$	16.672	\pm	0.031	$ & $	17.000	\pm	0.039	$ & $	17.241	\pm	0.037	$ & $	17.352	\pm	0.052	$ & $	17.788	\pm	0.038	$ & $	18.157	\pm	0.063	$ & $	18.671	\pm	0.234	$ \\
55983.09614	&	$	16.660	\pm	0.010	$ & $	16.987	\pm	0.011	$ & $	17.235	\pm	0.016	$ & $	17.343	\pm	0.027	$ & $	17.837	\pm	0.037	$ & $	18.087	\pm	0.058	$ & $	18.523	\pm	0.205	$ \\
55983.10140	&	$	16.666	\pm	0.011	$ & $	16.991	\pm	0.013	$ & $	17.224	\pm	0.016	$ & $	17.365	\pm	0.023	$ & $	17.781	\pm	0.037	$ & $	18.097	\pm	0.058	$ & $	18.568	\pm	0.225	$ \\
55984.01775	&	$	16.721	\pm	0.009	$ & $	17.061	\pm	0.006	$ & $	17.298	\pm	0.009	$ & $	17.400	\pm	0.013	$ & $	17.841	\pm	0.042	$ & $	18.133	\pm	0.072	$ & $	18.513	\pm	0.209	$ \\
55984.02319	&	$	16.711	\pm	0.006	$ & $	17.052	\pm	0.006	$ & $	17.278	\pm	0.010	$ & $	17.402	\pm	0.012	$ & $	17.823	\pm	0.038	$ & $	18.139	\pm	0.069	$ & $	18.828	\pm	0.335	$ \\
55984.02860	&	$	16.712	\pm	0.007	$ & $	17.053	\pm	0.005	$ & $	17.288	\pm	0.011	$ & $	17.415	\pm	0.012	$ & $	17.853	\pm	0.039	$ & $	18.091	\pm	0.064	$ & $	18.486	\pm	0.218	$ \\
55984.03398	&	$	16.706	\pm	0.008	$ & $	17.055	\pm	0.006	$ & $	17.289	\pm	0.009	$ & $	17.409	\pm	0.014	$ & $	17.867	\pm	0.035	$ & $	18.092	\pm	0.066	$ & $	18.494	\pm	0.187	$ \\
55984.03933	&	$	16.712	\pm	0.008	$ & $	17.055	\pm	0.005	$ & $	17.295	\pm	0.009	$ & $	17.417	\pm	0.012	$ & $	17.844	\pm	0.039	$ & $	18.276	\pm	0.084	$ & $	18.597	\pm	0.246	$ \\
56393.98839	&	$	16.743	\pm	0.013	$ & $	17.064	\pm	0.006	$ & $	17.300	\pm	0.012	$ & $	17.427	\pm	0.019	$ & $	17.910	\pm	0.054	$ & $	18.307	\pm	0.097	$ & $	18.711	\pm	0.213	$ \\
56393.99321	&	$	16.743	\pm	0.012	$ & $	17.067	\pm	0.006	$ & $	17.302	\pm	0.011	$ & $	17.448	\pm	0.017	$ & $	17.859	\pm	0.051	$ & $	18.488	\pm	0.115	$ & $	19.055	\pm	0.326	$ \\
56623.03304	&	$	16.929	\pm	0.024	$ & $	17.227	\pm	0.010	$ & $	17.453	\pm	0.012	$ & $	17.566	\pm	0.017	$ & $	18.022	\pm	0.039	$ & $	18.386	\pm	0.066	$ & $	19.124	\pm	0.270	$ \\
56623.34114	&	$	16.759	\pm	0.019	$ & $	17.092	\pm	0.009	$ & $	17.319	\pm	0.009	$ & $	17.457	\pm	0.014	$ & $	17.948	\pm	0.032	$ & $	18.219	\pm	0.051	$ & $	18.723	\pm	0.126	$ \\
56624.03286	&	$	--	--	--	$ & $	--	--	--	$ & $	--	--	--	$ & $	--	--	--	$ & $	17.982	\pm	0.058	$ & $	18.112	\pm	0.104	$ & $	--	--	--	$ \\
56624.04754	&	$	16.817	\pm	0.021	$ & $	17.119	\pm	0.008	$ & $	17.342	\pm	0.009	$ & $	17.462	\pm	0.020	$ & $	17.901	\pm	0.037	$ & $	18.254	\pm	0.059	$ & $	18.512	\pm	0.166	$ \\
56624.34930	&	$	16.743	\pm	0.018	$ & $	17.075	\pm	0.006	$ & $	17.307	\pm	0.009	$ & $	17.430	\pm	0.015	$ & $	17.864	\pm	0.033	$ & $	18.315	\pm	0.049	$ & $	18.466	\pm	0.095	$ \\
56631.04303	&	$	16.723	\pm	0.014	$ & $	17.039	\pm	0.007	$ & $	17.263	\pm	0.012	$ & $	17.390	\pm	0.018	$ & $	17.874	\pm	0.046	$ & $	18.167	\pm	0.078	$ & $	--	--	--	$ \\
56631.10349	&	$	16.714	\pm	0.007	$ & $	17.046	\pm	0.009	$ & $	17.283	\pm	0.014	$ & $	17.403	\pm	0.020	$ & $	17.853	\pm	0.038	$ & $	18.176	\pm	0.061	$ & $	--	--	--	$ \\
56631.23948	&	$	16.774	\pm	0.013	$ & $	17.100	\pm	0.006	$ & $	17.349	\pm	0.012	$ & $	17.467	\pm	0.017	$ & $	17.910	\pm	0.042	$ & $	18.202	\pm	0.065	$ & $	19.079	\pm	0.357	$ \\
56631.28399	&	$	16.805	\pm	0.008	$ & $	17.128	\pm	0.009	$ & $	17.356	\pm	0.017	$ & $	17.489	\pm	0.019	$ & $	17.995	\pm	0.059	$ & $	18.157	\pm	0.090	$ & $	18.252	\pm	0.270	$ \\
56632.34372	&	$	16.865	\pm	0.017	$ & $	17.172	\pm	0.018	$ & $	17.379	\pm	0.020	$ & $	17.515	\pm	0.027	$ & $	17.975	\pm	0.040	$ & $	18.245	\pm	0.054	$ & $	18.521	\pm	0.124	$ \\
56633.03212	&	$	16.861	\pm	0.009	$ & $	17.186	\pm	0.009	$ & $	17.422	\pm	0.011	$ & $	17.534	\pm	0.021	$ & $	18.017	\pm	0.038	$ & $	18.334	\pm	0.071	$ & $	--	--	--	$ \\
56633.11879	&	$	16.879	\pm	0.011	$ & $	17.201	\pm	0.011	$ & $	17.429	\pm	0.011	$ & $	17.548	\pm	0.018	$ & $	18.133	\pm	0.037	$ & $	18.475	\pm	0.061	$ & $	18.655	\pm	0.144	$ \\
56633.16571	&	$	16.910	\pm	0.009	$ & $	17.236	\pm	0.008	$ & $	17.465	\pm	0.011	$ & $	17.571	\pm	0.019	$ & $	18.045	\pm	0.037	$ & $	18.511	\pm	0.067	$ & $	19.002	\pm	0.199	$ \\
56633.22423	&	$	16.930	\pm	0.010	$ & $	17.261	\pm	0.006	$ & $	17.488	\pm	0.012	$ & $	17.616	\pm	0.020	$ & $	18.042	\pm	0.038	$ & $	18.488	\pm	0.061	$ & $	18.834	\pm	0.164	$ \\
56633.27159	&	$	16.914	\pm	0.013	$ & $	17.246	\pm	0.006	$ & $	17.480	\pm	0.011	$ & $	17.611	\pm	0.016	$ & $	18.154	\pm	0.037	$ & $	18.275	\pm	0.059	$ & $	18.784	\pm	0.144	$ \\
56633.32025	&	$	16.904	\pm	0.010	$ & $	17.225	\pm	0.010	$ & $	17.454	\pm	0.010	$ & $	17.557	\pm	0.013	$ & $	18.059	\pm	0.034	$ & $	18.284	\pm	0.049	$ & $	18.852	\pm	0.146	$ \\
56639.07203	&	$	16.771	\pm	0.015	$ & $	17.062	\pm	0.006	$ & $	17.270	\pm	0.020	$ & $	17.405	\pm	0.026	$ & $	17.834	\pm	0.033	$ & $	18.161	\pm	0.052	$ & $	18.641	\pm	0.167	$ \\
56639.18229	&	$	16.822	\pm	0.010	$ & $	17.125	\pm	0.007	$ & $	17.335	\pm	0.017	$ & $	17.455	\pm	0.026	$ & $	17.859	\pm	0.033	$ & $	18.142	\pm	0.044	$ & $	18.626	\pm	0.115	$ \\
56639.31330	&	$	16.791	\pm	0.016	$ & $	17.090	\pm	0.006	$ & $	17.288	\pm	0.019	$ & $	17.442	\pm	0.022	$ & $	17.898	\pm	0.035	$ & $	18.309	\pm	0.052	$ & $	18.538	\pm	0.122	$ \\
56639.33821	&	$	16.773	\pm	0.018	$ & $	17.070	\pm	0.007	$ & $	17.277	\pm	0.023	$ & $	17.407	\pm	0.026	$ & $	17.862	\pm	0.036	$ & $	18.289	\pm	0.051	$ & $	18.716	\pm	0.155	$ \\
\end{tabular}
\caption{Daten der Lichtkurven mit den Helligkeiten in Magnituden im AB-System und statistischer Unsicherheit}
\label{tab:Daten}
\end{table}

\clearpage

\listoffigures

\clearpage

\section*{Abschließende Worte}
Im Anschluss an diese Bachelor-Arbeit möchte ich an dieser Stelle meinem Themensteller Dr. Jochen Greiner danken, dass er mir ermöglichte ein solch interessantes Thema zu bearbeiten. Ebenso bin ich auch meinem Betreuer Arne Rau zu Dank verpflichtet. Die beiden nahmen sich stets außerordentlich viel Zeit für mich und waren immer offen für auch längere Diskussion.\\
Ein großes Dankeschön auch an Fabian Knust, der mir vor allem mit den XRbinary-Simulationen half.\\
Vielen Dank ebenso an Jan Bolmer, Jonny Elliott und Tassilo Schweyer die mir oft bei Computer- und Programmierproblemen unter die Arme griffen.\\
Insbesondere gilt mein Dank allen Kollegen des MPE für die freundliche Aufnahme und die wunderbare Atmosphäre während der Zeit meiner Bachelor-Arbeit.\\
Vor allem aber danke ich meinen Eltern, die mir es überhaupt ermöglicht haben zu studieren.

\end{document}